\documentclass[prd,aps,twocolumn,showpacs,tightenlines,floatfix]{revtex4}

\usepackage{graphicx}

\def\sumint{\hbox{$\sum$}\!\!\!\!\!\!\!\int}
\def\square{\vcenter{\vbox{\hrule height.4pt
          \hbox{\vrule width.4pt height8pt
          \kern8pt\vrule width.4pt}\hrule height.4pt}}}

\def\ranglec{\rangle_{\!\!c}}
\def\ranglex{\rangle_{\!\!x}}
\def\ranglecx{\rangle_{\!\!c,x}}
\newcommand{\beq}{\begin{equation}}
\newcommand{\eeq}{\end{equation}}
\newcommand{\bqa}{\begin{eqnarray}}
\newcommand{\eqa}{\end{eqnarray}}

\begin{document}

\title{Two-loop HTL Thermodynamics with Quarks}

\preprint{ DUKE-TH-02-221, TUW-03-06 }

\author{Jens O. Andersen}
\affiliation{Institute for Theoretical Physics, University of Utrecht,
       Leuvenlaan 4, 3584 CE Utrecht, The Netherlands}
\altaffiliation{Present address:  Nordita, Blegdamsvej 17, DK-2100 Copenhagen, Denmark}

\author{Emmanuel Petitgirard}
\affiliation{Physics Department, Ohio State University, Columbus OH 43210, USA}

\author{Michael Strickland}
\affiliation{Institut f\"ur Theoretische Physik, Technische Universit\"at Wien,
	Wiedner Hauptstrasse 8-10, A-1040 Vienna, Austria}
\affiliation{Physics Department, Duke University, Durham, NC 27701} 

\begin{abstract}
We calculate the quark contribution to the free energy of a hot quark-gluon plasma to
two-loop order using hard-thermal-loop (HTL) perturbation theory.  All ultraviolet divergences
can be absorbed into renormalizations of the vacuum energy and the HTL quark and gluon mass parameters.  The
quark and gluon HTL mass parameters are determined self-consistently by a variational
prescription.  Combining the quark contribution with the two-loop HTL perturbation theory
free energy for pure-glue we obtain the total two-loop QCD free energy.  Comparisons are
made with lattice estimates of the free energy for $N_f=2$ and with exact numerical results
obtained in the large-$N_f$ limit.
\end{abstract}
\maketitle
\newpage

\small

\section{Introduction}

The current generation of relativistic heavy-ion collision experiments
should exceed the energy density necessary for the formation of a
quark-gluon plasma.  It is therefore necessary to have a quantitative theoretical
framework which can be used to calculate the properties of a quark-gluon
plasma.  The usual line of reasoning is that since QCD is asymptotically 
free, its running coupling constant $\alpha_s$ becomes weaker as the 
temperature increases and therefore the behavior of hadronic matter at sufficiently 
high temperature should be calculable using perturbative methods.  Unfortunately, 
a straightforward perturbative expansion in powers of $\alpha_s$ does not seem 
to be of any quantitative use even at temperatures many orders of magnitude 
higher than those achievable in heavy-ion collisions.

The problem can be seen by looking at the perturbative expansion of 
the free energy ${\cal F}$ of a quark-gluon
plasma, whose weak-coupling expansion has been calculated completely 
through order $\alpha_s^{5/2}$ \cite{AZ-95,KZ-96,BN-96}
\bqa
{\cal F} &=& - {8 \pi^2\over45} T^4 \,
\left[ {\cal F}_0
\,+\, {\cal F}_2  {\alpha_s \over \pi}
\,+\, {\cal F}_3  \left( {\alpha_s\over \pi} \right)^{3/2}
\right.
\nonumber \\
&& \left.  \hspace{-1cm}
\,+\, {\cal F}_4  \left( {\alpha_s \over \pi} \right)^2
\,+\, {\cal F}_5  \left( {\alpha_s \over \pi} \right)^{5/2}
\,+\, O(\alpha_s^3 \log \alpha_s) \right] \,,
\label{freeg}
\eqa
%
%{\bf [jmp]}
with
\bqa
{\cal F}_0 &=& 1 + \textstyle{21\over 32}N_f  \, ,
\\
\label{correct}
{\cal F}_2 &=& - {15 \over 4} \left( 1 + \textstyle{5\over 12}N_f  \right)\;,
\\
{\cal F}_3 &=& 30 \left( 1 + \textstyle{1\over 6}N_f \right)^{3/2} \, ,
\\
{\cal F}_4 &=&  237.2 + 15.97 N_f - 0.413 N_f^2 \nonumber
\\ && \hspace{2mm} \nonumber
+ { 135 \over 2} \left( 1 + \textstyle{1\over 6}N_f  \right)
        \log \left[ {\alpha_s \over \pi}
        \left(1 + \textstyle{1\over 6}N_f \right) \right] 
\\ && \hspace{2mm}
-{165\over 8} \left(1+{5\over12} N_f\right)\left(1 -{2\over33} N_f\right)\log{\mu\over 2\pi T}
	\; ,
\\ \nonumber
{\cal F}_5 &=& -\left( 1 + \textstyle{1 \over 6}N_f\right)^{1/2}
	\Bigg[ 799.2 + 21.96 N_f + 1.926 N_f^2\Bigg] 
\\ && \hspace{2mm}
+{495\over 2} \left(1+{1\over6} N_f\right)\left(1 -{2\over33} N_f\right)\log{\mu\over 2\pi T}
\; ,
\eqa
%{\bf [jmp]}
where $\mu$ is the renormalization scale, $\alpha_s=\alpha_s(\mu)$ is the running
coupling constant in the $\overline{\mbox{MS}}$ scheme, and we have set $N_c=3$.
The coefficient of $\alpha_s^3 \log \alpha_s$ has recently
been computed \cite{KLRS-02}; however, since there are unknown 
perturbative and non-perturbative contributions at $O(\alpha_s^3)$ 
we do not include terms higher than $O(\alpha_s^{5/2})$ 
in Eq.~(\ref{freeg}).

\begin{figure}[t]
\includegraphics[width=8.5cm]{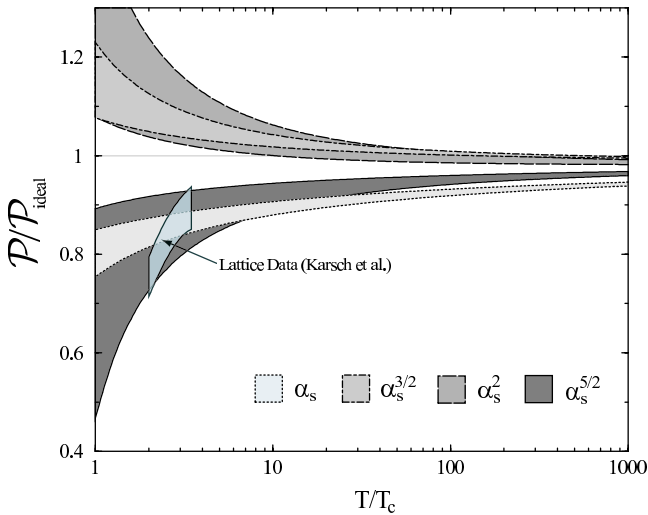}
\caption[a]{The perturbative free energy of QCD for $N_f=2$ massless quarks 
as a function of $T/T_c$.
The weak-coupling expansions through orders $\alpha_s$,
$\alpha_s^{3/2}$, $\alpha_s^2$, and $\alpha_s^{5/2}$
are shown as bands that correspond to varying
the renormalization scale $\mu$ by a factor of two around $2\pi T$.
Also shown is a lattice estimate by Karsch et al.~\cite{klp} for
the free energy.  The band indicates the estimated systematic
error of their result which is reported as (15$\pm$5)\%.  
}
\label{weakfig}
\end{figure}

In Fig.~\ref{weakfig}, the free energy with $N_f=2$ is shown as a
function of the temperature $T/T_c$,
 where $T_c$ is the critical temperature
for the deconfinement transition.  In the plot we have scaled the
free energy by the free energy of an ideal gas of quarks and gluons 
which for arbitrary $N_c$ and $N_f$ is
\beq
{\cal F}_{\rm ideal} = -{\pi^2 \over 45} T^4 
  \left( N_c^2-1 + {7\over 4} N_c N_f \right) \, .
\label{fideal}
\eeq
%{\bf [jmp]}
%
The weak-coupling expansions through
orders $\alpha_s$, $\alpha_s^{3/2}$, $\alpha_s^2$, and $\alpha_s^{5/2}$
are shown as bands that correspond to varying the renormalization scale,
$\mu$, by a factor of two around the central value $\mu=2\pi T$.
As successive terms in the weak-coupling expansion
are added, the predictions change wildly and the sensitivity to the
renormalization scale grows.
It is clear that a reorganization of the perturbation series
is essential if perturbative calculations are to be of any quantitative
use at temperatures accessible in heavy-ion collisions.

%%%%%%%%%%%%%%%%%%%%%%%%%%%%%%%%%%%%%%%%%%%%%%%%%%%%%%%%%%%%%%%%%%%%%%%
%%%%%%%%%%%%%%%%%%%%%%%%%%%%%%%%%%%%%%%%%%%%%%%%%%%%%%%%%%%%%%%%%%%%%%%

The free energy can also be calculated nonperturbatively using
lattice gauge theory \cite{Karsch}.
The thermodynamic functions for pure-glue QCD have been calculated with
high precision by Boyd et al.~\cite{lattice-0}. There have also been
calculations which include dynamical quarks~\cite{klp,lattice-Nf}.  In Fig.~1 we
have included the latest lattice estimate of Karsch et al.~\cite{klp} for
the free energy for $N_f=2$ flavors of light quarks.  The band indicates 
the estimated systematic error of their result which is reported as (15$\pm$5)\%.  
Note that the quarks in the simulations do have non-zero masses and that 
extrapolation to zero quark mass would require significant computing time.
Due to the difficulty associated with the inclusion of light/massless 
dynamical quarks on the lattice it is therefore desirable to have analytic 
methods which can be used to estimate the thermodynamic functions.

The only rigorous method available for reorganizing
perturbation theory in thermal QCD is {\it dimensional reduction}
to an effective 3-dimensional field theory \cite{KLRS,Kajantie-97}.
The coefficients of the terms in the effective lagrangian are calculated
using perturbation theory, but calculations within the
effective field theory are carried out nonperturbatively
using lattice gauge theory.
Dimensional reduction has the same limitations as ordinary
lattice gauge theory: it can be applied only to static quantities
and only at zero baryon number density.
Unlike in ordinary lattice gauge theory,
light dynamical quarks do not require any additional computer power,
because they only enter through the perturbatively calculated
coefficients in the effective lagrangian.
This method has been applied to the Debye screening mass
for QCD \cite{Kajantie-97} as well as the
pressure \cite{KLRS}.

There are some proposals for reorganizing
perturbation theory in QCD that are essentially just mathematical
manipulations of the weak coupling expansion.
The methods include {\it Pad\'e approximates} \cite{Pade},
{\it Borel resummation} \cite{Parwani},
and {\it self-similar approximates} \cite{Yukalov}.
These methods are used to construct more stable sequences
of successive approximations that agree with the weak-coupling expansion
when expanded in powers of $\alpha_s$.
These methods can only be applied to quantities
for which several orders in the weak-coupling expansion are known,
so they are limited in practice to the thermodynamic functions.

One promising approach for reorganizing perturbation theory
in thermal QCD is to use a variational framework.
The free energy ${\cal F}$ is expressed as the variational
minimum of a thermodynamic potential $\Omega(T,\alpha_s;m^2)$
that depends on one or more variational parameters
that we denote collectively by $m^2$:
\beq
{\cal F}(T,\alpha_s) \;=\; \Omega(T,\alpha_s;m^2)
\bigg|_{\partial\Omega/\partial m^2 = 0} \,.
\label{varf}
\eeq
%{\bf [mp]}
%
A particularly compelling variational formulation
is the {\it $\Phi$-derivable approximation}, in which the complete propagator
is used as an infinite set of variational parameters \cite{Phi}.
The $\Phi$-derivable thermodynamic potential $\Omega$ is the
2PI effective action, the sum of all diagrams that are
2-particle-irreducible with respect to the complete propagator \cite{CJT-74}.
The $n$-loop $\Phi$-derivable approximations, in which $\Omega$ is the
the sum of 2PI diagrams with up to $n$ loops, form a systematically
improvable sequence of variational approximations.
Until recently, $\Phi$-derivable approximations have proved to be
intractable for relativistic field theories except for simple cases
in which the self-energy is momentum-independent.
However there has been some recent progress in solving the
3-loop $\Phi$-derivable approximation for scalar field theories.
Braaten and Petitgirard have developed an analytic method
for solving the 3-loop $\Phi$-derivable approximation for the
massless $\phi^4$ field theory \cite{BP-01}.
Van Hees and Knoll have developed numerical methods for solving the
3-loop $\Phi$-derivable approximation for the massive $\phi^4$ field theory
\cite{vHK-01}. They also investigated renormalization issues
associated with the $\Phi$-derivable approximation.  These issues
have recently been studied in detail by Blaizot, Iancu, and 
Reinosa~\cite{BIR-03}.

The application of the $\Phi$-derivable approximation to QCD
was first discussed by McLerran and Freedman \cite{FM-77}.
One problem with this approach is that the
thermodynamic potential $\Omega$ is gauge dependent,
and so are the resulting thermodynamic functions.
The gauge dependence is the same order in $\alpha_s$
as the truncation error when evaluated off the stationary
point and twice the order in $\alpha_s$ when evaluated at
the stationary point~\cite{AS-02}.  
However the most serious problem is that even the application of
2-loop $\Phi$-derivable approximation to gauge theories 
has proved to be intractable.

The 2-loop $\Phi$-derivable approximation for QCD has been used as
the starting point for {\it HTL resummations} of the entropy by
Blaizot, Iancu and Rebhan \cite{BIR-99}
and of the pressure by Peshier \cite{Peshier-00}.
The thermodynamic potential $\Omega_{\rm 2-loop}$ is a functional
of the complete gluon propagator $D_{\mu\nu}(P)$.  However, in 
order to make the problem tractable the authors in Refs.~\cite{BIR-99}
and \cite{Peshier-00} were forced to make a variational ansatz for
the exact gluon propagator which they took as the HTL gluon propagator
in the infrared and free in the ultraviolet with an aribitrary momentum
scale separating the two momentum regions.  Using this ansatz they were able to
calculate the QCD thermodynamic functions; however, a first-principles calculation
of the corrections to their results for gauge theories would require the inclusion of 
exact vertices as well as exact propagators thus making the problem intractable.

The difficulties in calculating quantities using 
$\Phi$-derivable approximations in gauge theories 
motivates the use of simpler variational approximations.
One such strategy that involves a single variational parameter $m$
has been called {\it optimized perturbation theory} \cite{Stevenson-81},
{\it variational perturbation theory} \cite{varpert},
or the {\it linear $\delta$ expansion} \cite{deltaexp}.
This strategy was applied to the thermodynamics of the massless $\phi^4$
field theory by Karsch, Patkos and Petreczky under the name
{\it screened perturbation theory} \cite{KPP-97}.
The method has also been applied to spontaneously broken
field theories at finite temperature \cite{CK-98}.
The calculations of the thermodynamics of the massless $\phi^4$
field theory using screened perturbation theory
have been extended to 3 loops \cite{ABS-01}.
The calculations can be greatly simplified by using a double
expansion in powers of the coupling constant and $m/T$ \cite{AS-01}.

{\it HTL perturbation theory} (HTLpt) is an adaptation of this strategy
to thermal QCD \cite{ABS-99}.
The exactly solvable theory used as the starting point
is one whose propagators are the HTL quark and gluon propagators.
The variational mass parameters $m_D$ and $m_q$ are identified
with the Debye screening mass and the induced quark mass.
The one-loop free energy in HTLpt was calculated
for QCD in Ref.~\cite{ABS-99}
and for QCD with massless quarks in Ref.~\cite{ABS-00}.
At this order, the parameters $m_D$ and $m_q$ could not 
be determined variationally, so their perturbative limits were used.
The resulting thermodynamic functions had errors of order $\alpha_s$,
but the terms of order $\alpha_s^{3/2}$ associated with Debye screening
were correct.  A two-loop calculation is required to reduce the errors 
to order $\alpha_s^2$.  

In a previous paper we calculated the thermodynamic functions of pure-glue
QCD to next-to-leading order in HTLpt~\cite{ABPS-02}.  In that paper we showed that it was
possible to renormalize the resulting expressions for the thermodynamic
potential at next-to-leading order using only vacuum and mass counterterms and we also
showed that the corrections to the thermodynamic functions in going from leading-order to next-to-leading
order were small down to temperatures on the order of $10 \, T_c$.
In this paper we calculate the thermodynamic functions of QCD
to next-to-leading order in HTLpt including the contributions 
from quark and quark-gluon interaction diagrams.

We begin with a brief summary of HTL perturbation theory
including quarks in Section~\ref{HTLpt}.  In Section~\ref{diagrams},
we give the expressions for the one-loop and two-loop
diagrams for the thermodynamic potential.
In Section~\ref{scalarint},  we reduce those diagrams to scalar sum-integrals.
We are unable to compute those sum-integrals exactly, so in
Section~\ref{expand} we evaluate them by treating $m_D$ and $m_q$ as $O(g)$ quantities
and expand them in $m_D/T$ and $m_q/T$ keeping all terms that contribute up 
to $O(g^5)$.  The diagrams are
combined in Section~\ref{thermpot} to obtain the final result for the
two-loop thermodynamic potential up to $O(g^5)$.
In Section~\ref{freeenergy}, we present our numerical results for the
free energy of QCD at leading and next-to-leading order in HTLpt.
In Section~\ref{largeNf} we evaluate the free energy
in the large $N_f$ limit where exact numerical results have been
obtained~\cite{moore,ipp}.

There are several appendices that contain technical details
of the calculations.
In Appendix~\ref{app:rules},
we give the Feynman rules for HTL perturbation theory
in Minkowski space to facilitate the application of this formalism
to signatures of the quark-gluon plasma.
The most difficult aspect of these calculations was the evaluation of the
sum-integrals obtained from the expansion in $m_D/T$ and $m_q/T$.
We give the results for these sum-integrals in Appendix~\ref{app:sumint}.
The evaluation of some difficult thermal integrals
that were required to obtain the sum-integrals is described in
Appendix~\ref{app:int}.

\section{HTL perturbation theory}
\label{HTLpt}

The lagrangian density that generates the perturbative expansion for
QCD can be expressed in the form
\bqa\nonumber
{\cal L}_{\rm QCD}&=&
-{1\over2}{\rm Tr}\left(G_{\mu\nu}G^{\mu\nu}\right)
+i \bar\psi \gamma^\mu D_\mu \psi \nonumber \\
&&\hspace{9mm} +{\cal L}_{\rm gf}
+{\cal L}_{\rm ghost}
+\Delta{\cal L}_{\rm QCD}\;.
\label{L-QCD}
\eqa
%
%{\bf [mjp]}
The gauge potential is $A_\mu = A_\mu^a t^a$, with generators
$t^a$ of the fundamental representation of SU($N_c$) 
normalized so that ${\rm Tr}\,t^a t^b=\delta^{ab}/2$.
The field strength tensor is $G_{\mu\nu}= \partial_{\mu}A_{\nu}-\partial_{\nu}A_{\mu}
-ig[A_{\mu},A_{\nu}]$.  In the quark term there is an implicit
sum over the $N_f$ quark flavors and $D_\mu = \partial_\mu - i g A_\mu$ is
the covariant derivative for the fundamental representation.
The ghost term ${\cal L}_{\rm ghost}$ depends on the choice of the gauge-fixing term 
${\cal L}_{\rm gf}$. Two choices for the gauge-fixing term that depend on 
an arbitrary gauge parameter $\xi$ are the general covariant gauge
and the general Coulomb gauge:
\bqa
{\cal L}_{\rm gf}&=&
-{1\over\xi}{\rm Tr}\left(\left(\partial^{\mu}A_{\mu}\right)^2\right)
\hspace{1.1cm}\mbox{covariant}\;,
\label{Lgf-cov}
\\
&=&-{1\over\xi}{\rm Tr}\left(\left(\nabla\cdot {\bf A}\right)^2\right)
\hspace{1cm}\mbox{Coulomb}\;.
\label{Lgf-C}
\eqa
%
%{\bf [jmp]}
%
It is also convenient to introduce various 
invariants associated with the representations
of the SU($N_c$) gauge group.  Denoting the generators of the adjoint
representation as $(F^a)^{bc}=-i f^{abc}$ 
and generators of the fundamental representation
as $T^a$ we define the following group theory factors:
\bqa
\left[F^c F^c\right]^{ab} &=& f^{adc}f^{bdc} = c_A \, \delta^{ab} \, ,  \nonumber \\
{\rm Tr}\,F^a F^b &=& s_A \, \delta^{ab} \, , \nonumber \\
\delta^{aa} &=& d_A \, , \nonumber \\
\left[T^a T^a\right]_{ij} &=& c_F \, \delta_{ij} \, , \nonumber \\
{\rm Tr}\,T^a T^b &=& s_F \, \delta^{ab} \, , \nonumber \\
\delta^{ii} &=& d_F = s_F \, d_A/c_F \, .
\eqa
%{\bf [mjp]}
%
With the standard normalization 
\bqa
c_A = s_A &=& N_c \nonumber \, , \\
d_A &=& N_c^2-1 \nonumber \, , \\
c_F &=& (N_c^2-1)/(2 N_c) \nonumber \, , \\
s_F &=& N_f/2 \nonumber \, , \\
d_F &=& N_c N_f  \, .
\eqa
%{\bf [mjp]}

The perturbative expansion in powers of $g$
generates ultraviolet divergences.
The renormalizability of perturbative QCD guarantees that
all divergences in physical quantities can be removed by
renormalization of the coupling constant $\alpha_s = g^2/4 \pi$.
There is no need for wavefunction renormalization, because
physical quantities are independent of the normalization of
the field.  There is also no need for renormalization of the gauge
parameter, because physical quantities are independent of the
gauge parameter.

Hard-thermal-loop perturbation theory (HTLpt) is a reorganization
of the perturbation
series for thermal QCD. The lagrangian density is written as
\bqa
{\cal L}= \left({\cal L}_{\rm QCD}
+ {\cal L}_{\rm HTL} \right) \Big|_{g \to \sqrt{\delta} g}
+ \Delta{\cal L}_{\rm HTL}.
\label{L-HTLQCD}
\eqa
%
%{\bf [mjp]}
The HTL improvement term is
\bqa
\label{L-HTL}
{\cal L}_{\rm HTL}=-{1\over2}(1-\delta)m_D^2 {\rm Tr}
\left(G_{\mu\alpha}\left\langle {y^{\alpha}y^{\beta}\over(y\cdot D)^2}
	\right\rangle_{\!\!y}G^{\mu}_{\;\;\beta}\right)
	\nonumber \\
	+ (1-\delta)\,i m_q^2 \bar{\psi}\gamma^\mu \left\langle {y^{\mu}\over y\cdot D}
	\right\rangle_{\!\!y}\psi
	\, ,
\eqa
%
%{\bf [mp]}
where in the first term $D_{\mu}$ is the covariant derivative in the adjoint representation,
in the second term $D_\mu$ is the covariant derivative in the fundamental representation,
$y^{\mu}=(1,\hat{{\bf y}})$ is a light-like four-vector,
and $\langle\ldots\rangle_{ y}$
represents the average over the directions
of $\hat{{\bf y}}$.
The term~(\ref{L-HTL}) has the form of the effective lagrangian
that would be induced by
a rotationally invariant ensemble of colored sources with infinitely high
momentum. The parameter $m_D$ can be identified with the
Debye screening mass and the parameter $m_q$ can be identified as the
induced finite temperature quark mass.
HTLpt is defined by treating
$\delta$ as a formal expansion parameter.

The HTL perturbation expansion generates ultraviolet divergences.
In QCD perturbation theory, renormalizability constrains the ultraviolet
divergences to have a form that can be cancelled by the counterterm
lagrangian $\Delta{\cal L}_{\rm QCD}$.
We will demonstrate that renormalized perturbation theory can be implemented 
by including a counterterm lagrangian $\Delta{\cal L}_{\rm HTL}$ among 
the interaction terms in (\ref{L-HTLQCD}).
There is no proof that the HTL perturbation expansion is renormalizable,
so the general structure of the ultraviolet divergences is not known;
however, it was shown in our previous paper \cite{ABPS-02} that it was
possible to renormalize the next-to-leading order HTLpt prediction for the
free energy of pure-glue QCD using only a vacuum counterterm and
Debye mass counterterm.  Here we show that when quarks are
included it is also possible to renormalize the resulting expressions
using only vacuum, Debye mass, and quark mass counterterms.

The leading term in the delta expansion of the vacuum energy, ${\cal E}_0$,  
counterterm $\Delta{\cal E}_0$ was deduced in Ref.~\cite{ABS-99}
by calculating the free energy to leading order in $\delta$.
The ${\cal E}_0$ counterterm $\Delta{\cal E}_0$ must therefore have the form
\beq
\Delta {\cal E}_0 \;=\;
\left( {d_A\over128\pi^2\epsilon}
	+ O(\delta \alpha_s) \right) (1-\delta)^2 m_D^4\,.
\label{del-E0}
\eeq
%
%{\bf [mjp]}
To calculate the free energy to next-to-leading order in $\delta$,
we need the counterterm $\Delta {\cal E}_0$ to order $\delta$
and the counterterms $\Delta m_D^2$ and $\Delta m_q^2$ to order $\delta$.
We will show that there is a nontrivial cancellation of the ultraviolet
divergences if the mass counterterms have the form
\bqa
\label{delmd}
\Delta m_D^2 \;&=&\;
-{\alpha_s\over3\pi\epsilon}\left[{11\over4}c_A-s_F\right]m_D^2 \;, \\
\Delta m_q^2 \;&=& -{\alpha_s\over 3 \pi \epsilon}\left[{9\over8}{d_A\over c_A}\right] m_q^2
\;.
\label{delmq}
\eqa
%{\bf [mjp]}

Physical observables are calculated in HTLpt
by expanding them in powers of $\delta$,
truncating at some specified order, and then setting $\delta=1$.
This defines a reorganization of the perturbation series
in which the effects of
the $m_D^2$ and $m_q^2$ terms in~(\ref{L-HTL})
are included to all orders but then systematically subtracted out
at higher orders in perturbation theory
by the $\delta m_D^2$ and $\delta m_q^2$ terms in~(\ref{L-HTL}).
If we set $\delta=1$, the lagrangian (\ref{L-HTLQCD})
reduces to the QCD lagrangian (\ref{L-QCD}).
If the expansion in $\delta$ could be calculated to all orders,
all dependence on $m_D$ and $m_q$ should disappear when we set $\delta=1$.
However, any truncation of the expansion in $\delta$ produces results
that depend on $m_D$ and $m_q$.
Some prescription is required to determine $m_D$ and $m_q$
as a function of $T$ and $\alpha_s$.
We choose to treat both as variational parameters that should be
determined by minimizing the free energy.
If we denote the free energy truncated at some order in $\delta$ by
$\Omega(T,\alpha_s,m_D,m_q,\delta)$, our prescription is
\bqa
{\partial \ \ \over \partial m_D}\Omega(T,\alpha_s,m_D,m_q,\delta=1) &=& 0 \, ,
\nonumber \\
{\partial \ \ \over \partial m_q}\Omega(T,\alpha_s,m_D,m_q,\delta=1) &=& 0 \, .
\label{gap}
\eqa
%
%{\bf [mjp]}
Since $\Omega(T,\alpha_s,m_D,m_q,\delta=1)$ is a function of the
variational parameters $m_D$ and $m_q$, we will refer to it as the
{\it thermodynamic potential}.  We will refer to the variational equations
(\ref{gap}) as the {\it gap equations}.  The free energy ${\cal F}$
is obtained by evaluating the thermodynamic potential at the solution
to the gap equations~(\ref{gap}).  Other thermodynamic functions can then be
obtained by taking appropriate derivatives of ${\cal F}$ with respect to $T$.

\section{Diagrams for the thermodynamic potential}
\label{diagrams}
The thermodynamic potential at leading order in HTL perturbation theory
for an $SU(N_c)$ gauge theory with $N_f$ massless quarks is
\bqa
\Omega_{\rm LO}= d_A {\cal F}_{g}
+ d_F {\cal F}_q+\Delta_0{\cal E}_0\;,
\eqa
%{\bf [jmp] }   
where ${\cal F}_g$ is the contribution from each of the color states of the
gluon:
\bqa
{\cal F}_g & = & -{1 \over 2}\sumint_P
\left\{ (d-1) \log [-\Delta_T(P)] \;+\; \log \Delta_L(P) \right\}\;.
\label{Fg-def}
\eqa
%
%[{\bf jmp}]
The transverse and longitudinal HTL propagators
$\Delta_T(P)$ and $\Delta_L(P)$ 
are given in (\ref{Delta-T}) and (\ref{Delta-L}).
The quark contribution is
\bqa
\label{lq}
{\cal F}_q=-\sumint_{\{P\}}\log\det\left[P\!\!\!\!/-\Sigma(P)\right]\;,
\eqa
%{\bf [jmp]}
where $\Sigma(P)$ is the HTL fermion self-energy.
The leading order counterterm $\Delta_0{\cal E}_0$ 
was determined in Ref.~\cite{ABS-99}
\bqa
\label{count0}
\Delta_0{\cal E}_0={d_A\over128\pi^2\epsilon}m_D^4\;.
\eqa
%{\bf [jmp]}

%%%%%%%%%%%%%%%%%%%%%%%%%%%%%%%%%%%%%%%%%%%%%%%%%%%%%%%%%%%%%%%%%%%%%%%
\begin{figure}[t]
\includegraphics[width=6.5cm]{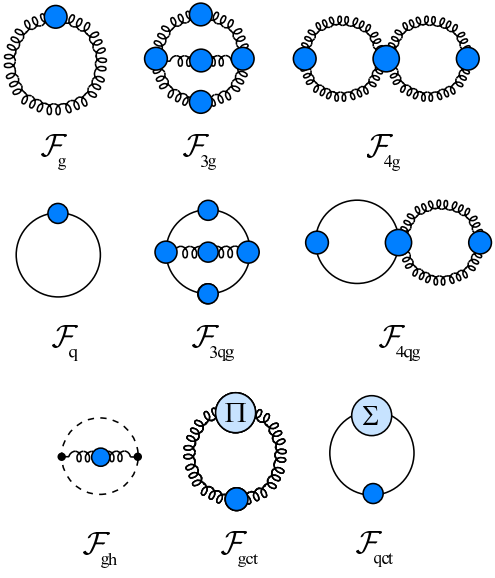}
\caption{Diagrams contributing through NLO in HTLpt.  Shaded circles indicate dressed HTL propagators and vertices.
}
\label{diagramfig}
\end{figure}
%%%%%%%%%%%%%%%%%%%%%%%%%%%%%%%%%%%%%%%%%%%%%%%%%%%%%%%%%%%%%%%%%%%%%%%

The thermodynamic potential at 
next-to-leading order in HTL perturbation theory
can be written as
\bqa\nonumber
\Omega_{\rm NLO}&=&\Omega_{\rm LO}+
d_A \left[{\cal F}_{3g}+{\cal F}_{4g}+{\cal F}_{gh}
+{\cal F}_{gct}
\right] \\ \nonumber
&&
+d_A s_F \left[{\cal F}_{3qg}+{\cal F}_{4qg}
\right] + d_f {\cal F}_{qct}
+\Delta_1{\cal E}_0
\\&&
+\Delta_1 m_D^2{\partial\over\partial m_D^2}
\Omega_{\rm LO}+\Delta_1 m_q^2{\partial\over\partial m_q^2}\Omega_{\rm LO}\;,
\label{OmegaNLO}
\eqa
where $\Delta_1{\cal E}_0$ and $\Delta_1m_D^2$ are the terms of order
$\delta$ in the vacuum energy density and mass counterterms.
The contributions from the two-loop diagrams with the
three-gluon and four-gluon vertices are
\bqa\nonumber
{\cal F}_{3g}&=&
{c_A\over12}g^2\sumint_{PQ}\Gamma^{\mu\lambda\rho}(P,Q,R)
\\&&\times
\Gamma^{\nu\sigma\tau}(P,Q,R)
\Delta^{\mu\nu}(P)\Delta^{\lambda\sigma}(Q)\Delta^{\rho\tau}(R)\;,
\label{F-3g}
\\ \nonumber
{\cal F}_{4g}&=&
{c_A\over8}g^2
\sumint_{PQ}\Gamma^{\mu\nu,\lambda\sigma}(P,-P,Q,-Q)
\\&&\times
\Delta^{\mu\nu}(P)\Delta^{\lambda\sigma}(Q)\;,
\label{F-4g}
\eqa
where $R=-Q-P$.
%{\bf [mpj] [[rewrote in terms of casimirs]] }

The contribution from the ghost diagram depends on the choice of gauge.
The expressions in the covariant and Coulomb gauges are
\bqa
{\cal F}_{gh}&=&
{c_A \over 2} g^2\sumint_{PQ}{1\over Q^2}{1\over R^2}Q^{\mu}R^{\nu}
\Delta^{\mu\nu}(P) \hspace{0.2cm}\mbox{covariant}\;,
\label{F-gh:cov}
\\ \nonumber
&=&
{c_A \over 2} g^2\sumint_{PQ}{1\over q^2}{1\over r^2}
\left(Q^{\mu}-Q\!\cdot\!n\;n^{\mu}\right)\left(R^{\nu}-R\!\cdot\!n\;
n^{\nu}\right)
\\&&\times
\Delta^{\mu\nu}(P)
\hspace{3.3cm}\mbox{Coulomb}\;.
\label{F-gh:C}
\eqa
%
%{\bf [mpj] [[rewrote in terms of casimirs]] }
The contribution from the HTL gluon counterterm diagram is
\bqa
{\cal F}_{\rm gct}&=& {1\over2}
\sumint_P\Pi^{\mu\nu}(P)\Delta^{\mu\nu}(P)\;.
\label{F-ct}
\eqa
%
%[{\bf jmp}]

The contributions from the two-loop diagrams with the quark-gluon 
three and four vertices are given by
\bqa\nonumber
{\cal F}_{3qg}&=&{1\over2}g^2\sumint_{P\{Q\}}\mbox{Tr}\left[\Gamma^{\mu}(P,Q,R)S(Q)\right.
\\&&
\left.
\hspace{1cm}
\times \Gamma^{\nu}(P,Q,R)
S(R)\right]\Delta^{\mu\nu}(P) 
\label{3qg}
\\  \nonumber
{\cal F}_{4qg}&=&{1\over2}g^2\sumint_{P\{Q\}}\mbox{Tr}\left[\Gamma^{\mu\nu}(P,-P,Q,Q)S(Q)\right]
\\&&
\hspace{1cm}
\times
\Delta^{\mu\nu}(P)\;,
\label{4qg}
\eqa
%{\bf [jmp]}
where $\mbox{Tr}$ implies taking the trace over $\gamma$-matrices.
The contribution from the HTL quark counterterm is
\bqa
\label{ctt}
{\cal F}_{\rm qct}
=%{1\over2}
-\sumint_{\{P\}}\mbox{Tr}\left[\Sigma(P)S(P)\right]\;.
\eqa
%{\bf [jmp]}

Provided that HTL perturbation theory is renormalizable,
the ultraviolet divergences at any order in $\delta$
can be cancelled by renormalizations of
the vacuum energy density ${\cal E}_0$,
the HTL mass parameters $m_D^2$ and $m_q^2$, 
and the coupling constant $\alpha_s$.
Renormalization of the coupling constant
does not enter until order $\delta^2$.
We will calculate the thermodynamic potential as a double expansion
in powers of $m_D/T$ and $m_q/T$, ad $g$
including all terms through 5$^{\rm th}$ order.
The $\delta \alpha_s$ term in $\Delta{\cal E}_0$
does not contribute until 6$^{\rm th}$ order in this expansion,
so the term of order $\delta$
in $\Delta{\cal E}_0$ can be obtained simply by expanding
Eq.~(\ref{count0}) to first order in $\delta$:
\bqa
\Delta_1{\cal E}_0 =
- {d_A\over 64\pi^2\epsilon} m_D^4 \;.
\label{del1e0}
\eqa
%
%{\bf [jmp]}
The remaining ultraviolet divergences must be removed by
renormalization of the mass parameters $m_D$ and $m_q$.
We will show below that, at order $\delta$, all remaining divergences
can be removed by the quark and Debye mass counterterms.
This provides nontrivial evidence for the renormalizability
of HTL perturbation theory at this order in $\delta$.

The sum of the 3-gluon, 4-gluon, and ghost contributions 
is gauge invariant.  By using the Ward identities, one can easily show
that the sum of these three diagrams is independent of the gauge parameter
$\xi$. With more effort, one can show the equivalence of the 
covariant gauge expression with $\xi=0$ and the Coulomb gauge expression
with $\xi=0$.  In a similar manner,
it can be shown that the sum of~(\ref{3qg}) and~(\ref{4qg}) is
independent of $\xi$ within the class of covariant and Coloumb gauges, as
well as the equivalence of the two with $\xi=0$.

\section{Reduction to scalar sum-integrals}
\label{scalarint}
The first step in calculating the quark contribution to the free energy
is to reduce the sum of the diagrams to scalar-sum-integrals.
The leading-order quark contribution 
can be rewritten as 
\bqa
{\cal F}_q=-2\sumint_{\{P\}}\log P^2-2\sumint_{\{P\}}
\log\left[{A_S^2-A_0^2\over P^2}\right]\;,
\label{qlead}
\eqa
%{\bf [jmp]}
where 
\bqa
A_0(P)&=&iP_0-{m_q^2\over iP_0}{\cal T}_P\;,\\ 
A_S(P)&=&|{\bf p}|+{m^2_q\over |{\bf p}|}\left[1-{\cal T}_P\right]\;.
\eqa
%{\bf [jmp]}
The HTL quark counterterm can be rewritten as
\bqa
\label{ct1}
{\cal F}_{qct}=-4\sumint_{\{P\}}
{P^2+m^2_q\over A_S^2-A_0^2}\;.
\eqa
%{\bf [jmp]}

\begin{widetext}
We proceed to simplify the sum of~(\ref{3qg}) and~(\ref{4qg}) in Landau gauge.
Using the Ward identities (\ref{qward1}) and (\ref{qward2}) the sum of (\ref{3qg}) and (\ref{4qg})
becomes
\bqa
{\cal F}_{3qg+4qg}&=&{1\over2}g^2\sumint_{P\{Q\}}\Bigg\{
\Delta_X(P)\mbox{Tr}\left[
\Gamma^{00}S(Q)
\right]
-\Delta_T(P)\mbox{Tr}\left[
\Gamma^{\mu}S(Q)\Gamma^{\mu}S(R')
\right]
\label{coll}
+\Delta_X(P)
\mbox{Tr}\left[
\Gamma^{0}S(Q)\Gamma^{0}S(R')
\right]\Bigg\}\;,
\eqa
%{\bf [jmp]}\\
%
where $S$ is the quark propagator, $\Delta_T$ is the transverse gluon propagator,
$\Delta_X$ is a combinaiton of the transverse and longitudinal gluon propagators
defined in (\ref{Delta-X}), and $R' = Q - P$.

Performing the traces of $\gamma$-matrices gives

\bqa
{\cal F}_{3qg+4qg}&=&
- g^2\sumint_{P\{Q\}}{1\over A_S^2(Q)-A_0^2(Q)}
\Bigg\{
2(d-1)\Delta_T(P){\hat{\bf q}\!\cdot\!\hat{\bf r}A_S(Q)A_S(R)-A_0(Q)A_0(R)
\over A_S^2(R)-A_0^2(R)} \nonumber \\
&& \;\; -2\Delta_X(P)
{A_0(Q)A_0(R)+A_S(Q)A_S(R)\hat{\bf q}\!\cdot\!\hat{\bf r}
\over A_S^2(R)-A_0^2(R) } 
-4m_q^2
\Delta_X(P)
\Bigg\langle
{A_0(Q)-A_s(Q)\hat{{\bf q}}\!\cdot\!\hat{{\bf y}}\over(P\!\cdot\!Y)^2
-(Q\!\cdot\!Y)^2}{1\over(Q\!\cdot\!Y)}\Bigg\rangle_{\!\!\bf \hat y} \nonumber \\
&& \;\;
+{8m_q^2\Delta_T(P)\over A_S^2(R)-A_0^2(R)}
\Bigg\langle
%\left[
{(A_0(Q)-A_S(Q)\hat{\bf q}\!\cdot\!\hat{\bf y})
(A_0(R)-A_S(R)\hat{\bf r}\!\cdot\!\hat{\bf y})
\over(Q\!\cdot\!Y)(R\!\cdot\!Y)}
\bigg\rangle_{\!\!\bf \hat y}
\nonumber \\
&& \;\;
+ {4 m_q^2\Delta_X(P)\over A_S^2(R)-A_0^2(R) }
\Bigg\langle
{2A_0(R)A_S(Q)\hat{\bf q}\!\cdot\!\hat{\bf y}-A_0(Q)A_0(R)-A_S(Q)A_S(R)
\hat{\bf q}\!\cdot\!\hat{\bf r}
\over(Q\!\cdot\!Y)(R\!\cdot\!Y)}
\Bigg\rangle_{\!\!\bf \hat y} \Bigg\} + O(g^2 m_q^4) \; ,
\eqa
%{\bf [mjp]}
where $A_0$ and $A_S$ are defined in (\ref{aodef}) and (\ref{asdef}), respectively.
\end{widetext}

\section{High-temperature Expansion}
\label{expand}
The free energy has been reduced to scalar sum-integrals.
If we tried to evaluate the 2-loop HTL free energy exactly,
there are terms 
that could at best be reduced to 5-dimensional integrals 
which would have to be evaluated numerically.
We will therefore evaluate the sum-integrals approximately
by expanding them in powers of $m_D/T$ and $m_q/T$.  We will carry out the 
expansion to high enough order  to include all terms 
through  order $g^5$ if $m_D$ and $m_q$ are taken to be of order $g$.

The free energy can be divided into contributions
from hard and soft momenta. We proceed to calculate the hard-hard
and hard-soft contributions. There is no soft-soft contribution
since one of the momenta in the loop is always fermionic and therefore hard.
\subsection{One-loop sum-integrals}
The one-loop sum-integrals include the leading quark contribution 
(\ref{lq}) and the HTL quark counterterm (\ref{ctt}).
The leading order free energy must be expanded to order $m_q^4$
to include all terms through order $g^5$ if $m_q$ is taken to be of order
$g$.
\subsubsection{Hard contributions}
The hard contribution from the LO gluon term~(\ref{Fg-def}) was given 
in~\cite{ABPS-02}
and reads
\bqa\nonumber
{\cal F}_g^{(h)}
&=& - {\pi^2 \over 45} T^4
+ {1 \over 24} \left[ 1
        + \left( 2 + 2{\zeta'(-1) \over \zeta(-1)} \right) \epsilon \right]
\\ && \nonumber
\times\left( {\mu \over 4 \pi T} \right)^{2\epsilon} m_D^2 T^2
%\nonumber\\&& 
- {1 \over 128 \pi^2}
\left( {1 \over \epsilon} - 7 + 2 \gamma + {2 \pi^2\over 3} \right)
\\ &&\times
\left( {\mu \over 4 \pi T} \right)^{2\epsilon} m_D^4 \,.
\label{Flo-h}
\eqa
%
%{\bf [jmp]}

The hard contribution from the HTL counterterm~(\ref{F-ct}) 
was given in~\cite{ABPS-02}
and reads
\bqa\nonumber
{\cal F}_{gct}^{(h)}
&=& - {1\over24} m_D^2 T^2
+ {1 \over 64 \pi^2}
\left( {1 \over \epsilon} - 7 + 2 \gamma + {2 \pi^2\over 3} \right)
\\&& \times
\left( {\mu \over 4 \pi T} \right)^{2\epsilon} m_D^4 \,.
\label{Fct-h}
\eqa
%
%{\bf [jmp]}

The sum-integrals over $P$ involve two momentum scales $m_q$ and $T$.
Since $P_0=(2n+1)\pi T$, the momentum is always hard.
We can therefore expand in powers of $m_q^2$. To second order in $m_q^2$,
we obtain
\bqa\nonumber
{\cal F}^{(h)}_q&=&
-2\sumint_{\{P\}}\log P^2-4m_q^2\sumint_{\{P\}}{1\over P^2}
\\&&\hspace{-1.6cm}
+2m_q^4\sumint_{\{P\}}
\left[{2\over P^4}
-{1\over p^2P^2}+{2\over p^2P^2}{\cal T}_P
-{1\over p^2P_0^2}\left({\cal T}_P\right)^2
\right]\;.
\eqa
%{\bf [jmp]}
Note that the function ${\cal T}_P$ cancels from the $m_q^2$ term.
The values of the sum-integrals are given in Appendix B.
Inserting those expressions, the hard 
quark contributions to the free energy reduce to
\bqa\nonumber
{\cal F}_q^{(h)}&=&-{7\pi^2\over180}T^4+{1\over6}
\left[1+\left(2-2\log2+2{\zeta^{\prime}(-1)\over\zeta(-1)}\right)\epsilon
\right]
\\&&\times
\left({\mu\over4\pi T}\right)^{2\epsilon}
m_q^2T^2
%\\ &&
+{1\over12\pi^2}\left(\pi^2-6\right) m_q^4\;.
\label{Quark1loop}
\eqa 
%{\bf [jmp]}
%
Note that this contribution is finite and so the leading order
counterterm $\Delta_0{\cal E}_0$ is the same as in the pure-glue case.
%\subsection{Counterterm}
The HTL quark counterterm is given in~(\ref{ct1}).
Expanding this term to second order in $m_q^2$ yields
\bqa\nonumber
{\cal F}^{(h)}_{\rm qct}&=&%{1\over2}
4m_q^2\sumint_{\{P\}}{1\over P^2}
-4m_q^4\sumint_{\{P\}}\left[{2\over P^4}
-{1\over p^2P^2}\right.
\\&&\left.
+{2\over p^2P^2}{\cal T}_P
-{1\over p^2P_0^2}\left({\cal T}_P\right)^2\right]\;.
\eqa
%{\bf [jmp]}\\
The values of the sum-integrals are given in Appendix B. Inserting those
expressions, the hard contributions to the HTL quark counterterm reduce to
\bqa
\label{count}
{\cal F}^{(h)}_{\rm qct}=-{1\over6}m_q^2T^2
-{1\over6\pi^2}\left(\pi^2-6\right)m_q^4\;.
\eqa
%{\bf [jmp]}\\
Note that the first term in Eq.~(\ref{count}) 
cancels the order-$\epsilon^0$ term in coefficient of $m_q^2$ in 
Eq.~(\ref{Quark1loop})
\subsubsection{Soft contributions}

The soft contribution comes from the $P_0=0$ term in the sum-integral.
At soft momentum $P=(0,{\bf p})$, the HTL self-energy functions
reduce to $\Pi_T(P) = 0$ and $\Pi_L(P) = m_D^2$.
The transverse term vanishes in dimensional regularization
because there is no momentum scale in the integral over ${\bf p}$.
Thus the soft contribution comes from the longitudinal term only.

The soft contribution to the leading order free energy~(\ref{Fg-def}) was
given in Ref.~\cite{ABPS-02} and reads
\bqa
{\cal F}_g^{(s)}
&=& - {1 \over 12 \pi}
\left[ 1 + {8 \over 3}\epsilon \right]
\left( {\mu \over 2 m_D} \right)^{2 \epsilon}
m_D^3 T \,.
\label{Flo-s}
\eqa
%
%{\bf [jmp]}

The soft contribution to the HTL gluon counterterm~({\ref{F-ct})
given in Ref.~\cite{ABPS-02} and reads
\bqa{
\cal F}_{gct}^{(s)}
&=& {1 \over 8 \pi}  m_D^3 T \,.
\label{Fct-s}
\eqa
%
%{\bf [jmp]}
There are no soft contribution from the leading-order quark 
term~(\ref{qlead})
or from the HTL quark counterterm~(\ref{ct1}).

\subsection{Two-loop sum-integrals}
The sum of the two-loop sum-integrals is given in~(\ref{coll}). 
Since these integrals
have an explicit factor of $g^2$, we need only to expand the sum-integrals
to order $m_q^2m_D/T^3$ and $m_D^3/T^3$ to include all terms through
order $g^5$.

The sum-integrals involve two momentum scales $m_q, m_D$ and $T$.
In order to expand them in powers of.., we separate them into
contributions from hard loop momenta and soft loop momenta.
This gives two separate regions which we will denote $(hh)$ and $(hs)$.
In the $(hh)$ region, all three momenta $P$, $Q$, and $R$ are hard.
In the $(hs)$ region, two of the three momenta are hard and the other soft. 

\subsubsection{Contributions from the (hh) region}
For hard momenta, the self-energies are suppressed by $m_D^2/T^2$
or $m_q^2/T^2$ relative to the propagators, so we can expand in powers
of $\Pi_T$, $\Pi_L$, and $\Sigma$.

The $(hh)$ contribution from~(\ref{F-3g})--(\ref{F-gh:cov}) was
given in Ref.~\cite{ABPS-02} and reads
\bqa\nonumber
{\cal F}_{3g+4g+gh}^{(hh)}
&=& {\pi^2 \over 12} {c_A \alpha_s \over 3 \pi} T^4
\;-\; {7 \over 96} \left[ {1\over\epsilon} + 4.621 \right]
{c_A \alpha_s \over 3 \pi}
\\&&
\left( {\mu \over 4 \pi T} \right)^{4\epsilon}
m_D^2 T^2 \,.
\label{F2loop-hh}
\eqa
%
%{\bf [mp]}
}

\begin{widetext}
The $(hh)$ contribution from~(\ref{3qg}) and~(\ref{4qg}) can be written as
\bqa\nonumber
{\cal F}_{3qg+4qg}^{(hh)}&=&(d-1)g^2\left[\sumint_{\{PQ\}}{1\over P^2Q^2}
-\sumint_{P\{Q\}}{2\over P^2Q^2}\right] 
+2m_D^2g^2\sumint_{P\{Q\}}\left[{1\over p^2P^2Q^2}{\cal T}_P+{1\over (P^2)^2Q^2}
- {d-2\over d-1}{1\over p^2P^2Q^2}
\right]
\\ \nonumber && \hspace{-6mm}
+m_D^2g^2\sumint_{\{PQ\}}
\left[ {d+1\over d-1}{1\over P^2Q^2r^2}
-{4d\over d-1}{q^2\over P^2Q^2r^4}-{2d\over d-1}
{P\!\cdot\!Q\over P^2Q^2r^4}\right]{\cal T}_R  \\ \nonumber
&& \hspace{-6mm}
+m_D^2g^2\sumint_{\{PQ\}}\left[ {3-d\over d-1}{1\over P^2Q^2R^2}+
{2d\over d-1}{P\!\cdot\! Q\over P^2Q^2r^4}
-{d+2\over d-1}
{1\over P^2Q^2r^2} 
+{4d\over d-1}{q^2\over P^2Q^2r^4}
-{4\over d-1}{q^2\over P^2Q^2r^2R^2} 
\right] \\ \nonumber
&&  \hspace{-6mm}
+2m_q^2g^2(d-1)\sumint_{\{PQ\}}\left[ {1\over P^2Q_0^2Q^2}
+{p^2-r^2\over P^2q^2Q_0^2R^2}
\right] {\cal T}_Q 
+2m_q^2g^2(d-1)\sumint_{P\{Q\}} \left[{2\over P^2(Q^2)^2}
-{1\over P^2Q_0^2Q^2}{\cal T}_Q\right]
\\ 
&& \hspace{-6mm}
+2m_q^2g^2(d-1)\sumint_{\{PQ\}}\left[ {d+3\over d-1}{1\over P^2Q^2R^2}
- {2\over P^2(Q^2)^2} 
+{r^2-p^2\over q^2P^2Q^2R^2}\right] .
\label{hhexpansion}
\eqa
%{\bf [mp]}
%
Using the expressions for the sum-integrals in Appendix B, this reduces
to 
\bqa
{\cal F}_{3qg+4qg}&=&{5\pi^2\over72}{\alpha_s\over\pi}T^4
-{1\over72}\left[{1\over\epsilon}+1.2963\right]{\alpha_s\over\pi}
\left({\mu\over4\pi T}\right)^{4\epsilon}m_D^2T^2
%\\ &&
+{1\over8}\left[{1\over\epsilon}+8.96751\right]{\alpha_s\over\pi}
\left({\mu\over4\pi T}\right)^{4\epsilon}m_q^2T^2
\;.
\label{Fquark2}
\eqa
%{\bf [mjp]}

\subsubsection{The (hs) contribution}

In the $(hs)$ region, the momentum $P$ is soft. 
The momenta $Q$ and $R$ are always hard. The function that multiplies 
the soft propagator $\Delta_T(0,{\bf p})$ or $\Delta_X(0,{\bf p})$
can be expanded in powers of the soft momentum ${\bf p}$. In the case
of $\Delta_T(0,{\bf p})$, the resulting integrals over ${\bf p}$\
have no scale and they vanish in dimensional regularization.
The integration measure $\int_{\bf p}$ scales like $m_D^3$,
the soft propagator $\Delta_X(0,{\bf p})$ scales like $1/m_D^2$,
and every power of $p$ in the numerator scales like $m_D$.

The $(hs)$ contribution from~(\ref{F-3g})--(\ref{F-gh:cov}) was
given in Ref.~\cite{ABPS-02} and reads
\bqa
{\cal F}_{3g+4g+gh}^{(hs)}
&=& - {\pi \over 2} {c_A \alpha_s \over 3 \pi} m_D T^3
%\\ \nonumber&& 
\;-\; {11 \over 32 \pi}
\left( {1 \over \epsilon} + {27 \over 11} + 2 \gamma \right)
%\\ && \times
{c_A \alpha_s \over 3 \pi}
\left( {\mu \over 4 \pi T} \right)^{2\epsilon}
\left( {\mu \over 2m_D} \right)^{2\epsilon}
        m_D^3 T \,.
\label{F2loop-hs}
\eqa
%
%{\bf [mp]}

The only terms that contribute through order $g^2 m_D^3 T$ 
and $m_q^2m_Dg^3T$ from ~(\ref{3qg}) and~(\ref{4qg}) are
\bqa\nonumber
{\cal F}_{3qg+4qg}^{(hs)}&=&g^2T\int_{\bf p}{1\over p^2+m^2_D}
\sumint_{\{Q\}}\left[
{2\over Q^2}-{4q^2\over(Q^2)^2}\right]
+2m_D^2g^2T\int_{\bf p}{1\over p^2+m_D^2}
\sumint_{\{Q\}}
\left[{1\over(Q^2)^2}
-(3+d){2\over d}{q^2\over(Q^2)^3}+{8\over d}{q^4\over 
(Q^2)^4}
\right]
\\ 
&&
-4m_q^2g^2T\int_{\bf p}{1\over p^2+m_D^2}
\sumint_{\{Q\}}\left[{3\over(Q^2)^2}
-{4q^2\over(Q^2)^3} -{4\over(Q^2)^2} {\cal T}_Q
-{2\over Q^2}\bigg\langle {1\over(Q\!\cdot\!Y)^2} \bigg\rangle_{\!\!\bf \hat y}
\right]
\;.
\eqa
%{\bf [jmp]}
\end{widetext}

In the terms that are already of order $g^2 m_D^3 T$, we can set $R=-Q$.
In the terms of order $g^2 m_D T^3$, we must expand the sum-integrand
to second order in ${\bf p}$.  After averaging over angles of ${\bf p}$,
the linear terms in ${\bf p}$ vanish and quadratic terms 
of the form $p^i p^j$ are replaced by $p^2 \delta^{ij}/d$.  
We can set $p^2 = -m_D^2$, because any factor
proportional to $p^2 + m_D^2$ will cancel the denominator of the 
integral over ${\bf p}$, leaving an integral with no scale.
This gives
\bqa\nonumber
{\cal F}_{3qg+4qg}^{(hs)}&=&-{1\over6}\alpha_sm_DT^3
+{\alpha_s\over24\pi^2}
\left[{1\over\epsilon}+1+2\gamma+4\log2\right]
\\ \nonumber&& \times
\left({\mu\over4\pi T}\right)^{2\epsilon}\left({\mu\over2m_D}\right)^{2\epsilon}m_D^3T
-{\alpha_s\over2\pi^2}m_q^2m_DT\;.
\\ &&
\label{F2loop-qhs}
\eqa
%{\bf [jmp]}

\subsubsection{The $(ss)$ contributions}
The $(ss)$ contribution from~(\ref{F-3g})--(\ref{F-gh:cov}) was
given in Ref.~\cite{ABPS-02} and reads
\bqa
{\cal F}_{3g+4g+gh}^{(ss)}
&=& {3 \over 16} \left[ {1 \over \epsilon} + 3 \right]
{c_A \alpha_s \over 3 \pi}
\left( {\mu \over 2 m_D} \right)^{4 \epsilon}
m_D^2 T^2 \,.
\label{F2loop-ss}
\eqa
%
%{\bf [jmp]}
There is no $(ss)$ contribution from the diagrams involving fermions.

\section{HTL-Improved Thermodynamics}
The free energy at second order in HTL perturbation theory 
defines a function $\Omega(T,\alpha_s,m_D,m_q,\delta=1)$.
We will refer to this function as the thermodynamic potential.
To obtain the free energy $F(T)$ as a function of the 
temperature, we need to specify a prescription for the 
mass parameter $m_D$ as a function of $T$ and $\alpha_s$.

\section{Thermodynamic Potential}
\label{thermpot}

In this section, we calculate the thermodynamic potential
$\Omega(T,\alpha_s,m_D,m_q,\delta=1)$ explicitly,
first to leading order in the $\delta$ expansion and then to
next-to-leading order.

\subsection{Leading order}
The complete expression for the leading order thermodynamic potential
is the sum of the contributions from 1-loop diagrams
and the leading term (\ref{count0})
in the vacuum energy counterterm.
The contributions from the 1-loop diagrams,
including all terms through order $g^5$, is the sum of
(\ref{Flo-h}), (\ref{Quark1loop}), and (\ref{Flo-s})
\bqa\nonumber
\Omega_{1-{\rm loop}} &=& - d_A {\pi^2 T^4\over45}
\left\{ 1 + {7\over4} {d_F\over d_A} - {15 \over 2} \hat m_D^2 - 30 {d_F \over d_A} \hat m_q^2
\right.\\ && \left. \hspace{-1.5cm}
+ 30 \hat m_D^3
+ {45 \over 8}
\left( {1\over \epsilon} + 2 \log {\hat \mu \over 2}
        - 7 + 2 \gamma + {2 \pi^2\over 3} \right)
        \hat m_D^4  \nonumber
\right.\\ && \left. 
	- 60 {d_F\over d_A}\left(\pi^2-6\right)\hat m_q^4
	\right\} \;,
\label{Omega-1loop}
\eqa
%{\bf [mjp]}
where $\hat m_D$, $\hat m_q$ and $\hat \mu$ are dimensionless variables:
\bqa
\hat m_D &=& {m_D \over 2 \pi T}  \;,
\\
\hat m_q &=& {m_q \over 2 \pi T}  \;,
\\
\hat \mu &=& {\mu \over 2 \pi T}  \;. 
\eqa
%
%{\bf [jmp]}
Adding the counterterm (\ref{count0}),
we obtain the thermodynamic potential at leading order in the
delta expansion:
\bqa\nonumber
\Omega_{\rm LO} &=& 
- d_A {\pi^2 T^4\over45}
\left\{ 1 + {7\over4} {d_F\over d_A} - {15 \over 2} \hat m_D^2 - 30 {d_F \over d_A} \hat m_q^2
\right.\\ && \left. \hspace{-0.5cm}
+ 30 \hat m_D^3
+ {45 \over 4}
\left( \log {\hat \mu \over 2}
        - {7\over2} + \gamma + {\pi^2\over 3} \right)
        \hat m_D^4  \nonumber
\right.\\ && \left. 
	- 60 {d_F\over d_A}\left(\pi^2-6\right)\hat m_q^4
	\right\} \;,
\label{Omega-LO}
\eqa
%{\bf [mjp]}
%

\subsection{Next-to-leading order}

The complete expression for the next-to-leading order correction to the thermodynamic potential
is the sum of the contributions from all 2-loop diagrams, the quark and gluon counterterms,
and renormalization counterterms.  The contributions from the 2-loop diagrams, including all 
terms though order $g^5$ is the sum of (\ref{F2loop-hh}), (\ref{Fquark2}), (\ref{F2loop-hs}), 
(\ref{F2loop-qhs}), and (\ref{F2loop-ss}) multiplied by the appropriate group structure 
constants listed in (\ref{OmegaNLO}):

\begin{widetext}
\bqa\nonumber
\Omega_{\rm 2-loop}&=&
- d_A {\pi^2 T^4\over45} {\alpha_s \over \pi} \Bigg\{ 
	-{5\over4}\left(c_A+{5\over2}s_F\right) 
	+ 15(c_A+s_F)\hat m_D
	- {55\over8}\bigg[
	\left(c_A - {4\over11}s_F \right)\left({1\over\epsilon}+4 \log{\hat\mu\over2}\right) 
\nonumber \\ &&
	- c_A \left({72\over11}\log\hat m_D-1.96869\right) - 0.4714 \, s_F \bigg] \hat m_D^2
	-{45\over2} s_F\left[{1\over\epsilon}+4 \log{\hat\mu\over2}+8.96751\right] \hat m_q^2 
	+ 180 \,s_F \hat m_D \hat m_q^2
\nonumber \\ &&
	+ {165\over4}\bigg[\left(c_A - {4\over11}s_F \right)\left({1\over\epsilon}
		+4 \log{\hat\mu\over2}-2\log\hat m_D \right)
	+ c_A \left({27\over11} + 2 \gamma \right)
	- {4\over11} s_F \left( 1 + 2 \gamma + 4\log 2\right)\bigg] \hat m_D^3
\Bigg\}
\;.
\label{Omega2loop}
\eqa
%{\bf [mjp]} 
%
The HTL gluon counterterm is the sum of~(\ref{Fct-h}) and~(\ref{Fct-s})
\bqa
\Omega_{\rm gct}&=&
-d_A {\pi^2 T^4\over45} \left\{
{15\over2}\hat{m}_D^2-45\hat{m}_D^3 
%\right. \\ && \left. \hspace{-5mm}
-{45\over4}\left({1\over\epsilon}+2\log{\hat{\mu}\over2}
-7+2\gamma+{2\pi^2\over3}\right)\hat{m}_D^4
\right\}
\;.
\label{OmegaGct}
\eqa
%{\bf [mjp]}
%
The HTL quark counterterm is given by~(\ref{count}) 
\bqa
\Omega_{\rm qct}&=&
-d_F {\pi^2 T^4\over45} \left\{
	30 \hat m_q^2 + 120 \left(\pi^2-6\right) \hat m_q^4 
	\right\} \;.
\label{OmegaQct}
\eqa
%{\bf [mjp]}
%
The ultraviolet divergences that remain after these 3 terms are added
can be removed by renormalization of the vacuum energy density ${\cal E}_0$
and the HTL mass parameter $m_D$.
The renormalization contributions at first order in $\delta$ are
\bqa
\Delta \Omega \;=\; \Delta_1{\cal E}_0
+ \Delta_1m_D^2 {\partial \ \ \over \partial m_D^2} \Omega_{\rm LO} 
+\Delta_1m_q^2 {\partial \ \ \over \partial m_q^2} \Omega_{\rm LO} 
\;.
\label{Omega-renorm}
\eqa
%{\bf [jmp]}

Using the results listed in Eqs.~(\ref{delmd}), (\ref{delmq}), and (\ref{del1e0}) the complete contribution from the counterterm at first order in $\delta$ is
\bqa
\Delta\Omega&=& -d_A {\pi^2 T^4\over45} \Bigg\{ {45\over4\epsilon} \hat m_D^4 
%	\nonumber \\ && \hspace{-1cm}
	+ {\alpha_s \over \pi} \Bigg[ {55\over 8}
	\left(c_A - {4\over 11}s_F\right)
	\left({1\over\epsilon}+2\log{\hat\mu\over 2} + 2 {\zeta'(-1)\over\zeta(-1)}+2 \right) \hat m_D^2
\nonumber \\ && \hspace{-4.5mm}
	- {165\over4} \left(c_A - {4\over 11}s_F\right)
		\left({1\over\epsilon}+2\log{\hat\mu\over 2} - 2 \log \hat m_D +2 \right) \hat m_D^3
%\nonumber \\ && \hspace{-7mm}
	+ {45 \over 2} s_F
		\left({1\over\epsilon}+2+2\log{\hat\mu\over 2} - 2\log2 + 2{\zeta'(-1)\over\zeta(-1)}\right) 
			\hat m_q^2 \Bigg] \Bigg\} \;. \nonumber
			\\
\label{OmegaVMct}
\eqa
%{\bf [mjp]}
%
Adding the contributions from the two-loop diagrams in~(\ref{Omega2loop}), the
HTL gluon and quark counterterms in~(\ref{OmegaGct}) and~(\ref{OmegaQct}), the
contribution from vacuum and mass renormalizations in~(\ref{OmegaVMct}), and
the leading order thermodynamic potential in~(\ref{Omega-LO}) we
obtain the complete expression for the QCD thermodynamic potential 
at next-to-leading order in HTLpt:

\bqa
\Omega_{\rm NLO}&=&
- d_A {\pi^2 T^4\over45} \Bigg\{ 
	1 + {7\over4} {d_F \over d_A} - 15 \hat m_D^3 
	- {45\over4}\left(\log\hat{\mu\over2}-{7\over2}+\gamma+{\pi^2\over3}\right)\hat m_D^4
	+ 60 {d_F \over d_A}\left(\pi^2-6\right)\hat m_q^4
\nonumber \\ && \hspace{-10mm}
	+ {\alpha_s\over\pi} \Bigg[ -{5\over4}\left(c_A + {5\over2}s_F\right) 
	+ 15 \left(c_A+s_F\right)\hat m_D
	-{55\over4}\bigg\{ c_A\left(\log{\hat\mu \over 2}- {36\over11}\log\hat m_D - 2.001\right)
		- {4\over11} s_F \left(\log{\hat\mu \over 2}-2.337\right)\!\!\bigg\} \hat m_D^2
\nonumber \\ && \hspace{-14mm}
	-45 \, s_F \left(\log{\hat\mu \over 2}+2.192\right)\hat m_q^2
	+{165\over2}\bigg\{ c_A\left(\log{\hat\mu \over 2}+{5\over22}+\gamma\right)
	- {4\over11} s_F \left(\log{\hat\mu \over 2}-{1\over2}+\gamma+2\log2\right)\!\!\bigg\} \hat m_D^3
%\nonumber \\ && \hspace{8cm}
	+ 180\,s_F\hat m_D \hat m_q^2 \Bigg]
\Bigg\} \;.
\label{Omega-NLO}
\eqa
%{\bf [mjp]} 
\end{widetext}

%%%%%%%%%%%%%%%%%%%%%%%%%%%%%%%%%%%%%%%%%%%%%%%%%%%%%%%%%%%%%%%%%%%%%%%
\begin{figure}[t]
\includegraphics[width=7.5cm]{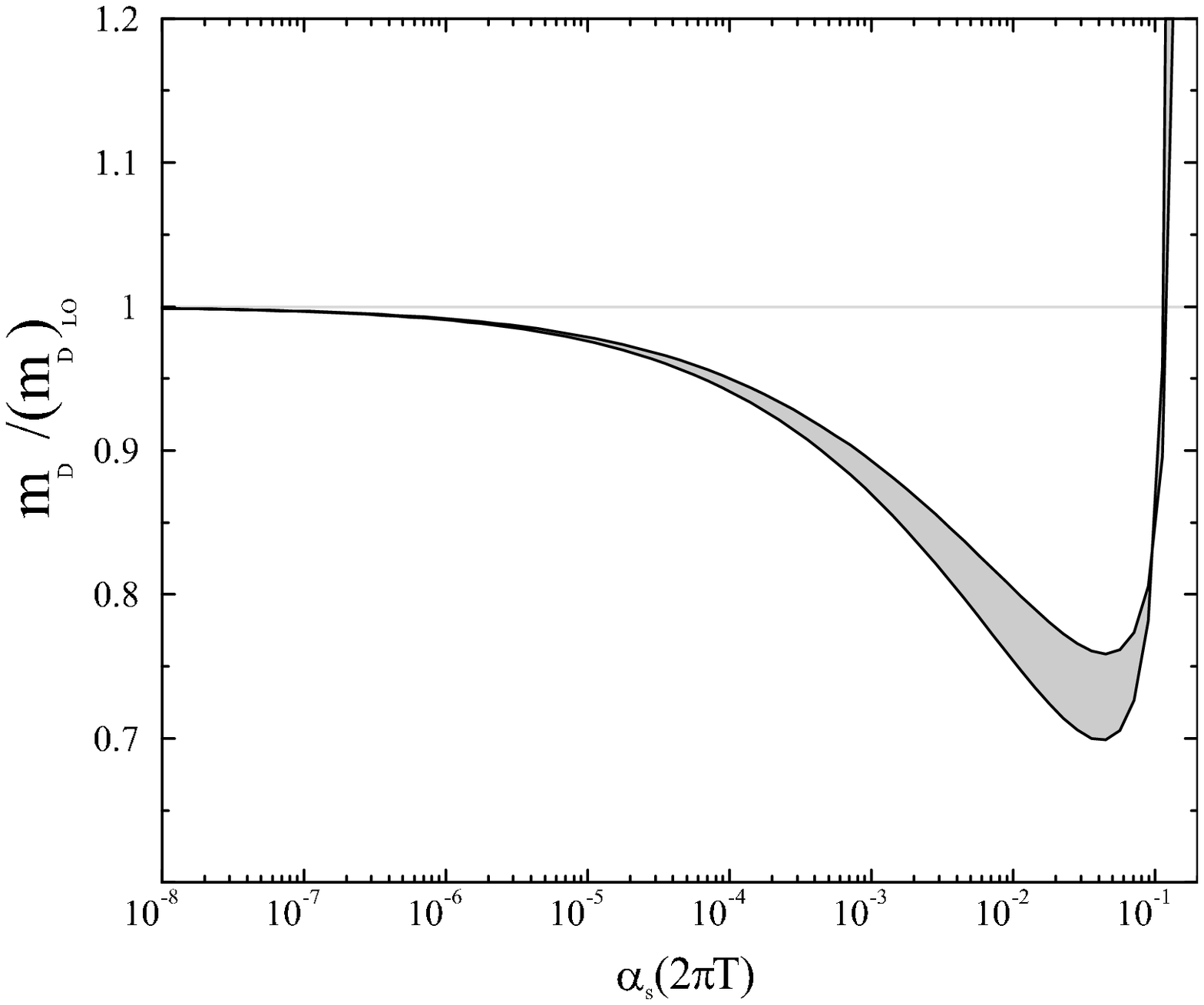}

\vspace{5mm}

\includegraphics[width=7.5cm]{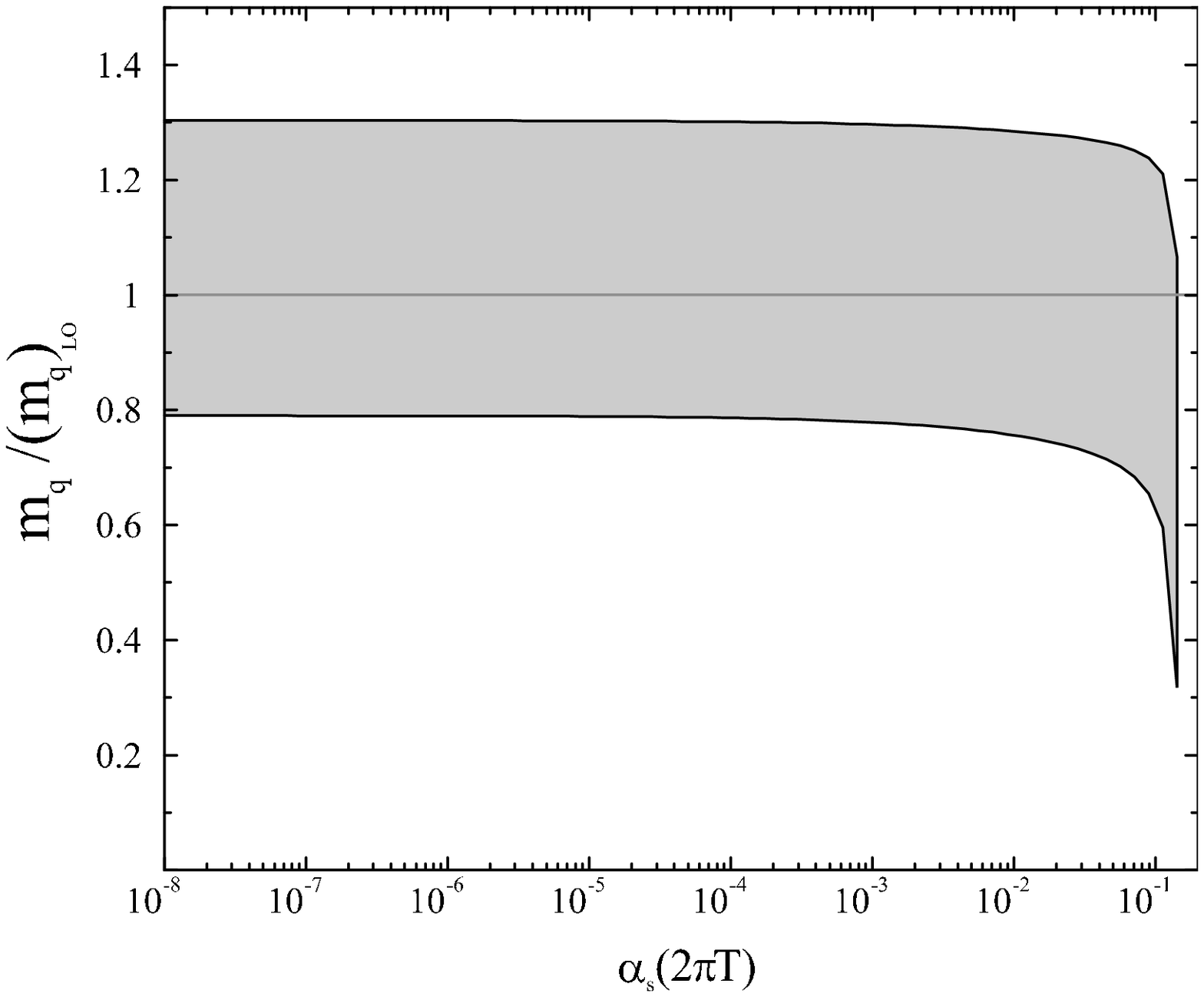}
\caption{Numerical solution of gap equations for (a) $m_D$ (\ref{mdgap}) and (b) $m_q$ (\ref{mqgap}) 
as a function of 
$\alpha_s(2\pi T)$ for $N_c=3$ and $N_f=3$.  The shaded band corresponds to varying 
the renormalization scale $\mu$ by a factor of two around $\mu = 2 \pi T$.}
\label{mdgapfig}
\end{figure}
%%%%%%%%%%%%%%%%%%%%%%%%%%%%%%%%%%%%%%%%%%%%%%%%%%%%%%%%%%%%%%%%%%%%%%%

\subsection{Gap Equation}

The quark and gluon mass parameters, $m_q$ and $m_D$, are determined variationally by
requiring that the derivative of $\Omega_{\rm NLO}$ with respect to each parameter taken
holding the other constant vanishes
\bqa
\label{gapeqs}
{\partial\over\partial m_q}\Omega_{\rm NLO}(T,\alpha_s,m_D,m_q)&=&0\;,\\
{\partial\over\partial m_D}\Omega_{\rm NLO}(T,\alpha_s,m_D,m_q)&=&0\;.
\eqa
%
%{\bf [mjp]} 
The first equation above results in the following equation for $m_q$ 
\beq
8 {d_F\over d_A} (\pi^2-6)\hat m_q^2 = 
	{\alpha_s s_F \over \pi} \Bigg[
	3 \left(\log{\hat\mu \over 2}+2.192\right) - 12 \hat m_D \Bigg] \, .
\label{mqgap}
\eeq
%{\bf [mjp]}
%
In the limit of small $\alpha_s$ the above gap equation does not go
to the perturbative limit for the quark mass which is $\hat m_{q,\rm pert}^2 = 
c_F \, \alpha_s/8\pi$.  The fact that $m_q$ does not go to the perturbative
value in the small $\alpha_s$ limit is due to the fact that the perturbative
limit of the quark gap equation results from terms which are $O(\alpha_s^2)$
and these terms are not included completely at NLO in HTLpt.  One might hope
that going to the next order in HTLpt would cure this problem; however, this
will in fact not happen since the fermion sector is infrared safe and therefore 
only even powers of $\hat m_q$ will appear at each order.  At NNLO all 
terms contributing at $O(\alpha_s^2)$ at NLO will be replaced by explicit 
powers of $\alpha_s$ and all $\hat m_q$ dependence will be pushed up to 
$O(\alpha_s^3)$.  This behavior will persist at all orders in HTLpt so that
at any order the weak-coupling limit of the gap equation quark mass will be
scale dependent.  In order to circumvent this problem we can consider
other possible prescriptions for choosing $\hat m_q$ which include 
requiring that $\hat m_q$ be equal to its perturbative value for all $\alpha_s$
or requiring that $\hat m_q$ be proportional to $\hat m_D$ with the proportionality 
constant fixed in the weak-coupling limit. 

Performing the derivative with respect to $m_D$ while holding $m_q$ fixed 
results in the following gap equation for $m_D$
\bqa
&& \hspace{-3mm} 45 \hat m_D^2 \left[ 1 + \left(\log\hat{\mu\over2}-{7\over2}+\gamma+{\pi^2\over3}\right)\hat m_D \right] 
\nonumber \\ && \hspace{-2mm}  
	= {\alpha\over\pi} \Bigg\{ 15 \left(c_A+s_F\right)
	-{55\over2}\bigg[ c_A\left(\log{\hat\mu \over 2}- {36\over11}\log\hat m_D - 3.637\right)
\nonumber \\ && \hspace{-2mm}
		- {4\over11} s_F \left(\log{\hat\mu \over 2}-2.337\right)\!\!\bigg] \hat m_D
%\nonumber \\ && \hspace{-2mm}
	+{495\over2}\bigg[ c_A\left(\log{\hat\mu \over 2}+{5\over22}+\gamma\right)
\nonumber \\ && \hspace{-2mm}
	- {4\over11} s_F \left(\log{\hat\mu \over 2}-{1\over2}+\gamma+2\log2\right)\!\!\bigg] \hat m_D^2
	+ 180\,s_F \hat m_q^2 \Bigg\} \;.
\label{mdgap}
\eqa
%{\bf [mjp]}
%
The last term in Eq.~(\ref{mdgap}) proportional to $\hat m_q^2$ 
can be written in terms of $\hat m_D$ using Eq.~(\ref{mqgap}).
In Fig.~\ref{mdgapfig} we plot the solutions to the gap equations for $m_D$ and
$m_q$ for $N_c=3$ and $N_f=2$.  The solution for $m_D$ goes to the perturbative
value in the limit of small $\alpha_s$, decreases below the perturbative value
as $\alpha_s$ increases, and becomes larger than the perturbative value at
$\alpha_s \sim 0.11$.  The solution for $m_q$ does not go to the perturbative
value in the limit of small $\alpha_s$ and is instead scale dependent even at
lowest order as dicussed above.  As $\alpha_s$ increases
the value of $m_q$ remains very flat regardless of the scale, changing 
significantly only near $\alpha_s \sim 0.10$.

\section{Free energy}
\label{freeenergy}

\begin{figure}[t]
\includegraphics[width=8cm]{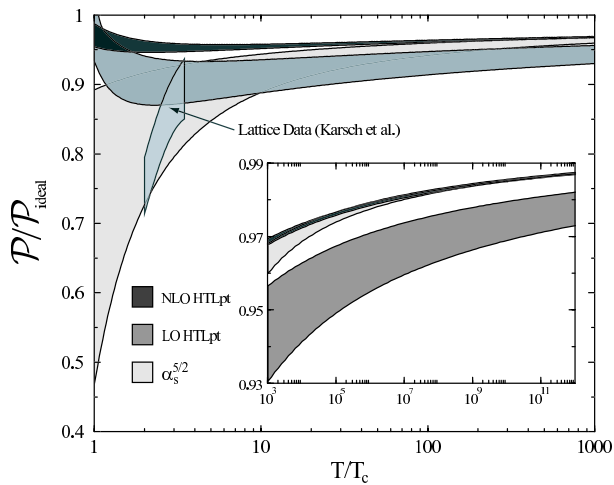}
\caption{LO and NLO HTLpt predictions for the free energy of QCD with $N_c=3$ and $N_f=2$ 
together with the perturbative prediction accurate to $g^5$.
The shaded bands correspond to varying the renormalization scale $\mu$ by a factor of 
two around $\mu = 2 \pi T$.
Also shown is a lattice estimate by Karsch et al.~\cite{klp} for
the free energy.  The band indicates the estimated systematic
error of their result which is reported as (15$\pm$5)\%.  
}
\label{freeenergyfig}
\end{figure}

The free energy is obtained by evaluating the leading and next-to-leading order 
thermodynamic potentials, (\ref{Omega-LO}) and (\ref{Omega-NLO}),
at the solution to the gap equations (\ref{mqgap}) and (\ref{mdgap}).  
In Fig.~\ref{freeenergyfig} we plot the leading and next-to-leading order HTLpt
predictions for the free energy of QCD with $N_c=3$ and $N_f=2$.  
We have studied the alternative prescriptions for the quark mass discussed
in the previous section and find that the NLO free energy obtained using these
prescriptions is numerically indistinguishable from that obtained
using the quark gap equations.  As can be
seen from this figure the corrections in going from LO to NLO are small over
the entire temperature range, especially when compared to convergence of
the perturbative result.  Additionally, the scale variation of the NLO HTLpt
result for the free energy is much smaller than the LO showing that the
partial resummation of the scale dependent logarithms reduces the scale
variation of the final results significantly.  

%%%%%%%%%%%%%%%%%%%%%%%%%%%%%%%%%%%%%%%%%%%%%%%%%%%%%%%%%%%%%%%%%%%%%%%
%%%%%%%%%%%%%%%%%%%%%%%%%%%%%%%%%%%%%%%%%%%%%%%%%%%%%%%%%%%%%%%%%%%%%%%

%%%%%%%%%%%%%%%%%%%%%%%%%%%%%%%%%%%%%%%%%%%%%%%%%%%%%%%%%%%%%%%%%%%%%%%
\begin{figure}[t]

\includegraphics[width=7.5cm]{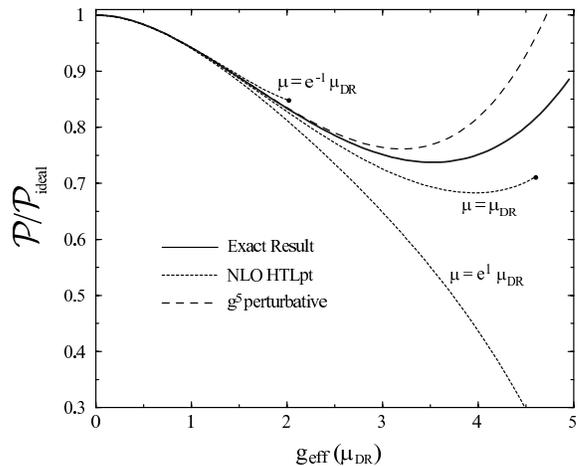}

\vspace{5mm}

\includegraphics[width=7.5cm]{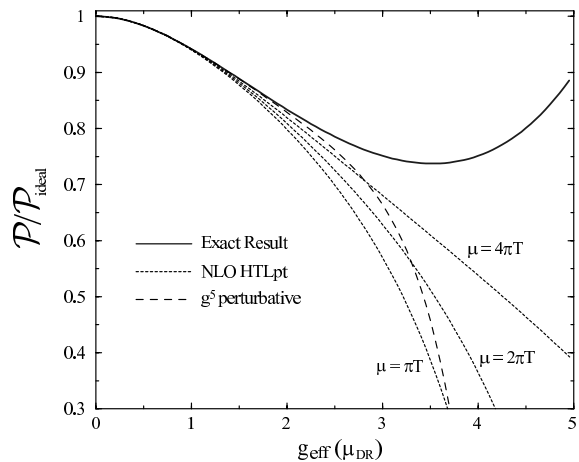}

\caption{
NLO HTLpt prediction for the $O(N_f^0)$ contribution to the
free energy, the numerical result of Ref.~\cite{ipp}, and the perturbative prediction accurate
to $g^5$ as a function of $g_{\rm eff}(\mu_{\rm DR}) = \sqrt{s_f} g(\mu_{\rm DR}) = 2 \pi \sqrt{s_f \alpha_s(\mu_{\rm DR})} $ where $\mu_{\rm DR} = \pi e^{-\gamma} T$.  Dots indicate the point at which there is
no longer a real-valued solution to the gap equation for $m_D$.
In (a) the renormalization scale $\mu$ is varied by a factor of $e$ around $\mu_{\rm DR}$.
In (b) the renormalization scale $\mu$ is varied by a factor of $2$ around $2\pi T$.
In both (a) and (b) the perturbative $g^5$ result is evaluated at the central scale.
}
\label{nffig}
\end{figure}
%%%%%%%%%%%%%%%%%%%%%%%%%%%%%%%%%%%%%%%%%%%%%%%%%%%%%%%%%%%%%%%%%%%%%%%

However, as was the case in pure-glue QCD \cite{ABPS-02}, 
the results seem to lie significantly above the 
lattice data which is available below $5\,T_c$.  There are several reasons
for why HTLpt might fail to describe the lattice data in this temperature range.  One
is that the hard modes are not resummed properly within HTLpt and that a description
using a $\Phi$-derivable approach which explicitly separates the hard and soft modes 
as done in Ref.~\cite{BIR-99} is better.  A second is that 
HTLpt discards some important physics like topological modes or the $Z_N$
symmetry of QCD near the phase transition.  

A third possibility is that the expansion
in $m_D/T$ and $m_q/T$ breaks down at these temperatures.  Numerically, 
$m_D/T \sim 1.2$ and $m_q/T \sim 0.5$ at $5\,T_c$ and  
$m_D/T \sim 1.6$ and $m_q/T \sim 0.6$ at $2\,T_c$, 
which casts doubt on the applicability of the expansion in this temperature
range. However, in the case of pure-glue we have been able to compare the
LO HTLpt result expanded to $O(\hat m_D^6)$ with the non-truncated LO expression
which is accurate to all orders in $\hat m_D$ and find that the expansions converge 
very rapidly.  Numerically we find that at $m_D/T=5$ truncations of the
LO order result accurate to $\hat m_D^4$ and $\hat m_D^6$ reproduce the exact 
result to 5\% and 0.2\%, respectively.  There have also been studies of the
convergence of the mass expansions of the three-loop free energy for a massless
scalar field theory using screened perturbation theory \cite{AS-01} and the 
$\Phi$-derivable approach \cite{BP-01} which demonstrated that mass expansions 
also converge very rapidly at NLO and NNLO.  This gives us some confidence that
the truncated NLO solutions are numerically reliable.  

%%%%%%%%%%%%%%%%%%%%%%%%%%%%%%%%%%%%%%%%%%%%%%%%%%%%%%%%%%%%%%%%%%%%%%%
\begin{figure}[t]

\includegraphics[width=7.2cm]{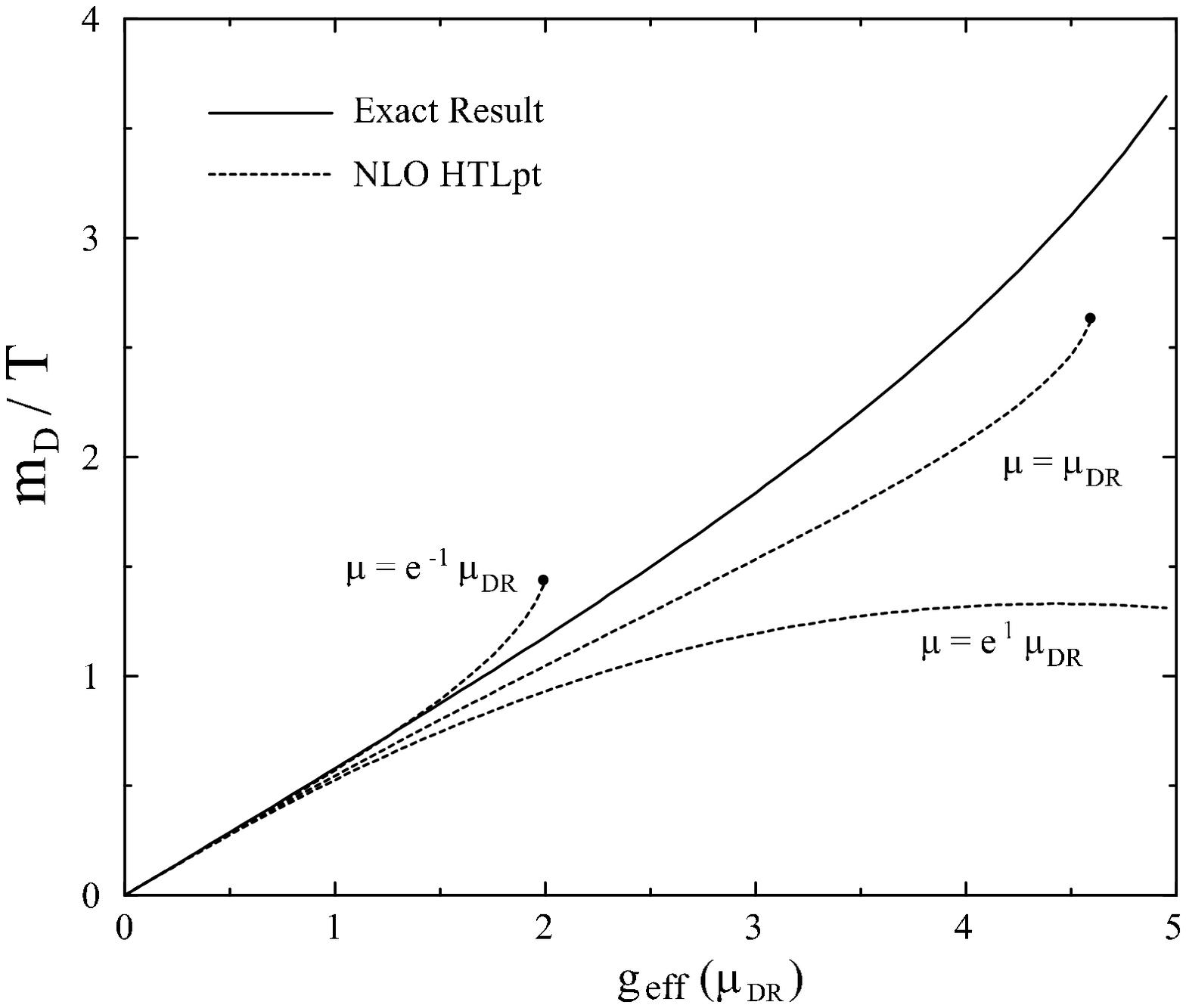}

\vspace{5mm}

\includegraphics[width=7.2cm]{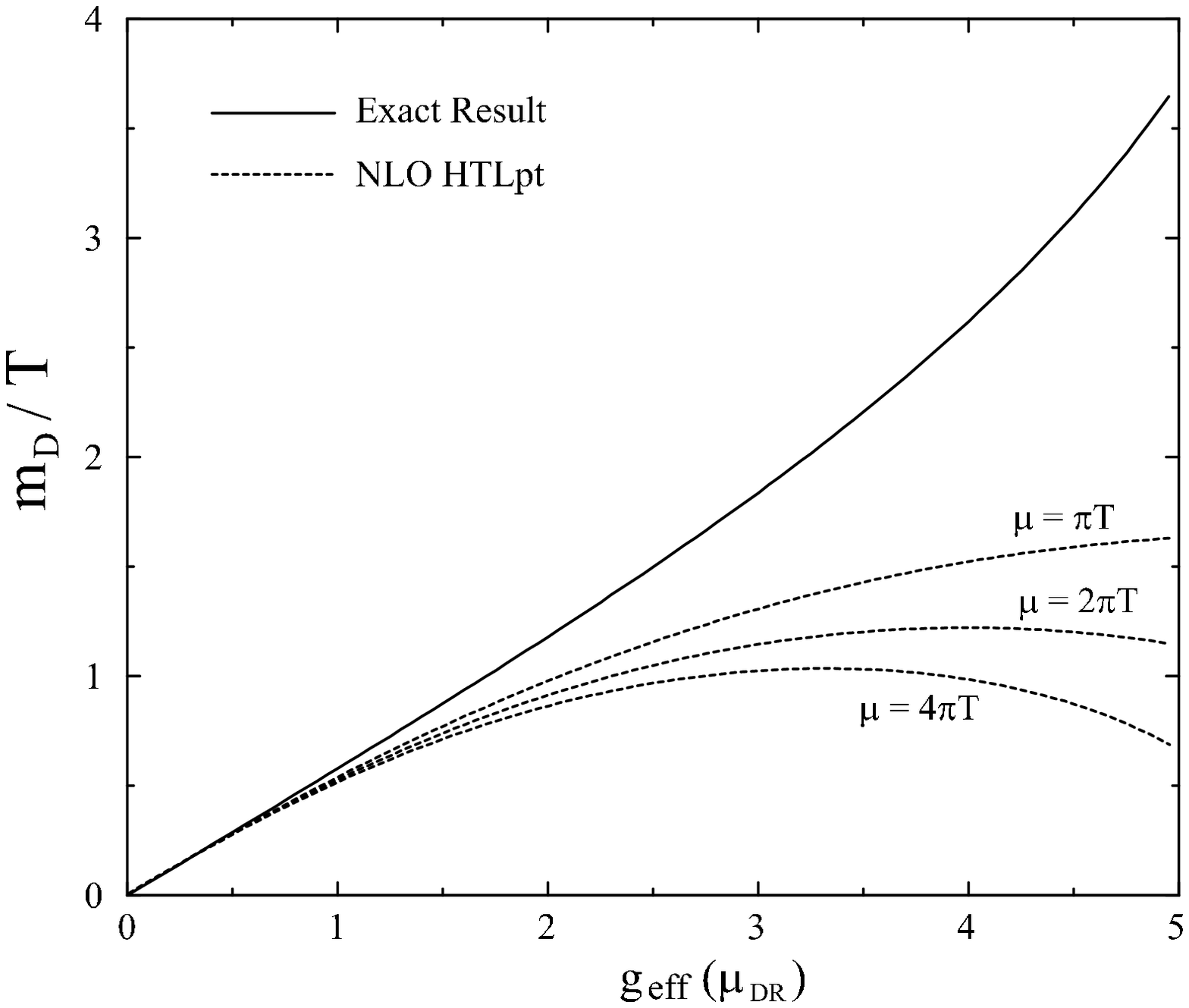}

\caption{
NLO HTLpt prediction for $m_D$ in the large $N_f$ limit 
and the exact numerical result \cite{tonyprivate} as a 
function of $g_{\rm eff}(\mu_{\rm DR}) = \sqrt{s_f} g(\mu_{\rm DR}) = 2 \pi \sqrt{s_f \alpha_s(\mu_{\rm DR})} $ where $\mu_{\rm DR} = \pi e^{-\gamma} T$.
In (a) the renormalization scale $\mu$ is varied by a factor of $e$ around $\mu_{\rm DR}$.
In (b) the renormalization scale $\mu$ is varied by a factor of $2$ around $2\pi T$.
}
\label{nffig2}
\end{figure}
%%%%%%%%%%%%%%%%%%%%%%%%%%%%%%%%%%%%%%%%%%%%%%%%%%%%%%%%%%%%%%%%%%%%%%%

\section{Large $N_f$}
\label{largeNf}

In the limit that $N_f$ is taken large while holding $g^2 N_f$ fixed it is possible
to solve for the $O(N_f^0)$ contribution to the free energy exactly \cite{moore,ipp}.  In
Fig.~\ref{nffig} we plot the NLO HTLpt prediction for the $O(N_f^0)$ contribution to the
free energy along with the numerical result of Ref.~\cite{ipp} and the perturbative prediction
which is accurate to $O(\alpha_s^{5/2})$.  In Fig.~\ref{nffig2} we plot the NLO HTLpt prediction 
for $m_D$ in the large $N_f$ limit and the exact numerical result \cite{tonyprivate}.
The HTLpt predictions for both the free energy and the Debye mass seem diverge from the
exact result around $g_{\rm eff} \sim 2$ regardless of the scale which is chosen; however, 
for both quantities, choosing the scale to be $\mu = \mu_{\rm DR} = \pi e^{-\gamma} T$ seems
to provide a reasonable reproduction of the exact results.  This result is comparable to
the performance of the $\Phi$-derivable approach in the large $N_f$ limit \cite{rebhan-03}. 

\section{Conclusions}

In this paper we have extended our previous HTLpt calculation of the thermodynamic functions
in pure-glue QCD to include the contribution of $N_f$ massless quarks.  We have
presented results for the leading- and next-to-leading-order HTLpt predictions for 
the QCD free energy for arbitrary $N_f$.  Using the NLO HTLpt expression for the
thermodynamic potential we were able to find variational solutions for both the
quark and gluon mass parameters allowing a first-principles prediction of the QCD
free energy.  As in the case of pure-glue we find that the NLO HTLpt prediction lies
significantly above the available lattice data below $5\,T_c$;  however, the problem
of oscillation of successive approximations and large scale dependence of the perturbative
results is eliminated by using this reorganization.  

The failure of HTLpt to describe
the lattice data in this region could be attributed to the failure of the expansion
performed in $\hat m_D$ and $\hat m_q$; however, a study of the convergence of the 
truncated LO expressions to a numerical evaluation of the exact LO expression shows 
that these expansions converge very rapidly.  Therefore, we are steered towards
the conclusion that a systematic description of QCD thermodynamics 
using HTLpt is not appropriate below $5\,T_c$.  The $\Phi$-derivable approach
seems to agree better with the lattice data in this range so perhaps
HTLpt is not resumming the hard modes properly and an explicit separation of
hard and soft scales is required.  However, we should point out that some authors
believe that a description of QCD thermodynamics near the phase transition in 
terms of Polyakov loops is necessary \cite{pisarski-00}. 

We have also compared the NLO HTLpt free energy and Debye mass 
with exact results which are 
available in the large $N_f$ limit.  This comparison shows
that, in the large $N_f$ limit, NLO HTLpt agrees with the exact result only out
to $g_{\rm eff} \sim 2$ and has large scale dependence after this point.  The
large scale dependence is not surprising given the fact that in the large $N_f$
limit the running of the coupling constant is driven by the presence of the
Landau singularity and even the exact results are sensitive to this beyond
$g_{\rm eff} \sim 5$.  The poor performance of NLO HTLpt, however, is comparable to recent
large $N_f$ predictions within the $\Phi$-derivable approach.
The failure of both approaches to agree better with the exact
result for large values of $g_{\rm eff}$ is an indication that a 
description of strongly-coupled QCD thermodynamics solely in terms 
of HTL quasiparticles is perhaps inappropriate.  However, it is possible
that the physics of large-$N_f$ QCD is so different from that of 
QCD with a small number of flavors that it cannot serve as a definitive
testing ground for the applicability of the quasiparticle approach 
to the physical case \cite{peshier-03}. 

\section*{Acknowledgments}
M.S. would like to thank A.Rebhan for discussions and comments.
J.O.A was supported by the Stichting voor Fundamenteel Onderzoek der Materie
(FOM), which is supported by the Nederlandse Organisatie voor Wetenschappelijk
Onderzoek (NWO).  M.S. was supported by US DOE grant DE-FG02-96ER40945 and 
by an FWF Der Wissenschaftsfonds Lise Meitner fellowship M689.

\appendix
\renewcommand{\theequation}{\thesection.\arabic{equation}}
\section{HTL Feynman Rules}
\label{app:rules}
In this appendix, we present Feynman rules for HTL perturbation theory 
in QCD.  We give explicit expressions for the propagators 
and for the quark-gluon 3 and 4 vertices.
The Feynman rules are given in Minkowski space to facilitate 
applications to real-time processes.
A Minkowski momentum is denoted $p = (p_0, {\bf p})$,
and the inner product is $p \cdot q = p_0 q_0 - {\bf p} \cdot {\bf q}$.
The vector that specifies the thermal rest frame 
is $n = (1,{\bf 0})$.

\subsection{Gluon Self-energy}

The HTL gluon self-energy tensor for a gluon of momentum $p$ is
\bqa
\label{a1}
\Pi^{\mu\nu}(p)=m_D^2\left[
{\cal T}^{\mu\nu}(p,-p)-n^{\mu}n^{\nu}
\right]\;.
\eqa
%
%{\bf [mpj]}
The tensor ${\cal T}^{\mu\nu}(p,q)$, which is defined only for momenta
that satisfy $p+q=0$, is
\bqa
{\cal T}^{\mu\nu}(p,-p)=
\left \langle y^{\mu}y^{\nu}{p\!\cdot\!n\over p\!\cdot\!y}
\right\rangle_{\bf\hat{y}} \;.
\label{T2-def}
\eqa
%
%{\bf [jmp]}
The angular brackets indicate averaging
over the spatial directions of the light-like vector $y=(1,\hat{\bf y})$.
The tensor ${\cal T}^{\mu\nu}$ is symmetric in $\mu$ and $\nu$
and satisfies the ``Ward identity''
\bqa
p_{\mu}{\cal T}^{\mu\nu}(p,-p)=
p\!\cdot\!n\;n^{\nu}\;.
\label{ward-t2}
\eqa
%
%{\bf [jmp]}
The self-energy tensor $\Pi^{\mu\nu}$ is therefore also
symmetric in $\mu$ and $\nu$ and satisfies
\bqa
p_{\mu}\Pi^{\mu\nu}(p)&=&0\;,\\
\label{contr}
g_{\mu\nu}\Pi^{\mu\nu}(p)&=&-m_D^2\;.
\eqa
%
%{\bf [jmp]}

The gluon self-energy tensor can be expressed in terms of two scalar functions,
the transverse and longitudinal self-energies $\Pi_T$ and $\Pi_L$,
defined by
\bqa
\label{pit2}
\Pi_T(p)&=&{1\over d-1}\left(
\delta^{ij}-\hat{p}^i\hat{p}^j
\right)\Pi^{ij}(p)\;, \\
\label{pil2}
\Pi_L(p)&=&-\Pi^{00}(p)\;,
\eqa
%
%{\bf [jmp]}
where ${\bf \hat p}$ is the unit vector
in the direction of ${\bf p}$.
In terms of these functions, the self-energy tensor is
\bqa
\label{pi-def}
\Pi^{\mu\nu}(p) \;=\; - \Pi_T(p) T_p^{\mu\nu}
- {1\over n_p^2} \Pi_L(p) L_p^{\mu\nu}\;,
\eqa
%
%{\bf [jmp]}
where the tensors $T_p$ and $L_p$ are
\bqa
T_p^{\mu\nu}&=&g^{\mu\nu} - {p^{\mu}p^{\nu} \over p^2}
-{n_p^{\mu}n_p^{\nu}\over n_p^2}\;,\\
L_p^{\mu\nu}&=&{n_p^{\mu}n_p^{\nu} \over n_p^2}\;.
\eqa
%
%{\bf [jmp]}
The four-vector $n_p^{\mu}$ is
\bqa
n_p^{\mu} \;=\; n^{\mu} - {n\!\cdot\!p\over p^2} p^{\mu}
\eqa
%
%{\bf [jmp]}
and satisfies $p\!\cdot\!n_p=0$ and $n^2_p = 1 - (n\!\cdot\!p)^2/p^2$.
The equation~(\ref{contr}) reduces to the identity
\bqa
(d-1)\Pi_T(p)+{1\over n^2_p}\Pi_L(p) \;=\; m_D^2 \;.
\label{PiTL-id}
\eqa
%
%{\bf [jmp]} 
We can express both self-energy functions in terms of the function
${\cal T}^{00}$ defined by (\ref{T2-def}):
\bqa
\Pi_T(p)&=& {m_D^2 \over (d-1) n_p^2}
\left[ {\cal T}^{00}(p,-p) - 1 + n_p^2  \right] \;,
\label{PiT-T}
\\
\Pi_L(p)&=& m_D^2
\left[ 1- {\cal T}^{00}(p,-p) \right]\;,
\label{PiT-L}
\eqa
%
%{\bf [jmp]}

In the tensor ${\cal T}^{\mu \nu}(p,-p)$ defined in~(\ref{T2-def}),
the angular brackets indicate the angular average over
the unit vector $\hat{\bf y}$.
In almost all previous work, the angular average in~(\ref{T2-def}) has been
taken in $d=3$ dimensions. For consistency of higher order radiative
corrections, it is essential to take the angular average in $d=3-2\epsilon$
dimensions and analytically continue to $d=3$ only after all poles in
$\epsilon$ have been cancelled.
Expressing the angular average as an integral over the cosine of an angle,
the expression for the $00$ component of the tensor is
\bqa
{\cal T}^{00}(p,-p) &=& {w(\epsilon)\over2}
\int_{-1}^1dc\;(1-c^2)^{-\epsilon}{p_0\over p_0-|{\bf p}|c} \;,
\label{T00-int}
\eqa
%
%{\bf [jmp]}
where the weight function $w(\epsilon)$ is
\bqa
w(\epsilon)={\Gamma(2-2\epsilon)\over\Gamma^2(1-\epsilon)}2^{2\epsilon}
= {\Gamma({3\over2}-\epsilon)
        \over \Gamma({3\over2}) \Gamma(1-\epsilon)} \;.
\label{weight}
\eqa
%
%{\bf [jmp]}
The integral in (\ref{T00-int}) must be defined so that it is analytic
at $p_0=\infty$.
It then has a branch cut running from $p_0=-|{\bf p}|$ to $p_0=+|{\bf p}|$.
If we take the limit $\epsilon\rightarrow 0$, it reduces to
\begin{eqnarray}
{\cal T}^{00}(p,-p) &=&
{p_0 \over 2|{\bf p}|}
                \log {p_0 +|{\bf p}| \over p_0-|{\bf p}|}\;,
\end{eqnarray}
%
%{\bf [jmp]}
which is the expression that
appears in the usual HTL self-energy functions.

\label{app:prop}

The Feynman rule for the gluon propagator is
\bqa
i \delta^{a b} \Delta_{\mu\nu}(p) \;,
\eqa
%
%{\bf [jm]}
where the gluon propagator tensor $\Delta_{\mu\nu}$
depends on the choice of gauge fixing.
We consider two possibilities that introduce an arbitrary
gauge parameter $\xi$:  general covariant gauge and
general Coulomb gauge.
In both cases, the inverse propagator reduces in the
limit $\xi\rightarrow\infty$ to
\bqa
\Delta^{-1}_{\infty}(p)^{\mu\nu}&=&
-p^2 g^{\mu \nu} + p^\mu p^\nu - \Pi^{\mu\nu}(p)\;.
\label{delta-inv:inf0}
\eqa
%
%{\bf [jmp]}
This can also be written
\bqa
\Delta^{-1}_{\infty}(p)^{\mu\nu}&=&
- {1 \over \Delta_T(p)}       T_p^{\mu\nu}
+ {1 \over n_p^2 \Delta_L(p)} L_p^{\mu\nu}\;,
\label{delta-inv:inf}
\eqa
%
%{\bf [jmp]}
where $\Delta_T$ and $\Delta_L$ are the transverse and longitudinal
propagators:
\bqa
\Delta_T(p)&=&{1 \over p^2-\Pi_T(p)}\;,
\label{Delta-T:M}
\\
\Delta_L(p)&=&{1 \over - n_p^2 p^2+\Pi_L(p)}\;.
\label{Delta-L:M}
\eqa
%
%{\bf [jmp]}
The inverse propagator for general $\xi$ is
\bqa
\Delta^{-1}(p)^{\mu\nu}&=&
\Delta^{-1}_{\infty}(p)^{\mu\nu}-{1\over\xi}
p^{\mu}p^{\nu}\hspace{0.2cm}\mbox{covariant}\;,
\label{Delinv:cov}
\\ \nonumber
&=&\Delta^{-1}_{\infty}(p)^{\mu\nu}-{1\over\xi}
\left(p^{\mu}-p\!\cdot\!n\;n^{\mu}\right)
\left(p^{\nu}-p\!\cdot\!n\;n^{\nu}\right)
\\ && 
\hspace{3.7cm}\mbox{Coulomb}\;.
\label{Delinv:C}
\eqa
%
%{\bf [jmp]}
The propagators obtained by inverting the tensors in~(\ref{Delinv:C})
and~(\ref{Delinv:cov}) are
\bqa
\Delta^{\mu\nu}(p)&=&-\Delta_T(p)T_p^{\mu\nu}
+\Delta_L(p)n_p^{\mu}n_p^{\nu}
- \xi {p^{\mu}p^{\nu} \over (p^2)^2}
\nonumber \\
&& \hspace{3.7cm}\mbox{covariant}\;,
\label{D-cov}
\\
&=&-\Delta_T(p)T_p^{\mu\nu}
+\Delta_L(p)n^{\mu}n^{\nu}-\xi{p^{\mu}p^{\nu}\over\left(n_p^2p^2\right)^2}
\nonumber \\
&& \hspace{3.7cm}
\mbox{Coulomb}\;.
\label{D-C}
\eqa
%
%{\bf [jmp]}

It is convenient to define the following combination of propagators:
\bqa
\Delta_X(p) &=& \Delta_L(p)+{1\over n_p^2}\Delta_T(p) \;.
\label{Delta-X}
\eqa
%
%{\bf [jmp]}
Using (\ref{PiTL-id}), (\ref{Delta-T:M}), and (\ref{Delta-L:M}),
it can be expressed in the alternative form
\bqa
\Delta_X(p) &=&
\left[ m_D^2 - d \, \Pi_T(p) \right] \Delta_L(p) \Delta_T(p) \;,
\label{Delta-X:2}
\eqa
%
%{\bf [jmp]}
which shows that it vanishes in the limit $m_D \to 0$.
In the covariant gauge, the propagator tensor can be written
\bqa\nonumber
&&\Delta^{\mu\nu}(p) =
\left[ - \Delta_T(p) g^{\mu \nu} + \Delta_X(p) n^\mu n^\nu \right]
\\ \nonumber&& 
- {n \!\cdot\! p \over p^2} \Delta_X(p)
        \left( p^\mu n^\nu  + n^\mu p^\nu \right)
\\
&&
+ \left[ \Delta_T(p) + {(n \!\cdot\! p)^2 \over p^2} \Delta_X(p)
        - {\xi \over p^2} \right] {p^\mu p^\nu \over p^2} \;.
\label{gprop-TC}
\eqa
%
%{\bf [jmp]}
This decomposition of the propagator into three terms
has proved to be particularly convenient for explicit calculations.
For example, the first term satisfies the identity
\bqa
&&\nonumber
\left[- \Delta_T(p) g_{\mu \nu} + \Delta_X(p) n_\mu n_\nu \right]
\Delta^{-1}_{\infty}(p)^{\nu\lambda}  =
\\
&&
{g_\mu}^\lambda - {p_\mu p^\lambda \over p^2}
%\\&&\\hspace{-0.5cm}
+ {n \!\cdot\! p \over n_p^2 p^2} {\Delta_X(p) \over \Delta_L(p)}
        p_\mu n_p^\lambda \;.
\label{propid:2}
\eqa
%
%{\bf [jmp]}
\subsection{Quark Self-energy}
The HTL self-energy of a quark with momentum $p$ is given by
\bqa
\label{selfq}
\Sigma(P)=m_q^2/\!\!\!\!{\cal T}(p)
\;,
\eqa
%{\bf [jmp]}
where
\bqa
\label{deftf}
{\cal T}^{\mu}(p)=
\left\langle{y^{\mu}\over p\cdot y}
%p_0-{\bf p}\cdot\hat{\bf y}}
\right\rangle_{\hat{\bf y}}
\;.
\eqa
%{\bf [jmp]}
Expressing the angular average as an integral over the cosine of an angle,
the expression is
\bqa 
\label{def-tf}
{\cal T}^{\mu}(p)={w(\epsilon)\over2}
\int_{-1}^1dc\;(1-c^2)^{-\epsilon}{y^{\mu}\over p_0-|{\bf p}|c}\;,
\eqa
%{\bf [jmp]}
The integral in (\ref{def-tf}) must be defined so that it is analytic 
at $p_0=\infty$.
It then has a branch cut running from $p_0=-|{\bf p}|$ to $p_0=+|{\bf p}|$.  
In three dimensions, this reduces to
\bqa\nonumber
\Sigma(P)&=&
{m_q^2\over 2|{\bf p}|}\gamma_0\log{p_0+|{\bf p}|\over p_0-|{\bf p|}}
\\&&
+{m_q^2\over |{\bf p}|}{\bf \gamma}\cdot \hat{\bf p}
\left(1-{p_0\over 2|{\bf p}|}\log{p_0+|{\bf p}|\over p_0-|{\bf p|}}\right)\;.
\eqa
%{\bf [jmp]}

\subsection{Quark Propagator}
The Feynman rule for the quark propagator is 
\bqa
i\delta^{ab}S(p)\;.
\eqa
%{\bf [jmp]}
The quark propagator can be written as
\bqa
\label{qprop}
S(p)={1\over/\!\!\!p-\Sigma(p)}\;,
\eqa
%{\bf [jmp]}
where the quark self-energy is given by~(\ref{selfq}).
The inverse quark propagator can be written as
\bqa
S^{-1}(p)=/\!\!\!p-\Sigma(p)\;.
\eqa
%{\bf [jmp]}
This can be written as
\bqa
S^{-1}(p)=/\!\!\!\!{\cal A}(p)\;,
\eqa
%{\bf [jmp]}
where we have organized $A_0(p)$ and $A_S(p)$ into:
\bqa
\label{qself}
A_{\mu}(p)=(A_0(p),A_S(p)\hat{\bf p})\;.
\eqa
%{\bf [jmp]}
The functions $A_0(p)$ and $A_S(p)$ are defined
\bqa
\label{aodef}
A_0(p)&=&p_0-{m_q^2\over p_0}{\cal T}_p\;,\\
A_S(p)&=& |{\bf p}|+{m_q^2\over |{\bf p}|}\left[1-{\cal T}_p\right]\;.
\label{asdef}
\eqa
%{\bf [jmp]}
\subsection{Quark-gluon vertex}
The quark-gluon vertex with outgoing gluon momentum $p$, 
incoming fermion momentum $q$, and outgoing quark momentum $r$, 
Lorentz index $\mu$ and color index $a$ is
\bqa
\label{3qgv}
\Gamma^{\mu}_a(p,q,r)
=gt_a\left(\gamma^{\mu}-m_q^2\tilde{{\cal T}}^{\mu}(p,q,r)\right)\;.
\eqa
%{\bf [jm]}
The tensor in the HTL correction term is only defined for $p-q+r=0$:
\bqa
\tilde{{\cal T}}^{\mu}(p,q,r)
=\left\langle
y^{\mu}\left({y\!\!\!/\over q\!\cdot\!y\;\;r\!\cdot\!y}\right)
\right\rangle_{\hat{\bf y}}\;.
\label{T3-def}
\eqa
%{\bf [jm]}
This tensor is even under the permutation of $q$ and $r$.
It satisfies the ``Ward identity''
\bqa
p_{\mu}\tilde{\cal T}^{\mu}(p,q,r)=
\tilde{\cal T}^{\mu}(q)-\tilde{\cal T}^{\mu}(r)\;.
\eqa
%{\bf [jm]}
The quark-gluon vertex therefore satisfies
the Ward identity
\bqa
p_{\mu}\Gamma^{\mu}(p,q,r)=S^{-1}(q)-S^{-1}(r)\;.
\label{qward1}
\eqa
%{\bf [jm]}

\subsection{Quark-gluon four-vertex}
We define the quark-gluon four-point vertex with outgoing gluon 
        momenta $p$ and $q$, incoming fermion momentum $r$, and outgoing
        fermion momentum $s$.  Generally this vertex has both adjoint and
        fundamental indices, however, for this calculation we will only 
        need the quark-gluon four-point vertex traced over the adjoint 
        color indices.  In this case
\bqa\nonumber
\delta^{ab} \Gamma^{\mu\nu}_{abij}(p,q,r,s) &=& 
    - g^2 m_q^2 c_F \delta_{ij} \tilde{\cal T}^{\mu\nu}(p,q,r,s)  \\
&    \equiv &g^2 c_F \delta_{ij} \Gamma^{\mu\nu}  \, ,
\label{4qgv}
\eqa
%{\bf [m]}
where $c_F = (N_c^2-1)/(2 N_c)$.
There is no tree-level term. The tensor in the 
HTL correction term is only defined for $p+q-r+s=0$
\bqa\nonumber
\tilde{{\cal T}}^{\mu\nu}(p,q,r,s)
&=&\left\langle
y^{\mu}y^{\nu}\left({1\over r\!\cdot\!y}+{1\over s\!\cdot\!y}\right)
\right.\\&&\times \left.
{y\!\!\!/\over[(r-p)\!\cdot\!y]\;[(s+p)\!\cdot\!y]}
\right\rangle\;.
\label{T4-def}
\eqa
%{\bf [mj]}
This tensor is symmetric in $\mu$ and $\nu$ and is traceless.
It satisfies the Ward identity:
\bqa
p_{\mu}\Gamma^{\mu\nu}(p,q,r,s)=\Gamma^{\nu}(q,r-p,s)-\Gamma^{\nu}(q,r,s+p)\;.
\label{qward2}
\eqa
%{\bf [mj]}
\subsection{HTL Quark Counterterm}
The Feynman rule for the insertion of an HTL quark counterterm into a quark
propagator is
\bqa
i\delta^{ab}\Sigma(p)\;,
\eqa
where $\Sigma(p)$ is the HTL quark self-energy given in~(\ref{qself}).
\subsection{Imaginary-time formalism}
\label{app:ITF}

In the imaginary-time formalism,
Minkoswski energies have discrete imaginary values
$p_0 = i (2 \pi n T)$
and integrals over Minkowski space are replaced by sum-integrals over
Euclidean vectors $(2 \pi n T, {\bf p})$.
We will use the notation $P=(P_0,{\bf p})$ for Euclidean momenta.
The magnitude of the spatial momentum will be denoted $p = |{\bf p}|$,
and should not be confused with a Minkowski vector.
The inner product of two Euclidean vectors is
$P \cdot Q = P_0 Q_0 + {\bf p} \cdot {\bf q}$.
The vector that specifies the thermal rest frame
remains $n = (1,{\bf 0})$.

The Feynman rules for Minkowski space given above can be easily
adapted to Euclidean space.  The Euclidean tensor in a given
Feynman rule is obtained from the corresponding Minkowski tensor
with raised indices by replacing each Minkowski energy $p_0$
by $iP_0$, where $P_0$ is the corresponding Euclidean energy,
and multipying by $-i$ for every $0$ index.
This prescription transforms $p=(p_0,{\bf p})$ into $P=(P_0,{\bf p})$,
$g^{\mu \nu}$ into $- \delta^{\mu \nu}$,
and $p\!\cdot\!q$ into $-P\!\cdot\!Q$.
The effect on the HTL tensors defined in (\ref{T2-def}),
(\ref{T3-def}), and (\ref{T4-def}) is equivalent to
substituting $p\!\cdot\!n \to - P\!\cdot\!N$ where $N = (-i,{\bf 0})$,
$p\!\cdot\!y \to -P\!\cdot\!Y$ where $Y = (-i,{\bf \hat y})$,
and $y^\mu \to Y^\mu$.
For example, the Euclidean tensor corresponding to (\ref{T2-def}) is
\bqa
{\cal T}^{\mu\nu}(P,-P)=
\left \langle Y^{\mu}Y^{\nu}{P\!\cdot\!N \over P\!\cdot\!Y}
\right\rangle \;.
\label{T2E-def}
\eqa
%
%%{[\bf jmp]}
The average is taken over the directions of the unit vector ${\bf \hat y}$.

Alternatively, one can calculate a diagram
by using the Feynman rules for Minkowski momenta,
reducing the expressions for diagrams to scalars,
and then make the appropriate substitutions,
such as $p^2 \to -P^2$, $p \cdot q \to - P \cdot Q$,
and $n \cdot p \to i n \cdot P$.
For example, the propagator functions (\ref{Delta-T:M})
and (\ref{Delta-L:M}) become
\bqa
\Delta_T(P)&=&{-1 \over P^2 + \Pi_T(P)}\;,
\label{Delta-T}
\\
\Delta_L(P)&=&{1 \over p^2+\Pi_L(P)}\;.
\label{Delta-L}
\eqa
%
%%{[\bf jmp]}
The expressions for the HTL self-energy functions $\Pi_T(P)$
and $\Pi_L(P)$ are given by
(\ref{PiT-T}) and (\ref{PiT-L}) with $n_p^2$ replaced by
$n_P^2 = p^2/P^2$ and ${\cal T}^{00}(p,-p)$ replaced by
\bqa
{\cal T}_P &=& {w(\epsilon)\over2}
        \int_{-1}^1dc\;(1-c^2)^{-\epsilon}{iP_0\over iP_0-pc} \;.
\label{TP-def}
\eqa
%
%%{[\bf jmp]}
Note that this function differs by a sign from the 00 component
of the Euclidean tensor corresponding
to~(\ref{T2-def}):
\bqa
{\cal T}^{00}(P,-P) = - {\cal T}^{00}(p,-p)\bigg|_{p_0 \to iP_0}
                    = - {\cal T}_P \;.
\eqa
%
%{\bf [jmp]}
A more convenient form for calculating sum-integrals
that involve the function ${\cal T}_P$ is
\bqa
{\cal T}_P &=&
        \left\langle {P_0^2 \over P_0^2 + p^2c^2} \right\ranglec \, ,
\label{TP-int}
\eqa
%
%%{[\bf jmp]}
where the angular brackets represent an average over $c$ defined by
\begin{equation}
\left\langle f(c) \right\rangle_{\!c} \equiv w(\epsilon) \int_0^1 dc \,
(1-c^2)^{-\epsilon} f(c)
\label{c-average}
\end{equation}
%{\bf [jmp]}
%
and $w(\epsilon)$ is given in~(\ref{weight}).

\section{Sum-integrals}
\setcounter{equation}{0}
\label{app:sumint}

In the imaginary-time formalism for thermal field theory, 
the 4-momentum $P=(P_0,{\bf p})$is Euclidean with $P^2=P_0^2+{\bf p}^2$. 
The Euclidean energy $p_0$ has discrete values:
$P_0=2n\pi T$ for bosons and $P_0=(2n+1)\pi T$ for fermions,
where $n$ is an integer. 
Loop diagrams involve sums over $P_0$ and integrals over ${\bf p}$. 
With dimensional regularization, the integral is generalized
to $d = 3-2 \epsilon$ spatial dimensions.
We define the dimensionally regularized sum-integral by
\bqa
  \hbox{$\sum$}\!\!\!\!\!\!\int_{P}& \;\equiv\; &
  \left(\frac{e^\gamma\mu^2}{4\pi}\right)^\epsilon\;
  T\sum_{P_0=2n\pi T}\:\int {d^{3-2\epsilon}p \over (2 \pi)^{3-2\epsilon}}\;,\\ 
  \hbox{$\sum$}\!\!\!\!\!\!\int_{\{P\}}& \;\equiv\; &
  \left(\frac{e^\gamma\mu^2}{4\pi}\right)^\epsilon\;
  T\sum_{P_0=(2n+1)\pi T}\:\int {d^{3-2\epsilon}p \over (2 \pi)^{3-2\epsilon}}\;,
\label{sumint-def}
\eqa
%[{\bf jmp}]
where $3-2\epsilon$ is the dimension of space and $\mu$ is an arbitrary
momentum scale. 
The factor $(e^\gamma/4\pi)^\epsilon$
is introduced so that, after minimal subtraction 
of the poles in $\epsilon$
due to ultraviolet divergences, $\mu$ coincides 
with the renormalization
scale of the $\overline{\rm MS}$ renormalization scheme.

\subsection{One-loop sum-integrals}
The simple one-loop sum-integrals required in our calculations
can be derived from the formulas
\bqa\nonumber
\sumint_{P}{p^{2m}\over(P^2)^n}
&=&
\left({\mu\over4\pi T}\right)^{2\epsilon}
{2\Gamma({3\over2}+m-\epsilon)\Gamma(n-{3\over2}-m+\epsilon)
\over\Gamma(n)\Gamma(2-2\epsilon)}
\\\nonumber
&&\;\times\,
\Gamma(1-\epsilon)\zeta(2n-2m-3+2\epsilon)
e^{\epsilon\gamma}
\\ &&
\;\;\;\;\times\,T^{4+2m-2n}(2\pi)^{1+2m-2n}\;,
\\
\sumint_{\{P\}}{p^{2m}\over(P^2)^n}&=&
(2^{2n-2m-d}-1)\sumint_{P}{p^{2m}\over(P^2)^n}\;. 
\eqa
%{\bf [jmp]}
%
The specific bosonic one-loop sum-integrals needed are
\bqa\nonumber
\sumint_{P}{1\over P^2}
&=&{T^2\over12}
\left({\mu\over4\pi T}\right)^{2\epsilon}
\Bigg[1+\left(
2+2{\zeta^{\prime}(-1)\over\zeta(-1)}
\right)\epsilon%+O(\epsilon^2)
\\ && \hspace{-0.7cm}
+\left(4+{\pi^2\over4}
+4{\zeta^{\prime}(-1)\over\zeta(-1)}
+2{\zeta^{\prime\prime}(-1)\over\zeta(-1)}
\right)\epsilon^2
\Bigg]\;, 
\\
\sumint_P {p^2 \over (P^2)^2} &=& {1 \over 8} T^2 \;,
\\\nonumber
\sumint_P {1 \over (P^2)^2} &=&
{1 \over (4\pi)^2} \left({\mu\over4\pi T}\right)^{2\epsilon} 
\\ &&  
  \; \times\,\left[ {1 \over \epsilon} + 2 \gamma
       + \left( {\pi^2 \over 4} - 4 \gamma_1 \right) \epsilon \right] \;,
\\\nonumber
\sumint_P {1 \over p^2 P^2} &=&
{1 \over (4\pi)^2} \left({\mu\over4\pi T}\right)^{2\epsilon} 
2 \left[ {1\over\epsilon} + 2 \gamma + 2 
\right.\\ && \left.       
+ \left( 4 + 4 \gamma + {\pi^2 \over 4} - 4 \gamma_1 \right) 
                \epsilon \right]\;.
 \label{ex1}
\eqa
%{\bf [jmp]}
The specific fermionic one-loop sum-integrals needed are
\bqa
\sumint_{\{P\}}\log P^2&=&{7\pi^2\over360}T^4\;,\\ \nonumber
\sumint_{\{P\}}{1\over P^2}
&=&-{T^2\over24}
\left({\mu\over4\pi T}\right)^{2\epsilon}
\nonumber \\
&& \hspace{-12mm}
\times \bigg[1
+\bigg(
2-2\log2
+2{\zeta^{\prime}(-1)\over\zeta(-1)}
\bigg)\epsilon \nonumber
\\ \nonumber&& \hspace{-11mm}
+\left(4+{\pi^2\over4}
-4\log2-2\log^22
\right.\\ &&\left. \hspace{-7mm}
+4(1-\log2){\zeta^{\prime}(-1)\over\zeta(-1)}
+2{\zeta^{\prime\prime}(-1)\over\zeta(-1)}
\right)
\epsilon^2\bigg]\;,
\label{simple1}
\\ %\nonumber
\sumint_{\{P\}}{1\over(P^2)^2}&=&
{1\over(4\pi)^2}\left({\mu\over4\pi T}\right)^{2\epsilon}
\bigg[
{1\over\epsilon}+2\gamma
%\\&&
+4\log2
\bigg],\hspace{5mm} \\ \nonumber
\sumint_{\{P\}}{p^2\over(P^2)^2}&=&-{T^2\over16}
\left({\mu\over4\pi T}\right)^{2\epsilon}
\\&&
\times \bigg[1+\bigg({4\over3}-2\log2
+2{\zeta^{\prime}(-1)\over\zeta(-1)}\bigg)\epsilon%+O(\epsilon^2)
\bigg],
\\\nonumber
\sumint_{\{P\}}{p^2\over(P^2)^3}&=&{1\over(4\pi)^2}
\left({\mu\over4\pi T}\right)^{2\epsilon}
\\&&
\times \, {3\over4}
\bigg[{1\over\epsilon}+2\gamma-{2\over3}
+4\log2\bigg] ,\\ \nonumber
\sumint_{\{P\}}{p^4\over(P^2)^3}&=&-{5T^2\over64}
\left({\mu\over4\pi T}\right)^{2\epsilon}
\\ && \hspace{-3mm}
\times \bigg[
1+\bigg({14\over15}-2\log2
+2{\zeta^{\prime}(-1)\over\zeta(-1)}
\bigg)\epsilon
\bigg]\;,\\ \nonumber
\sumint_{\{P\}}{p^4\over(P^2)^4}&=&{1\over(4\pi)^2}
\left({\mu\over4\pi T}\right)^{2\epsilon}
\\&& \hspace{3mm}
\times {5\over8}\bigg[
{1\over\epsilon}+2\gamma-{16\over15}
+4\log2
\bigg] \;,
\\ \nonumber
\sumint_{\{P\}}{1\over p^2P^2}&=&{1\over(4\pi)^2}
\left({\mu\over4\pi T}\right)^{2\epsilon}
2\bigg[
{1\over\epsilon}+2+2\gamma+4\log2
\\\nonumber&& \hspace{-5mm}
+\bigg(4+8\log2+4\log^22+4\gamma(1+2\log2)
\\ &&\hspace{26mm}
+{\pi^2\over4}-4\gamma_1
\bigg)\epsilon
\bigg]\;.
\eqa
%
%{\bf [jmp]}

The errors are all of one order higher in $\epsilon$  than 
the smallest term shown.
The number $\gamma_1$ is the first Stieltjes gamma constant
defined by the equation
\begin{equation}
\label{zeta}
\zeta(1+z) = {1 \over z} + \gamma - \gamma_1 z + O(z^2)\;.
\end{equation}
%[{\bf jmp}]
\subsection{One-loop HTL sum-integrals}

We also need some more difficult one-loop sum-integrals 
that involve the HTL function 
defined in (\ref{def-tf}).

The specific bosonic sum-integrals needed are
\bqa\nonumber
\sumint_P {1 \over p^4} {\cal T}_P &=&
{1 \over (4\pi)^2} \left({\mu\over4\pi T}\right)^{2\epsilon}
\\&& \times
(-1)\left[ 
{1 \over \epsilon} + 2 \gamma 
+ 2\log2\right] \
\;,
\label{exa}
\\ \nonumber
\sumint_P {1 \over p^2 P^2} {\cal T}_P &=&
{1 \over (4\pi)^2} \left({\mu\over4\pi T}\right)^{2\epsilon}
\\ && \hspace{-12mm} \times
\left[ 
2 \log2 \left({1 \over \epsilon} + 2 \gamma \right)
+ 2 \log^2 2 + {\pi^2 \over 3}  \right] \;, \\
\sumint_P {1 \over (P^2)^2} {\cal T}_P &=&
{1 \over (4\pi)^2} \left({\mu\over4\pi T}\right)^{2\epsilon}
{1 \over 2}
\left[ 
{1 \over \epsilon} + 2 \gamma + 1 \right] . \hspace{3mm}
%\\
%\sumint_P {1 \over p^4} ({\cal T}_P)^2 &=&
%{1 \over (4\pi)^2} \left({\mu\over4\pi T}\right)^{2\epsilon}
%\left( - {2 \over 3} \right)
%\left[ (1+ 2 \log 2) \left( {1 \over \epsilon} + 2 \gamma \right)
%\right.
%\nonumber\\
%&& \hspace{5cm} \left.
%       - {4 \over 3} + {22 \over 3} \log 2 + 2 \log^2 2 \right] \;.
%\label{sumint-T:5}
\eqa
%[{\bf jmp}]

The specific fermionic sum-integrals needed are
\bqa\nonumber
\sumint_{\{P\}} {1 \over (P^2)^2} {\cal T}_P &=&
{1 \over (4\pi)^2} \left({\mu\over4\pi T}\right)^{2\epsilon}
\\ && \hspace{-2mm} \times
{1 \over 2}
\bigg[ {1 \over \epsilon} + 2 \gamma + 1 
+ 4\log2\bigg] \;,
\label{ht1} \\ \nonumber
\sumint_{\{P\}}{1\over p^2P^2}{\cal T}_P&=&
{2\over(4\pi)^2}\left({\mu\over4\pi T}\right)^{2\epsilon}
\\&& \hspace{-15mm} \times
\left[ \log2 \left({1 \over \epsilon} + 2 \gamma \right) 
        + 5 \log^2 2 + {\pi^2 \over 6}  \right] \;, \\ \nonumber
\sumint_{\{P\}}{1\over P^2P_0^2}{\cal T}_P&=&
{1\over(4\pi)^2}\left({\mu\over4\pi T}\right)^{2\epsilon}
\\ && \nonumber \hspace{-11mm}
\times\Bigg[ {1\over \epsilon^2}
+2(\gamma+2\log2){1\over \epsilon}
+{\pi^2\over4}
\\&& \hspace{-4mm}
+4\log^22
+8\gamma\log2-4\gamma_1 \Bigg] \;, \\ \nonumber
\sumint_{\{P\}}{1\over p^2P_0^2}\left({\cal T}_P\right)^2&=&
{4\over(4\pi)^2}\left({\mu\over4\pi T}\right)^{2\epsilon}
\\ && \hspace{-7mm} \times
\bigg[\log2\left({1\over\epsilon}+2\gamma\right)
+5\log^22
\bigg]\;,
\label{htlf} \\
\nonumber
\sumint_{\{P\}}{1\over P^2}
\bigg\langle {1\over(Q\!\cdot\!Y)^2} \bigg\rangle_c
%\langle{1\over(P\!\cdot\!Y)^2}\rangle_{\!\!\bf \hat y}
&=&
{1\over(4\pi)^2}\left({\mu\over4\pi T}\right)^{2\epsilon} \hspace{2.2cm}
\\&&
\hspace{-12mm} \times(-1)\bigg[{1\over\epsilon}-1+2\gamma
+4\log2\bigg]\;.
\label{cp1ly} 
\eqa
%[{\bf jp} [m] mod last]
The errors are all of order $\epsilon$

It is straightforward to calculate the sum-integrals
(\ref{ht1})--(\ref{htlf}) using the
representation (\ref{TP-int}) of the function ${\cal T}_P$.
For example, the sum-integral (\ref{exa}) can be written
\bqa
\sumint_P {1 \over p^4} {\cal T}_P =
\sumint_P {1 \over p^4}
        \left\langle {P_0^2\over P_0^2  + p^2c^2} \right\ranglec \, ,
\label{example}
\eqa
%
%[{\bf jmp}]
where the angular brackets denote an average over $c$
as defined in (\ref{c-average}).
\bqa
\sumint_P {1 \over p^4} {\cal T}_P =\sumint_{P}{1\over p^4}
\left[1-\left\langle{p^2c^2\over P_0^2+p^2c^2}\right\rangle_c\right] .
\eqa
%[{\bf jmp}]
The first term in the square brackets vanishes with dimensional regularization,
while after rescaling the momentum by ${\bf p}\rightarrow{\bf p}/c$, the
second term reads
\bqa
\sumint_P {1 \over p^4} {\cal T}_P =-
\left\langle c^{1+2\epsilon}\right\rangle_c
\sumint_{P}{1\over p^2P^2}\;.
\eqa
%[{\bf jmp}]
Evaluating the average over $c$,
using the expression (\ref{ex1}) for the sum-integral,
and expanding in powers of $\epsilon$, we obtain the result
(\ref{exa}).
Following the same strategy, all the sum-integrals
(\ref{ht1})--(\ref{htlf}) can be reduced to linear
combinations of simple sum-integrals
with coefficients that are averages over $c$.
The only difficult integral is the double average over $c$
that arises from (\ref{htlf}):
\bqa\nonumber
\left\langle{c_1^{1+2\epsilon}-c_2^{1+2\epsilon}\over c_1^2-c_2^2}
\right\rangle_{c_1,c_2}
=2\log2+2\left(\log^22-2\log2\right)\epsilon\;.
\\ &&\eqa
%{\bf [jmp]}

\subsection{Simple two-loop sum-integrals}
The simple two-loop sum-integrals that are needed are
\bqa
\sumint_{\{PQ\}}{1\over P^2Q^2R^2}&=&0 \; , \\ \nonumber
\sumint_{\{PQ\}}{1\over P^2Q^2r^2}&=&
{T^2 \over (4 \pi)^2} \left({\mu\over4\pi T}\right)^{4\epsilon}
\\&& \hspace{-16mm} \times
\left(-{1\over6}\right)\left[{1\over\epsilon}
+4
-2\log2+4{\zeta^{\prime}(-1)\over\zeta(-1)}
\right]
\label{two1}
\; , \\ \nonumber
\sumint_{\{PQ\}}{q^2\over P^2Q^2r^4}&=&
{T^2 \over (4 \pi)^2} \left({\mu\over4\pi T}\right)^{4\epsilon} \hspace{2.4cm}
\\&& \hspace{-26mm} \times
\left(-{1\over12}\right)\left[{1\over\epsilon}+
{11\over3}
+2\gamma-2\log2+2{\zeta^{\prime}(-1)\over\zeta(-1)}
\right]\;, 
\label{two2}
\\ \nonumber
\sumint_{\{PQ\}}{q^2\over P^2Q^2r^2R^2}&=&
{T^2 \over (4 \pi)^2} \left({\mu\over4\pi T}\right)^{4\epsilon}
\\&& 
\times
\left(-{1\over72}\right)
\bigg[{1\over\epsilon}
-7.002
\bigg]
\label{two3}
\;,\\ \nonumber
\sumint_{\{PQ\}}{P\cdot Q\over P^2Q^2r^4}&=&
{T^2 \over (4 \pi)^2} \left({\mu\over4\pi T}\right)^{4\epsilon}
\\&&
\hspace{-5mm} \times\left(-{1\over36}\right)
\left[1-6\gamma+6{\zeta^{\prime}(-1)\over\zeta(-1)}
\right]
\label{twolast}
\;, \\ \nonumber
\sumint_{\{PQ\}}{p^2\over q^2P^2Q^2R^2}&=&
{T^2\over(4\pi)^2}\left({\mu\over4\pi T}\right)^{4\epsilon}
\\ &&
\times
\left({5\over72}\right)
\left[
{1\over\epsilon}
+9.5667
\right]
\label{ntwo1}
\;,\\ \nonumber
\sumint_{\{PQ\}}{r^2\over q^2P^2Q^2R^2}&=&
{T^2\over(4\pi)^2}\left({\mu\over4\pi T}\right)^{4\epsilon}
\\&&
\times
\left(-{1\over18}\right)
\left[
{1\over\epsilon}
+8.1420\right]\;,
\label{ntwo2}
\eqa
%{\bf [jmp]}
where $R=-(P+Q)$ and $r=|{\bf p}+{\bf q}|$.  The corrections are all of order
$\epsilon$. 
To motivate the integration formula we will use to evaluate the two-loop
sum-integrals, we first present the analogous integration formula
for one-loop sum-integrals.  In a one-loop sum-integral,
the sum over $P_0$ can be replaced by a contour integral in $p_0 = -i P_0$:
\begin{eqnarray}\nonumber
\sumint_P F(P) &=& \lim_{\eta \to 0^+}
\int {d p_0 \over 2 \pi i} \int_{\bf p}
\left[ F(-i p_0,{\bf p})  - F(0,{\bf p}) \right]
\\ &&
\hspace{2.5cm}\times
e^{\eta p_0} n(p_0) \;,
\end{eqnarray}
%
%{\bf [m]}
where $n(p_0) = 1/(e^{\beta p_0} - 1)$ is the Bose-Einstein thermal
distribution and the contour runs
from $-\infty$ to $+\infty$ above the real axis and
from  $+\infty$ to $-\infty$ below the real axis.
This formula can be expressed in a more convenient form by
collapsing the contour onto the real axis
and separating out those terms  with the exponential convergence
factor $n(|p_0|)$.  The remaining terms run along contours from
$-\infty \pm i \varepsilon$ to 0 and have the convergence factor
$e^{\eta p_0}$.  This allows the contours to be deformed so that
they run from 0 to $\pm i \infty$ along the imaginary $p_0$ axis,
which corresponds to real values of $P_0 = -i p_0$.
Assuming that $F(-i p_0,{\bf p})$ is a real function of $p_0$,
i.e. that it satisfies
$F(-i p_0^*,{\bf p})= F(-i p_0,{\bf p})^*$,
the resulting formula for the sum-integral is
\begin{eqnarray}\nonumber
\sumint_P F(P) &=&
\int_P F(P)
\\ &&
\hspace{-5mm} + \int_p \epsilon(p_0) n(|p_0|) \,
2 {\rm Im} F(-i p_0+ \varepsilon,{\bf p}) \;,
\label{int-1loop}
\end{eqnarray}
%
%{\bf [m]}
where $\epsilon(p_0)$ is the sign of $p_0$.
The first integral on the right side is over the $(d+1)$-dimensional
Euclidean vector $P = (P_0,{\bf p})$
and the second is over the $(d+1)$-dimensional
Minkowskian vector $p = (p_0,{\bf p})$.

The two-loop sum-integrals can be evaluated by using a
generalization of the one-loop formula (\ref{int-1loop}):
\begin{widetext}
\begin{eqnarray} \nonumber
\sumint_{\{PQ\}} F(P) G(Q) H(R) &=&
\int_{PQ} F(P) G(Q) H(R)
- \int_p \epsilon(p_0) n_F(|p_0|) \, 2 \, {\rm Im} F(-i p_0+ \varepsilon,{\bf p}) 
	\, {\rm Re} \int_Q G(Q) H(R)\bigg|_{P_0 = -ip_0 + \varepsilon}
\\ && \nonumber
\hspace{-16mm}
- \int_p \epsilon(p_0) n_F(|p_0|) \, 2 \, {\rm Im} G(-i p_0+ \varepsilon,{\bf p}) 
	\, {\rm Re} \int_Q H(Q) F(R)\bigg|_{P_0 = -ip_0 + \varepsilon}
\\ && \nonumber
\hspace{-16mm}
+ \int_p \epsilon(p_0) n_B(|p_0|) \, 2 \, {\rm Im} H(-i p_0+ \varepsilon,{\bf p}) 
	\, {\rm Re} \int_Q F(Q) G(R)\bigg|_{P_0 = -ip_0 + \varepsilon}
\\ && \nonumber 
\hspace{-16mm}
+ \int_p \epsilon(p_0) n_F(|p_0|) \, 2 \, {\rm Im} F(-i p_0+ \varepsilon,{\bf p}) \,
        \int_q \epsilon(q_0) n_F(|q_0|) \, 2 \, {\rm Im} G(-i q_0+ \varepsilon,{\bf q}) 
	\, {\rm Re} H(R)\bigg|_{R_0 = i (p_0 + q_0)+ \varepsilon}
\\ && \nonumber
\hspace{-16mm}
- \int_p \epsilon(p_0) n_F(|p_0|) \, 2 \, {\rm Im} G(-i p_0+ \varepsilon,{\bf p}) \,
        \int_q \epsilon(q_0) n_B(|q_0|) \, 2 \, {\rm Im} H(-i q_0+ \varepsilon,{\bf q}) 
	\, {\rm Re} F(R)\bigg|_{R_0 = i (p_0 + q_0)+ \varepsilon}
\\ &&  
\hspace{-16mm}
- \int_p \epsilon(p_0) n_B(|p_0|) \, 2 \, {\rm Im} H(-i p_0+ \varepsilon,{\bf p}) \,
        \int_q \epsilon(q_0) n_F(|q_0|) \, 2 \, {\rm Im} F(-i q_0+ \varepsilon,{\bf q}) 
	\, {\rm Re} G(R)\bigg|_{R_0 = i (p_0 + q_0)+ \varepsilon}
\;.
\label{int-2loop}
\end{eqnarray}
%{\bf [jmp]}
\end{widetext}
This formula can be derived in 3 steps.
First, express the sum over $P_0$ as the sum of two contour integrals over
$p_0$, one that encloses the real axis ${\rm Im}\, p_0 = 0$
and another that encloses the line ${\rm Im} \, p_0= - {\rm Im} \, q_0$.
Second, express the the sum over $q_0$ as a contour integral
that encloses the real-$q_0$ axis.  The resulting terms can be combined into the expression
(\ref{int-2loop}).
The integrals of the imaginary parts that enter into our calculation can be
reduced to
\begin{eqnarray}\nonumber
&&\int_p \epsilon(p_0) n(|p_0|) \, 2 {\rm Im}
        {1 \over P^2}\bigg|_{P_0=-i p_0+ \varepsilon}
        f(-i p_0 + \varepsilon,{\bf p})
\\&&        
= \int_{\bf p} {n(p)\over p} {1 \over 2} \sum_{\pm}
        f(\pm i p + \varepsilon,{\bf p}) \label {impart1} \, , \\
\nonumber
&&\int_p \epsilon(p_0) n(|p_0|) \,  2 {\rm Im}
        {\cal T}_P\bigg|_{P_0=-i p_0+ \varepsilon}
        f(-i p_0 + \varepsilon,{\bf p}) \nonumber \\
        &&
=  - \int_{\bf p} p\,n(p) {1 \over 2} \sum_{\pm}
        \left\langle c^{-3+2\epsilon}
        f(\pm i p + \varepsilon,{\bf p}/c) \right\ranglec \;.
\end{eqnarray}
%
%{\bf [jm]}
The latter equation is obtained by inserting the expression (\ref{TP-int})
for ${\cal T}_P$, using (\ref{impart1}), and then making the change of
variable ${\bf p} \rightarrow {\bf p}/c$ to put the thermal integral into
a standard form.

As a simple illustration, we apply the formula (\ref{int-2loop})
to the sum-integral (\ref{two1}).  The nonvanishing terms are
\begin{eqnarray}
\sumint_{\{PQ\}} {1 \over P^2 Q^2 r^2} &=&
-2 \int_p n_F(|p_0|) \, 2 \pi \delta(p_0^2 - p^2) \int_Q {1 \over Q^2 r^2}
\nonumber
\\ \nonumber
&& \;+\; \int_p n_F(|p_0|) \, 2 \pi \delta(p_0^2 - p^2)
\\ &&\times
         \int_q n_F(|q_0|) \, 2 \pi \delta(q_0^2 - q^2) {1 \over r^2}   \,.
\end{eqnarray}
%
%{\bf [jmp]}

The delta functions can be used to evaluate the integrals over
$p_0$ and $q_0$.  The integral over $Q$ is given in (\ref{int4:1})
up to corrections of order $\epsilon$.
This reduces the sum-integral to
\begin{eqnarray}\nonumber
\sumint_{\{PQ\}} {1 \over P^2 Q^2 r^2} &=&
-{4 \over (4 \pi)^2} \left[ {1 \over \epsilon} + 4 - 2 \log 2 \right]
\mu^{2 \epsilon} 
\\ &&
\hspace{-2.2cm} \times \int_{\bf p} {n_F(p) \over p} p^{-2 \epsilon}
+ \int_{\bf p q} {n_F(p) n_F(q) \over p q} {1 \over r^2} \,.
\end{eqnarray}
%
%{\bf [jmp]}
The momentum integrals are evaluated in (\ref{int-th:1}) and
(\ref{int-th:2}).  Keeping all terms that contribute through
order $\epsilon^0$, we get the result (\ref{two1}).
The sum-integral (\ref{two2}) can be evaluated in the same way:
\begin{eqnarray}\nonumber
\sumint_{\{PQ\}} {q^2 \over P^2 Q^2 r^4} &=&
-{2 \over (4 \pi)^2 }
\left[ {1 \over \epsilon} - 2 \log 2 \right]
\mu^{2 \epsilon} \int_{\bf p} {n_F(p) \over p} p^{-2 \epsilon}
\\ &&
\;+\; \int_{\bf p q} {n_F(p) n_F(q) \over p q} {q^2 \over r^4} \,.
\end{eqnarray}
%
%{\bf [jmp]}
The sum-integral (\ref{twolast}) can be reduced to a linear combination of
(\ref{two1}) and (\ref{two2}) by expressing the numerator
in the form $P\!\cdot\!Q = P_0 Q_0 + (r^2 - p^2 - q^2)/2$
and noting that the $P_0 Q_0$ term vanishes upon summing over $P_0$ or
$Q_0$.

The sum-integral (\ref{two3}) is a little more difficult.
After applying the formula (\ref{int-2loop}) and using the delta
functions to integrate over $p_0$, $q_0$, and $r_0$,
it can be reduced to
\begin{eqnarray}\nonumber
&&\sumint_{\{PQ\}} {q^2\over P^2 Q^2 r^2 R^2} =
\int_{\bf p}{n_B(p)\over p}\int_{Q}{q^2\over p^2Q^2R^2}\bigg|_{P =-i p}
\\ && \nonumber
-\int_{\bf p}{n_F(p)\over p}\int_{Q}{1\over Q^2R^2}
\left({q^2\over r^2}+{p^2\over q^2}\right)\bigg|_{P =-i p}
\\ && \nonumber
+\int_{\bf pq}{n_F(p)n_F(q)\over pq}{p^2\over r^2}
{r^2-p^2-q^2\over\Delta(p,q,r)}
\\ 
&&
-\int_{\bf pq}{n_F(p)n_B(q)\over pq}\left({p^2\over q^2}+{r^2\over q^2}\right)
{r^2-p^2-q^2\over\Delta(p,q,r)} \, ,
\label{sumint:q/PQrR}
\end{eqnarray}
%
%{\bf [jmp]}
where $\Delta(p,q,r)$ is the triangle function that is negative
when $p$, $q$, and $r$ are the lengths of 3 sides of a triangle:
\begin{equation}
\Delta(p,q,r) = p^4 + q^4 + r^4 - 2 (p^2 q^2 + q^2 r^2 + r^2 p^2) \,.
\label{triangle}
\end{equation}
%{\bf [jmp]}
%
After using (\ref{int4:4})--(\ref{int4:6}) to integrate over $Q$,
the first term on the right side of (\ref{sumint:q/PQrR})
is evaluated using (\ref{int-th:1}).
The 2-loop thermal integrals on the right side of (\ref{sumint:q/PQrR})
are given in (\ref{int-thT:1})--(\ref{int-thT:4}).
Adding together all the terms, we get the final result (\ref{two3}).
The sum-integrals~(\ref{ntwo1}) and~(\ref{ntwo2}) are evaluated in a similar
manner.

\subsection{Two-loop HTL sum-integrals}
We also need some more difficult two-loop sum-integrals
that involve the functions ${\cal T}_P$ defined in (\ref{def-tf})
\begin{widetext}
\bqa
\sumint_{\{PQ\}}{1\over P^2Q^2r^2}{\cal T}_R&=&
{T^2 \over (4 \pi)^2} \left({\mu\over4\pi T}\right)^{4\epsilon}
	\left(-{1\over48}\right)
	\Bigg[{1\over\epsilon^2} + \left( 2 +12\log 2 + 4 {\zeta'(-1) \over \zeta(-1)} \right)
	{1\over\epsilon} +136.362 \Bigg]\;, 
\label{htlf1}
\\ 
\sumint_{\{PQ\}} {q^2\over P^2Q^2r^4}{\cal T}_R &=&
{T^2 \over (4 \pi)^2} \left({\mu\over4\pi T}\right)^{4\epsilon} \left(-{1\over576}\right)
	\Bigg[{1\over\epsilon^2} 
	+\left({26\over3}+52\log2+4{\zeta'(-1) \over \zeta(-1)}\right){1\over\epsilon}
	+446.438 \Bigg]\;,
\label{htlf2}
\\ 
\sumint_{\{PQ\}}{P\!\cdot\!Q\over P^2Q^2r^4}{\cal T}_R &=&
	{T^2 \over (4 \pi)^2} \left({\mu\over4\pi T}\right)^{4\epsilon} \left(-{1\over96}\right)
	\Bigg[{1\over\epsilon^2} +\left(4\log2+4{\zeta'(-1)\over\zeta(-1)} \right) 
	{1\over\epsilon} +69.174 \Bigg] \; , 
\label{htl3}
\\ 
\sumint_{\{PQ\}}{r^2-p^2\over P^2q^2Q^2_0R^2}{\cal T}_Q&=&
	-{T^2\over(4\pi)^2}\left({\mu\over4\pi T}\right)^{4\epsilon}{1\over8}
	\Bigg[{1\over\epsilon^2} +\left(2+2\gamma +{10\over3}\log2
	+2{\zeta^{\prime}(-1)\over\zeta(-1)}\right){1\over\epsilon}
	+46.8757 \Bigg] \; .
\label{com2l}
\eqa
%{\bf [jmp]}
\end{widetext}
The errors are all of order $\epsilon$.
To calculate the sum-integral~(\ref{htlf1}), we begin by using the 
representation~(\ref{TP-int}) of the function ${T}_R$:
\bqa\nonumber
\sumint_{\{PQ\}}{1\over P^2Q^2r^2}{\cal T}_R&=&
\sumint_{\{PQ\}}{1\over P^2Q^2r^2}
\\&&\hspace{-1.5cm}
-\sumint_{\{PQ\}}{1\over P^2Q^2}
\left\langle{c^2\over R_0^2+r^2c^2}\right\rangle_c\;.
\label{sumint2:6bb}
\eqa
The first sum-integral on the right hand side is given by~(\ref{two1}).
To evaluate the second sum-integral, we apply the sum-integral 
formula~(\ref{int-2loop}):
\bqa \nonumber
\sumint_{\{PQ\}} {1 \over P^2 Q^2 (R_0^2 + r^2 c^2)} && 
\\\nonumber 
        && \hspace{-3.7cm} = -\int_{\bf p} {n_F(p) \over p}
        2 {\rm Re} \int_Q {1 \over Q^2 (R_0^2+r^2c^2)}
        \bigg|_{P_0 = - i p + \varepsilon}
\\ && \nonumber \hspace{-3.7cm}
        + c^{-3+2\epsilon}
\int_{\bf p} {n_B(p) \over p}
        \int_Q {1 \over Q^2 R^2} \bigg|_{P \to (-i p,{\bf p}/c)}
        \, 
\\
&&\nonumber \hspace{-3.7cm}
  + \int_{\bf pq} {n_F(p) n_F(q) \over p q}
        {\rm Re} {r^2 c^2 - p^2 - q^2 \over
        \Delta(p+i\varepsilon,q,r c)}
\\ && \hspace{-3.7cm}
        - 2 c^{-3+2\epsilon} \, 
\int_{\bf pq} {n_F(p) n_B(q) \over p q}
{\rm Re}
        { r_c^2 - p^2 - q^2 \over \Delta(p+i\varepsilon,q,r_c)} 
\; ,
\label{sumint2:6b}
\eqa
%
%{\bf [jmp]}
where $r_c = |{\bf p} + {\bf q}/c|$.
In the terms on the right side with a single thermal integral,
the appropriate averages over $c$ of the integrals over $Q$
are given in~(\ref{int4:8.1}) and~(\ref{int4HTL:1}). 

The subsequent integrals over ${\bf p}$ are special cases 
of (\ref{int-th:1}) and (\ref{int-th:2}):
\bqa\nonumber
\int_{\bf p} n_B(p) 
\, p^{-1-2\epsilon}& =&
2^{8 \epsilon}
{(1)_{-4\epsilon} ({1\over2})_{2\epsilon}
        \over (1)_{-2\epsilon} ({3\over2})_{-\epsilon}} \,
{\zeta(-1+4\epsilon) \over \zeta(-1)} \hspace{12mm}
\\ && \hspace{5mm} \times
(e^\gamma \mu^2)^\epsilon (4 \pi T)^{-4\epsilon} \, {T^2 \over 12} \, , 
\label{bs}
\\ 
\int_{\bf p} n_F(p) 
\, p^{-1-2\epsilon} &=&
\left[1-2^{-1+4\epsilon}\right]
\int_{\bf p} n_B(p) 
\, p^{-1-2\epsilon} \; ,
\label{int-th:-1}
\eqa
%
%{\bf [jm]}
This yields
\bqa\nonumber
&& \hspace{-4mm} -2\int_{\bf p} {n_F(p) \over p}
 \, {\rm Re} \int_Q {1 \over Q^2}
        \left\langle {c^2 \over R_0^2 + r^2 c^2} \right\ranglec
                \bigg|_{P_0 = - i p + \varepsilon}
\\ && \hspace{-2mm}
+ \int_{\bf p} {n_B(p)\over p} \left\langle c^{-1+2\epsilon}
        \int_Q {1 \over Q^2 R^2} \bigg|_{P \to (-i p,{\bf p}/c)}
        \right\ranglec\, 
\nonumber
\\ \nonumber
&& \;=\; {T^2 \over (4 \pi)^2} \left({\mu\over4\pi T}\right)^{4\epsilon}
{1\over 48}
\Bigg[{1 \over \epsilon^2}
\\ && \hspace{2mm}
- \left( 6 - 12 \log 2 - 4 {\zeta'(-1) \over \zeta(-1)} \right)
        {1 \over \epsilon}
+ 70.122\Bigg] \;.
\label{intave:1}
\eqa
%{\bf [jmp]}

For the two terms in (\ref{sumint2:6bb}) with a double thermal integral,
the averages weighted by $c^2$ are given in~(\ref{f1}) 
and~(\ref{intHTL:4x}).
Adding them to (\ref{intave:1}), the final result is
\bqa\nonumber
&&\sumint_{\{PQ\}} {1 \over P^2 Q^2}
        \left\langle {c^2 \over R_0^2 + r^2 c^2} \right\ranglec
=  {T^2 \over (4 \pi)^2} \left({\mu\over4\pi T}\right)^{4\epsilon}
\left({1\over 48}\right)\\ &&
\hspace{-1mm}\times
%\\ &&%\nonumber\times
\Bigg[ {1 \over \epsilon^2}
- \left( 6 - 12 \log 2 - 4 {\zeta'(-1) \over \zeta(-1)} \right)
        {1 \over \epsilon}
+ 51.9307
\Bigg] \; .
\label{sumintave:2}
\eqa
%
%{\bf [jmp]}
Inserting this into (\ref{sumint2:6bb}),
we obtain the final result (\ref{htlf1}).

The sum-integral (\ref{htlf2}) is evaluated in a similar way to
(\ref{htlf1}).
Using the representation (\ref{TP-int}) for ${\cal T}_R$, we get
\bqa\nonumber
\sumint_{\{PQ\}} {q^2 \over P^2 Q^2 r^4} {\cal T}_R &=&
        \sumint_{\{PQ\}} {q^2 \over P^2 Q^2 r^4}
\\ &&
\hspace{-14mm}
        - \sumint_{\{PQ\}} {q^2 \over P^2 Q^2 r^2} \left\langle
        {c^2 \over R_0^2 + r^2 c^2} \right\ranglec \; .
\label{sumint2:7a}
\eqa
%
%{\bf [jmp]}
The first sum-integral on the right hand side is given by~(\ref{two2}).
To evaluate the second sum-integral, we apply the sum-integral
formula~(\ref{int-2loop}):
\bqa\nonumber
\sumint_{\{PQ\}} {q^2 \over P^2 Q^2 r^2 (R_0^2+r^2c^2)} &&
\\\nonumber
&& \hspace{-3cm}
= -\int_{\bf p} {n_F(p) \over p} {\rm Re}
        \int_Q {p^2+q^2 \over Q^2 r^2 (R_0^2+r^2c^2)}
        \bigg|_{P_0 = -i p + \varepsilon}
\\ && \hspace{-3cm} \nonumber
+c^{-1+2\epsilon}
\int_{\bf p} {n_B(p) \over p}p^{-2}\;
        \int_Q {q^2 \over Q^2 R^2}\bigg|_{P \to (-i p,{\bf p}/c)} 
\nonumber
\\ \nonumber
&& \hspace{-3cm}
+ \int_{\bf pq} {n_F(p) n_F(q) \over p q}
        {q^2 \over r^2} \, {\rm Re}
        {r^2 c^2 - p^2 - q^2 \over \Delta(p+i\varepsilon,q,r c)}
\\\nonumber
  &&      \hspace{-3cm} - c^{-1 + 2\epsilon} 
\int_{\bf pq} {n_F(p) n_B(q) \over p q}
{p^2 + r_c^2 \over q^2} \,
\\ && \hspace{-2cm} \times        
{\rm Re} { r_c^2-p^2-q^2
        \over \Delta(p+i\varepsilon,q,r_c)} \, .
\label{sumint3:6b}
\eqa

%
%{\bf [jmp]}
In the terms on the right side with a single thermal integral,
the weighted averages over $c$ of the integrals over $Q$
are given in~(\ref{int4:7a}),~(\ref{int4HTL:3}),
and~(\ref{int4HTL:4}):
After using (\ref{int-th:-1}) to evaluate the thermal integral, we obtain
\bqa\nonumber
&& -\int_{\bf p} {n_F(p) \over p}{\rm Re} \int_Q {p^2+q^2 \over Q^2 r^2}
        \left\langle {c^2 \over R_0^2+r^2c^2} \right\ranglec
        \bigg|_{P_0 = -i p + \varepsilon}
\\\nonumber && 
+ \int_{\bf p} {n_B(p) \over p}
{1 \over p^2} \left\langle c^{1+2\epsilon}
        \int_Q {q^2 \over Q^2 R^2} \bigg|_{P \to (-i p,{\bf p}/c)}
        \right\ranglec 
\\
\nonumber
&& \;=\;   {T^2 \over (4 \pi)^2} \left({\mu\over4\pi T}\right)^{4\epsilon}
\left({1\over576}\right)\left[ {1\over\epsilon^2}
\right.\\&&\left.
-\left({34\over3}-36\log 2-
4 {\zeta'(-1) \over \zeta(-1)}\right){1\over\epsilon}+229.354
\right]\, ,
\label{intave:2}
\eqa
%
%{\bf [jmp]}

For the two terms in (\ref{sumint3:6b}) with a double thermal integral,
the averages weighted by $c^2$ are given in (\ref{f3}),
(\ref{intHTL:6x}), and (\ref{intHTL:7x}).
Adding them to (\ref{intave:2}), the final result is
\bqa\nonumber
&&\sumint_{\{PQ\}} {q^2 \over P^2 Q^2 r^2}
        \left\langle {c^2 \over R_0^2 + r^2 c^2} \right\ranglec
= {T^2 \over (4 \pi)^2} \left({\mu\over4\pi T}\right)^{4\epsilon}
\left({1\over 576}\right)
\\ \nonumber&&%\times
\times\Bigg[ {1 \over \epsilon^2}
%\right.\nonumber\\&&  \left.
- \left({118\over3} - 52 \log2 - 4 {\zeta^{\prime}(-1)\over\zeta(-1)} \right)
{1\over\epsilon} 
%\\ &&\hspace{5cm}
+ 91.002
\Bigg]\;.
\\ &&
\label{sumintHTL:c2}
\eqa
%
%{\bf [jmp]}

To evaluate (\ref{htl3}), we use the expression (\ref{TP-int})
for ${\cal T}_R$ and the identity $P\!\cdot\!Q = (R^2-P^2-Q^2)/2$
to write it in the form
\bqa
&&\sumint_{\{PQ\}} {P\!\cdot\!Q \over P^2 Q^2 r^4} {\cal T}_R =
\sumint_{\{PQ\}} {P\!\cdot\!Q \over P^2 Q^2 r^4}
\nonumber\\&&
- \sumint_{\{P\}} {1\over P^2} \sumint_R {1 \over r^4} {\cal T}_R
- {1\over2} \langle c^2 \rangle_c \sumint_{\{PQ\}} {1 \over P^2 Q^2 r^2}
\nonumber\\&&
- {1\over2} \sumint_{\{PQ\}} {1 \over P^2 Q^2}
        \left\langle {c^2(1-c^2) \over R_0^2+r^2c^2} \right\ranglec \,.
\label{sumint2:8:2}
\eqa
%
%{\bf [jmp]}
The sum-integrals in the first 3 terms on the right side of
(\ref{sumint2:8:2}) are given in~(\ref{simple1}), (\ref{exa}),
(\ref{two1}), and (\ref{twolast}).  The last sum-integral
before the average weighted by $c$ is given in (\ref{sumint2:6b}).
The average weighted by $c^2$ is given in (\ref{sumintave:2}).
The average weighted by $c^4$ can be computed in the same way.
In the integrand of the single thermal integral,
the weighted averages over $c$ of the integrals over $Q$ are given 
in~(\ref{int4:8.2}) and~(\ref{int4HTL:2}):
%
%\bqa&&
%\left\langle c^4 \left(
%2 {\rm Re} \int_Q {1 \over Q^2 (R_0^2+r^2c^2)}
%        \bigg|_{P_0 = - i p + \varepsilon}
%+ c^{-3+2\epsilon} \int_Q {1 \over Q^2 R^2}
%        \bigg|_{P \to (-i p,{\bf p}/c)} \right) \right\ranglec
%\nonumber
%\\
%&& \;=\;  {1 \over (4 \pi)^2} \mu^{2\epsilon} p^{-2 \epsilon}
%\left[ \left({23\over6}-4\log 2\right){1\over\epsilon}+{104\over9}-\pi^2
%-3\log 2+8\log^2 2 \right] \, ,
%\eqa
%
After using (\ref{int-th:-1}) to evaluate the thermal integral, we obtain
\bqa\nonumber
&&
-2\int_{\bf p} {n_F(p) \over p}
{\rm Re} \int_Q {1 \over Q^2}
        \left\langle {c^4 \over R_0^2 + r^2 c^2} \right\ranglec
                \bigg|_{P_0 = - i p + \varepsilon}
\\ &&
+ \int_{\bf p}{n_B(p)\over p}\left\langle c^{1+2\epsilon}
        \int_Q {1 \over Q^2 R^2} \bigg|_{P \to (-i p,{\bf p}/c)}
        \right\ranglec 
\nonumber
\\\nonumber
&& \;=\;  {T^2 \over (4 \pi)^2} \left({\mu\over4\pi T}\right)^{4\epsilon}
\bigg[-\left({7\over72}-{1\over6}\log 2\right){1\over\epsilon}
+0.2150%4969
\bigg]\, ,
 \\ &&\
\label{intave:3}
\eqa
%
%%{[\bf jmp]}
For the two terms with a double thermal integral,
the averages weighted by $c^4$ are given in (\ref{f2}) and
(\ref{intHTL:5x}).
Adding them to (\ref{intave:3}), we obtain
\bqa\nonumber
&&\sumint_{\{PQ\}} {1 \over P^2 Q^2}
\left\langle {c^4 \over R_0^2 + r^2 c^2} \right\ranglec=
{T^2 \over (4 \pi)^2} \left({\mu\over4\pi T}\right)^{4\epsilon} \hspace{1cm}
\\ &&\hspace{2cm}\times
\left[-\left({7\over72}-{1\over6}\log 2\right){1\over\epsilon}+0.1359%48
\right]\, ,
\eqa
%
%{\bf [jmp]}

We finally need to evaluate~(\ref{com2l}). Applying~(\ref{int-2loop}) gives
\bqa\nonumber
\sumint_{\{PQ\}}{r^2-p^2\over P^2q^2Q^2_0R^2}{\cal T}_Q
&=&
\left[\int_{\bf p}{n_B(p)\over p}+\int_{\bf p}{n_F(p)\over p}
\right]
\\\nonumber&&\hspace{-1cm}\times
{\rm Re}\int_{Q}\left\langle {p^2-q^2\over Q^2r^2(R_0^2+r^2c^2)}\right\ranglec\bigg|_{P_0=-i p} 
\\ \nonumber 
&&\hspace{-3.4cm}
+\int_{\bf pq}{n_F(p)n_F(q)\over pq}{\rm Re}\left\langle
{r_c^2-p^2\over q^2}{r^2_c-p^2-q^2\over\Delta(p+i\epsilon,q,r_c)}c^{-1+2\epsilon}\right\ranglec
\\ \nonumber
&&\hspace{-3.4cm}
+\int_{\bf pq}{n_B(p)n_F(q)\over pq}{\rm Re} \left\langle
{r_c^2-p^2\over q^2}{r^2_c-p^2-q^2\over\Delta(p+i\epsilon,q,r_c)}
c^{-1+2\epsilon} \right\ranglec
\\
&&\hspace{-3.4cm}
+\int_{\bf pq}{n_F(p)n_B(q)\over pq}{\rm Re} \left\langle
{p^2-q^2\over r^2}{r^2c^2-p^2-q^2\over\Delta(p+i\epsilon,q,rc)} \right\ranglec \;.
\eqa
%{\bf[jmp]}

In the terms on the right side, with a single thermal factor, the weighted
average is given in Eq.~(\ref{last4d}), After using Eq.~(\ref{int-th:-1})
to evaluate the thermal integral, we obtain
\bqa\nonumber
&&\left[\int_{\bf p}{n_B(p)\over p}+\int_{\bf p}{n_F(p)\over p}
\right]\int_{Q}\left\langle{p^2-q^2\over Q^2r^2(R_0^2+r^2c^2)}\right\ranglec
\\ && \hspace{4cm}
={T^2\over(4\pi)^2}%\left({\mu\over4\pi T}\right)^{4\epsilon}
\left({\pi^2\over24}\right)\;.
\label{yiha}
\eqa
%{\bf [jmp]}
The terms with two thermal factors are given in Eqs.~(\ref{ff4}),~(\ref{lll})
and~(\ref{llll}). Adding them to~(\ref{yiha}), we finally obtain~(\ref{com2l}).

\section{Integrals}
\label{app:int}

Dimensional regularization can be used to 
regularize both the ultraviolet divergences and infrared divergences
in 3-dimensional integrals over momenta. 
The spacial dimension is generalized to  $d = 3-2\epsilon$ dimensions.
Integrals are evaluated at a value of $d$ for which they converge and then
analytically continued to $d=3$.
We use the integration measure
\begin{equation}
  \int_{\bf p} \;\equiv\;
  \left(\frac{e^\gamma\mu^2}{4\pi}\right)^\epsilon\,
  \int {d^{3-2\epsilon}p \over (2 \pi)^{3-2\epsilon}}\,.
\end{equation}
%%{[\bf jmp]}%

\subsection{3-dimensional integrals}

We require one integral that does not involve the 
Bose-Einstein distribution function.
The momentum scale in these integrals is set by the mass
$m=m_D$.
The one-loop integral is
\bqa
%\int_{\bf p} \log(p^2+m^2) & = & 
%- {m^3\over 6\pi}  \;,
%\label{int-3:1}
%\\ 
\int_{\bf p} {1 \over p^2+m^2} & = & 
- {m\over 4\pi} \left( {\mu \over 2 m} \right)^{2 \epsilon}
\left[1 + 2 \epsilon  \right] \,.
\label{bi3}
\eqa
%
%%{[\bf jmp]}
The error is one order higher in $\epsilon$ than the smallest term shown.  

\subsection{Thermal integrals}
The thermal integrals involve the Fermi-Dirac distribution
$n_F(p) = 1/(e^{\beta p} +1)$.
The one-loop integrals can all be obtained from the general
formula
\bqa\nonumber
\int_{\bf p}{n_F(p)\over p}p^{2\alpha}
&=&\left(1-2^{-1-2\alpha+2\epsilon}\right)
{\zeta(2+2\alpha-2\epsilon)\over4\pi^2}
\\&& \hspace{-15mm} \times
{\Gamma(2+2\alpha-2\epsilon)\Gamma({1\over2})\over\Gamma({3\over2}-\epsilon)}
\left(e^{\gamma}\mu^2\right)^{\epsilon}
T^{2+2\alpha-2\epsilon}\;.
\label{int-th:1}
\eqa
%{\bf [jmp]}
The simple two-loop thermal integrals are
\bqa
\label{int-th:2}
\int_{\bf p q}{n_F(p)n_F(q)\over pq}{1\over r^2}
&=&{T^2\over(4\pi)^2}
%\left({\mu\over4\pi T}\right)^{4\epsilon}
{1\over3}\left[1-\log2\right]\;,\\\nonumber
\int_{\bf p q}{n_F(p)n_F(q)\over pq}{q^2\over r^4}
&=&{T^2\over(4\pi)^2}\left({\mu\over4\pi T}\right)^{4\epsilon}
\\&&
\hspace{-3.2cm}\times\left(-{1\over36}\right)
\bigg[
5+6\gamma+6\log2-6{\zeta^{\prime}(-1)\over\zeta(-1)}
+3.076\epsilon\bigg]\;,
\label{int-th:3} \\ \nonumber
\int_{\bf p q}{n_B(p)n_F(q)\over pq}{p^2\over r^4}
&=&{T^2\over(4\pi)^2}\left({\mu\over4\pi T}\right)^{4\epsilon}\hspace{2.5cm}
\\&&
\hspace{-3.2cm}\times\left(-{1\over36}\right)
\bigg[
7-6\gamma-18\log2+6{\zeta^{\prime}(-1)\over\zeta(-1)}
+29.508\epsilon
\bigg]\;, 
\label{int-th:4}\\ \nonumber
\int_{\bf p q}{n_B(p)n_F(q)\over pq}{q^2\over r^4}
&=&{T^2\over(4\pi)^2}\left({\mu\over4\pi T}\right)^{4\epsilon}
\\&&
\hspace{-3.2cm}\times\left({1\over18}\right)
\bigg[
1-6\gamma-12\log2+6{\zeta^{\prime}(-1)\over\zeta(-1)}
+31.134\epsilon
\bigg]\;.
\label{int-th:5}
\eqa
%{\bf [jmp]}
We also need some more complicated 2-loop thermal integrals that involve
the triangle function defined in Eq.~(\ref{triangle}):
\begin{widetext}
\bqa
\int_{\bf p q}{n_F(p)\over p}{n_F(q)\over q}
{r^4\over q^2\Delta(p,q,r)}
&=&
{T^2\over(4\pi)^2}\left({\mu\over4\pi T}\right)^{4\epsilon}
\left(-{7\over96}\right)
\Bigg[
{1\over\epsilon^2}+
\left(
{22\over7}+2\gamma
+2\log2
+2{\zeta^{\prime}(-1)\over\zeta(-1)}
-{7\over20}\zeta(3)
\right){1\over\epsilon}
%\nonumber \\ && \hspace{7cm}
+47.2406
\Bigg]\!, 
\label{int-thT:1}
\hspace{6mm} \\ 
\int_{\bf p q}{n_F(p)\over p}{n_F(q)\over q}
{r^2\over\Delta(p,q,r)}
&=&{T^2\over(4\pi)^2}\left({\mu\over4\pi T}\right)^{4\epsilon}
\left(-{1\over48}\right)\left[
{1\over\epsilon^2}
+2
\left(1+\gamma+\log2+{\zeta^{\prime}(-1)\over\zeta(-1)}\right){1\over\epsilon} 
\nonumber \right. \\ && \left. \hspace{-12mm}
+4+4\gamma+{\pi^2\over2}+4\gamma\log2-6\log^22+4\log2
-4\gamma_1
+4(1+\gamma+\log2){\zeta^{\prime}(-1)\over\zeta(-1)}
+2{\zeta^{\prime\prime}(-1)\over\zeta(-1)}
\right]\;,
\label{int-thT:2}\\
\int_{\bf p q}{n_F(p)\over p}{n_F(q)\over q}
{p^4\over q^2\Delta(p,q,r)} 
&=&{T^2\over(4\pi)^2}\left({\mu\over4\pi T}\right)^{4\epsilon}
\left({49\,\zeta(3)\over1920}\right)
\left[
{1\over\epsilon}
+2+2\log2+2{\zeta^{\prime}(-3)\over\zeta(-3)}
+2{\zeta^{\prime}(3)\over\zeta(3)}
\right]\;,
\label{int-thT:3}\\
\int_{\bf p q}{n_F(p)\over p}{n_F(q)\over q}
{p^2(p^2+q^2)\over r^2\Delta(p,q,r)}
&=&
{T^2\over(4\pi)^2}\left({\mu\over4\pi T}\right)^{4\epsilon}
\left(-{1\over96}\right)
\left[{1\over\epsilon^2}
+\left({26\over3}+10\gamma-6{\zeta^{\prime}(-1)\over\zeta(-1)}
+10\log2\right){1\over\epsilon}
+41.1586
\right]\;, 
\label{int-thT:4} \\
\int_{\bf pq} {n_F(p)\over p}{n_F(q)\over q}{p^2\over\Delta(p,q,r)}
&=&{T^2\over(4\pi)^2}\left({\mu\over4\pi T}\right)^{4\epsilon}
\left(-{1\over96}\right)
\Bigg[
{1\over\epsilon^2}
+2\left(1+\gamma+\log2
%\right.\\ &&%\hspace{4cm}\left.
+{\zeta^{\prime}(-1)\over\zeta(-1)}
\right){1\over\epsilon}
+37.0573\Bigg]\;,
\\ 
\int_{\bf pq} {n_F(p)\over p}{n_B(q)\over q}{p^2\over\Delta(p,q,r)}
&=&{T^2\over(4\pi)^2}\left({\mu\over4\pi T}\right)^{4\epsilon}
\left({1\over96}\right)
\Bigg[
{1\over\epsilon^2}
+2\left(1+\gamma-\log2
%\right.\\&&%\hspace{4cm}\left.
+{\zeta^{\prime}(-1)\over\zeta(-1)}
\right){1\over\epsilon}
+19.2257\Bigg]\;,\\ 
\int_{\bf pq} {n_F(p)\over p}{n_B(q)\over q}{p^4\over q^2\Delta(p,q,r)}
&=&{T^2\over(4\pi)^2}\left({\mu\over4\pi T}\right)^{4\epsilon}
\left(-{7\,\zeta(3)\over1920}\right)
\Bigg[{1\over\epsilon}
+2-{2\over7}\log2+2{\zeta^{\prime}(-3)\over\zeta(-3)}
+2{\zeta^{\prime}(3)\over\zeta(3)}
\Bigg]\;,
\\ 
\int_{\bf pq} {n_F(p)\over p}{n_B(q)\over q}{r^4\over q^2\Delta(p,q,r)}
&=&{T^2\over(4\pi)^2}\left({\mu\over4\pi T}\right)^{4\epsilon}
\left({1\over24}\right)
\Bigg[
{1\over\epsilon^2}
+\left(4+2\gamma-5\log2-{7\zeta(3)\over80}
+2{\zeta^{\prime}(-1)\over\zeta(-1)}
\right){1\over\epsilon}
+18.1551\Bigg]\;,
\\
\int_{\bf pq} {n_F(p)\over p}{n_B(q)\over q}{r^2\over\Delta(p,q,r)}
&=&{T^2\over(4\pi)^2}\left({\mu\over4\pi T}\right)^{4\epsilon}
\left(-{1\over96}\right)
\Bigg[
{1\over\epsilon^2}
+2\left(1+\gamma+5\log2
%\right.\\&&%\hspace{4cm}\left.
+{\zeta^{\prime}(-1)\over\zeta(-1)}
\right){1\over\epsilon}
+84.2513\Bigg]
\;. 
\eqa
%{\bf [jmp]}
\end{widetext}

The most difficult thermal integrals to evaluate involve both the
triangle function and the HTL average defined in
(\ref{c-average}). There are 2 sets of these integrals.
The first set is
\bqa\nonumber
&&
\int_{\bf pq}{n_F(p)n_F(q)\over pq}\mbox{Re}\Bigg\langle
c^2{r^2c^2-p^2-q^2\over\Delta(p+i\varepsilon,q,rc)}
\Bigg\rangle_c
\\&&\hspace{2cm}={T^2\over(4\pi)^2}\left[1.458\times10^{-2}\right] \;, 
\label{f1}
\\\nonumber
&&\int_{\bf pq}{n_F(p)n_F(q)\over pq}\mbox{Re}\Bigg\langle
c^4{r^2c^2-p^2-q^2\over\Delta(p+i\varepsilon,q,rc)}
\Bigg\rangle_c
\\&&\hspace{2cm}={T^2\over(4\pi)^2}\left[
1.7715\times10^{-2}
\right] \;, 
\label{f2}
\\\nonumber
&&\int_{\bf pq}{n_F(p)n_F(q)\over pq}\mbox{Re}\Bigg\langle
{q^2\over r^2}c^2{r^2c^2-p^2-q^2\over\Delta(p+i\varepsilon,q,rc)}
\Bigg\rangle_c
\\&&={T^2\over(4\pi)^2}\left[ -1.1578\times10^{-2} \right] \;,
\label{f3}
\\\nonumber
&&\int_{\bf pq }
{n_B(p)\over p}{n_F(q)\over q}
\mbox{Re}\left\langle
{p^2-q^2\over r^2}{r^2c^2-p^2-q^2\over\Delta(p+i\epsilon,q,rc)}
\right\rangle
\\&&
\hspace{2cm} = {T^2\over(4\pi)^2}\left[0.17811\right]\;.
\label{ff4}
\eqa
%{\bf [jmp]}

The second set of these integrals involve the variable
$r_c = |{\bf p} + {\bf q}/c|$:
\bqa\nonumber
&& \int_{\bf pq} {n_F(p) n_B(q) \over p q}
{\rm Re} \left\langle c^{-1+2\epsilon}
        {r_c^2 - p^2 - q^2 \over \Delta(p+i\varepsilon,q,r_c)} \,
        \right\ranglec 
\\ &&
\hspace{1.8cm} = {T^2 \over (4\pi)^2}\left[0.19678\right] \; , 
\label{intHTL:4x}
\\\nonumber
&& \int_{\bf pq} {n_F(p) n_B(q) \over p q}
{\rm Re} \left\langle c^{1+2\epsilon}
        {r_c^2 - p^2 - q^2 \over \Delta(p+i\varepsilon,q,r_c)} \,
        \right\ranglec 
\\&&
\hspace{1.8cm} = {T^2 \over (4\pi)^2} %\left({\mu\over4\pi T}\right)^{4\epsilon}
%\left(- {1 \over 24} \right) 
\left[4.8368\times10^{-2}\right] \; ,
\label{intHTL:5x}
\\ \nonumber
&& \int_{\bf pq} {n_F(p) n_B(q) \over p q}{p^2 \over q^2}
{\rm Re} \left\langle c^{1+2\epsilon}
        {r_c^2 - p^2 - q^2 \over \Delta(p+i\varepsilon,q,r_c)} \,
        \right\ranglec 
\\&&
\hspace{1.8cm} = {T^2 \over (4\pi)^2} \left({\mu\over4\pi T}\right)^{4\epsilon}
{1 \over 96}
\left[{1\over\epsilon}
+7.7702
\right] \, ,
\label{intHTL:6x}
\\\nonumber
&& \int_{\bf pq} {n_F(p) n_B(q) \over p q}
{\rm Re} \left\langle c^{1+2\epsilon} {r_c^2 \over q^2}
        {r_c^2 - p^2 - q^2\over \Delta(p+i\varepsilon,q,r_c)} \,
        \right\ranglec 
\\&&
\hspace{4mm} = {T^2 \over(4\pi)^2} \left({\mu\over4\pi T}\right)^{4\epsilon}
{11-8\log2\over288}\left[
{1\over\epsilon}+7.79693
\right] \;,
\label{intHTL:7x}\\ \nonumber
&&\int_{\bf pq}{n_F(p)\over p}{n_F(p)\over q}
\mbox{Re}\left\langle
{r_c^2-p^2\over q^2}{r^2_c-p^2-q^2\over\Delta(p+i\epsilon,q,r_c)}
c^{-1+2\epsilon}
\right\ranglec
\\\nonumber&&
\hspace{3mm} = -{T^2\over(4\pi)^2}\left({1\over24}\right)\left[
{1\over\epsilon^2}
+\left(2+2\gamma\right.\right.
\left.\left.+2\log2
+2{\zeta^{\prime}(-1)\over\zeta(-1)}
\right){1\over\epsilon}
\right.\\ && \left.
\hspace{3cm}+40.316
\right]\;,
\label{lll}
\\ \nonumber
&&\int_{\bf pq}{n_B(p)\over p}{n_F(p)\over q}
\mbox{Re}\left\langle
{r_c^2-p^2\over q^2}{r^2_c-p^2-q^2\over\Delta(p+i\epsilon,q,r_c)}
c^{-1+2\epsilon}
\right\ranglec
\\\nonumber&&
\hspace{3mm}=-{T^2\over(4\pi)^2}\left({1\over12}\right)\bigg[
{1\over\epsilon^2}
+\left(2+2\gamma
+4\log2
+2{\zeta^{\prime}(-1)\over\zeta(-1)}
\right){1\over\epsilon}
\\ &&
\hspace{3cm}+52.953\bigg]\;.
\label{llll}
\eqa
%
%{\bf [jmp]}
The simplest way to evaluate integrals like 
(\ref{int-th:2})--(\ref{int-th:5}) whose integrands factor into
separate functions of $p$, $q$, and $r$  is to Fourier transform
to coordinate space where they reduce to an integral over a single
coordinate ${\bf R}$:
\bqa
\int_{\bf pq} f(p) \, g(q) \, h(r) &=&
\int_{\bf R} \tilde f(R) \, \tilde g(R) \, \tilde h(R) \,.
\label{int-fgh}
\eqa
%%{[\bf jmp]}

The Fourier transform is
\bqa
\tilde{f}(R)=\int_{\bf p}e^{i{\bf p}\cdot {\bf R}}f(p)\;,
\eqa
%[{\bf jmp}]
and the dimensionally regularized coordinate integral is 
\bqa
\int_{\bf R} &=&
\left( {e^\gamma \mu^2 \over 4 \pi} \right)^{-\epsilon}
\int d^{3-2 \epsilon}R \,.
\eqa
%[{\bf jmp}]
%
The Fourier transforms we need are
\bqa
\int_{\bf p} p^{2 \alpha} \,  e^{i {\bf p} \cdot {\bf R}} &=&
{1 \over 8\pi} 
{\Gamma({3\over2} + \alpha - \epsilon)
        \over \Gamma({1\over2}) \Gamma(-\alpha)}
\left( e^\gamma \mu^2 \right)^\epsilon 
\nonumber \\ && \hspace{1.2cm}
\times \left( {2 \over R} \right)^{3 + 2 \alpha - 2\epsilon} \;, 
\\ \nonumber
\int_{\bf p} {n(p) \over p} \,  p^{2 \alpha} \,
        e^{i {\bf p} \cdot {\bf R}} &=&
{1 \over 4\pi} {1 \over \Gamma({1\over2})}
\left( e^\gamma \mu^2 \right)^\epsilon 
\left( {2 \over R} \right)^{{1\over2} - \epsilon}
\\ &&
\hspace{-1.2cm}\times\int_0^\infty dp \, p^{2 \alpha + {1\over2} - \epsilon} n(p)
        J_{{1\over2}-\epsilon}(pR) \;. 
\label{fourier-n}
\eqa
%[{\bf jmp}]
If $\alpha$ is an even integer, the Fourier transform (\ref{fourier-n})
is particularly simple in the limit $d \to 3$:
\bqa
&&\hspace{-0.7cm}
\int_{\bf p}{n_F(p)\over p}e^{i{\bf p}\cdot {\bf R}}=
{T\over 4\pi R}\left(
{1\over x}-\mbox{csch}x
\right)
\;, \\ %\nonumber
&&
\hspace{-1cm}
\int_{\bf p}{n_F(p)\over p}p^2e^{i{\bf p}\cdot {\bf R}}=
{\pi T^3\over2R}
\left(
\mbox{csch}^3x+{1\over2}\mbox{csch}x
%\right.\\ && \left.
-{1\over x^3}
\right)
\;,
\eqa
%[{\bf jmp}]
where $x=\pi RT$

We can use these simple expressions only if the integral
over the coordinate ${\bf R}$ in (\ref{int-fgh})
converges for $d=3$. Otherwise, we must first make subtractions
on the integrand to make the integral convergent.

The integrals~(\ref{int-th:2})--(\ref{int-th:5}) 
can be evaluated directly by applying
the Fourier transform formula (\ref{int-fgh}) in the limit $\epsilon \to 0$.

The integrals (\ref{int-thT:1})--(\ref{int-thT:3}) can be evaluated by first
averaging over angles.
The triangle function can be expressed as
\begin{equation}
\Delta(p,q,r) = - 4 p^2 q^2 (1 - \cos^2 \theta) \;,
\label{triangle-theta}
\end{equation}
%
%{[\bf jmp]}
where $\theta$ is the angle between ${\bf p}$ and ${\bf q}$.
For example, the angle average for (\ref{int-thT:1}) is
\bqa\nonumber
\left\langle {r^4 \over \Delta(p,q,r)} 
\right\rangle_{\!\!{\bf \hat p}\cdot{\bf \hat q}}
&=& -{w(\epsilon) \over 8} \int_{-1}^{+1} dx \, (1-x^2)^{-1-\epsilon} 
\\&&\times
        {(p^2 + q^2 + 2 p q x)^2 \over p^2 q^2} \;.
\label{ang-ave:1}
\eqa
%
%{[\bf jmp]}
After integrating over $x$ and inserting the result into 
(\ref{int-thT:1}), the integral reduces to
\bqa\nonumber
&&
\int_{\bf pq} {n_F(p) \over p} \, {n_F(q) \over q} \, 
        {r^4 \over q^2\Delta(p,q,r)} \hspace{3.5cm} \\
&& \hspace{2mm} = \int_{\bf pq} {n_F(p) \over p} \, {n_F(q) \over q} 
\left( {1 - 2 \epsilon \over 8 \epsilon} \, {p^2 \over q^4}
        + {7 - 6 \epsilon \over 8 \epsilon} \, {1 \over q^2} \right) \,.
\eqa
%
%{\bf [jmp]}
The integrals over ${\bf p}$ and ${\bf q}$ factor into separate integrals
that can be evaluated using (\ref{int-th:1}).
After averaging over angles, the integrals
(\ref{int-thT:2}) and (\ref{int-thT:3}) reduce to
\bqa\nonumber
&& \int_{\bf pq} {n_F(p) \over p} \, {n_F(q) \over q} 
        {r^2 \over \Delta(p,q,r)} \\
&&\hspace{1.2cm}={1 - 2 \epsilon \over 4 \epsilon}
\int_{\bf p} {n_F(p) \over p} 
\int_{\bf q} {n_F(q) \over q} \, {1 \over  q^2} \;,
\\\nonumber
&& \int_{\bf pq} {n_F(p) \over p} \, {n_F(q) \over q} 
        {p^4 \over q^2 \Delta(p,q,r)}  \\
&&\hspace{1.2cm} = {1 - 2 \epsilon \over 8 \epsilon}
\int_{\bf p} {n_F(p) \over p} \, p^2
\int_{\bf q} {n_F(q) \over q} \, {1 \over  q^4} \;.
\eqa
%
%{\bf [jmp]}
        
The integral~(\ref{int-thT:4}) can be evaluated by using the
identity
\begin{equation}
\left\langle {p^2+q^2 \over r^2\Delta(p,q,r)} 
\right\rangle_{\!\!{\bf \hat p}\cdot{\bf \hat q}}
= {1 \over 2 \epsilon} \left\langle {1 \over r^4} 
\right\rangle_{\!\!{\bf \hat p}\cdot{\bf \hat q}}
+ {1-2\epsilon \over 8\epsilon} {1 \over p^2 q^2} \;.
\label{ang-ave:2}
\end{equation}
%
%{[\bf jmp]}
The identity can be proved by expressing the angular averages 
in terms of integrals over the cosine of the angle between 
${\bf p}$ and ${\bf q}$ as in (\ref{ang-ave:1}), 
and then integrating by parts. Inserting the identity (\ref{ang-ave:2})
into (\ref{int-thT:4}), the integral reduces to 
\bqa
\nonumber
\int_{\bf pq} {n_F(p) \over p} \, {n_F(q) \over q} \, 
        {p^2(p^2+q^2) \over r^2 \Delta(p,q,r)} &=&
{1 \over 2 \epsilon}
\int_{\bf pq} {n_F(p) \over p} \, {n_F(q) \over q} \, {p^2 \over r^4} 
\\ &&
\hspace{-2.3cm} \,+\, {1 - 2 \epsilon \over 8 \epsilon}
\int_{\bf p} {n_F(p)\over p}\int_{\bf q}{n_F(q) \over q} \, {1 \over  q^2} \,.
\eqa
%
%{\bf [jmp]}
The integral in the first term on the right is given in (\ref{int-th:3}),
while the second term can be evaluated using (\ref{int-th:1}).

The integral~(\ref{f1}) %~ and~(\ref{f2}) 
can be evaluated directly in three 
dimensions by first averaging over $c$ and $x$, and then integrate 
the resulting functions numerically over $p$ and $q$.

To evaluate the weighted averages over $c$ of the thermal integrals
in Eqs.(\ref{f2})--~(\ref{ff4}), we first isolate
the divergent parts, which come from the region $p-q \to 0$.
We write the product of thermal functions
in the form
\bqa
\nonumber
n_F(p) n_F(q) &=&
\left( n_F(p) n_F(q) - {s^2 n_F^2(s) \over p q}  \right)
\\&& \hspace{1cm}
+  {s^2 n^2_F(s) \over p q}  \; ,
\label{nnsub-1}
\eqa
%
%{\bf [jmp]}
where $s= (p+q)/2$.  In the difference term, the HTL average over
$c$ and the angular average over $x = \hat {\bf p} \cdot \hat {\bf q}$
can be calculated in three dimensions:
\begin{eqnarray}\nonumber
{\rm Re} \left\langle c^4 {r^2 c^2 - p^2 - q^2 \over
	\Delta(p+i\varepsilon,q,r c)}  \right\ranglecx &=&
{2(p^2 + q^2) \over 3 (p^2-q^2)^2}
\\
&& \hspace{-4cm}
+ {1 \over 12 p q} \log {p+q \over |p-q|}
-{(3p^2+q^2)(p^2+3q^2)\over6(p^2-q^2)^3}\log(p/q)\;,
\nonumber \\
\\ \nonumber
{\rm Re} \left\langle c^2 {q^2 \over r^2} {r^2 c^2 - p^2 - q^2 \over
        \Delta(p+i\varepsilon,q,r c)}  \right\ranglecx &=&
{q^2 \over 3 (p^2 - q^2)^2}
\\ \nonumber && \hspace{-4.5cm}
\times\left[ 2 - {1 \over 2} \log{|p^2-q^2| \over p q} \right.
%\nonumber
\\
&& \hspace{-4cm}
\left. \;-\; {p^2+q^2 \over 4 p q} \log {p+q \over |p-q|}
        - {p^2 + q^2 \over p^2 - q^2} \log(p/q) \right]\, ,
\\ \nonumber
\mbox{Re}\left\langle
{p^2-q^2\over r^2}{r^2c^2-p^2-q^2\over\Delta(p+i\epsilon,q,rc)}
\right\rangle_{c,x}
&=&{1\over4pq(p^2-q^2)}
\\&&\hspace{-4.5cm}\times
\left[
-(p^2+q^2)\log{p+q\over|p-q|}
-2pq\log{|p^2-q^2|\over pq}
\right]\;.
\end{eqnarray}
%
%{\bf [jmp]}
The remaining 2-dimensional integral over $p$ and $q$
can then be evaluated numerically:
\begin{eqnarray}\nonumber
&& \hspace{-1.3cm} \int_{\bf pq}\left( {n_F(p) n_F(q) \over p q}
        - {s^2 n^2_F(s) \over p^2 q^2}  \right) 
\\ 
 && \hspace{-1cm} 
\times {\rm Re} \left\langle c^4 {r^2 c^2 - p^2 - q^2 \over
	\Delta(p+i\varepsilon,q,r c)}  \right\ranglec 
 = {T^2 \over (4\pi)^2} \left[ 8.980\times10^{-3}\right] \,, 
\\ \nonumber 
\label{xxxx}
&& \hspace{-1.3cm} \int_{\bf pq}\left( {n_F(p) n_F(q) \over p q}
        - {s^2 n^2_F(s) \over p^2 q^2}  \right){q^2 \over r^2} 
\\ 
 && \hspace{-1cm} 
\times {\rm Re} \left\langle c^2 {r^2 c^2 - p^2 - q^2 \over
        \Delta(p+i\varepsilon,q,r c)}  \right\ranglec  
 = {T^2 \over (4\pi)^2} \left[ 7.792\times10^{-3}\right] \,, 
\label{intHTL:3f}
\\ \nonumber && \hspace{-1.3cm} \int_{\bf pq }\left(
	{n_B(p)\over p}{n_F(q)\over q} -{s^2n_B(s)n_F(s)\over p^2q^2}
	\right) 
\\ && \hspace{-1cm} \times
\mbox{Re}\left\langle
{p^2-q^2\over r^2}{r^2c^2-p^2-q^2\over\Delta(p+i\epsilon,q,rc)}
\right\rangle_c
={T^2\over(4\pi)^2}\left[0.17811\right]\,. 
\label{extra}
\eqa
%{\bf [jmp]}

The integrals involving the $n^2_F(s)$ term in (\ref{nnsub-1})
are divergent, so the HTL average over
$c$ and the angular average over $x = \hat {\bf p} \cdot \hat {\bf q}$
must be calculated in $3-2\epsilon$ dimensions.
The first step in the calculation of the $n^2(s)$ term
is to change variables from ${\bf p}$ and ${\bf q}$ to
$s = (p+q)/2$, $\beta = 4pq/(p+q)^2$,
and $x= \hat{\bf p} \cdot \hat{\bf q}$:
\begin{eqnarray}&&\hspace{-5mm}\nonumber
\int_{\bf pq} {s^2 n^2_F(s) \over p^2 q^2} \, f(p,q,r) =
{64 \over (4\pi)^4}
\left[ (e^\gamma \mu^2)^\epsilon
        {\Gamma({3\over2}) \over \Gamma({3\over2}-\epsilon)} \right]^2
\\ &&
        \times \int_0^\infty ds \, s^{1-4 \epsilon}n^2_F(s)
s^2 \int_0^1 d \beta \, \beta^{-2 \epsilon} (1-\beta)^{-1/2}
\nonumber\\&&\hspace{8mm}\times  
\Big\langle f(s_+,s_-,r) + f(s_-,s_+,r) \Big\rangle_{\!\!x} \, ,
\label{int:n2f}
\end{eqnarray}
%
%{\bf [jmp]}
where $s_\pm = s[1 \pm \sqrt{1-\beta}]$
and $r=s [4 - 2\beta(1-x)]^{1/2}$.
The 2 terms inside the average over $x$
come from the regions $p>q$ and $p<q$, respectively.
The integral over $s$ is easily evaluated:
\bqa\nonumber
&& \hspace{-8mm} \int_0^\infty ds \, s^{1-4 \epsilon}n^2_F(s) \;=\;
\Gamma(2-4\epsilon)
\left[-(1-2^{4\epsilon})\zeta(1-4\epsilon)
\right.\\ &&\left.
\hspace{8mm} +(1-2^{-1+4\epsilon})\zeta(2-4\epsilon) \right] T^{2-4\epsilon}\; ,
\label{intth:n2}\\ \nonumber 
&& \hspace{-8mm} \int_0^\infty ds \, s^{1-4 \epsilon}n_F(s)n_B(s) 
\\ &&
\hspace{8mm} = 2^{-2+4\epsilon}
\Gamma(2-4\epsilon)
\zeta(2-4\epsilon)T^{2-4\epsilon}\; .
\label{intth:n2b}
\eqa
%{\bf [jmp]}
%
It remains only to evaluate the averages over $c$ and $x$ and
the integral over $\beta$.

The first step in the calculation of the $n_F^2(s)$ term
of~(\ref{f2}) is to decompose the integrand into 2 terms:
\begin{equation}
{r^2 c^2 - p^2 - q^2 \over
	\Delta(p+i\varepsilon,q,r c)} \;=\;
-{1 \over 2} \sum_\pm {1 \over (p+i\varepsilon \pm q)^2 - r^2 c^2}\;.
\end{equation}
%
%{\bf [jmp]}
The weighted averages over $c$ gives a hypergeometric function:
\begin{eqnarray}\nonumber
&&\hspace{-1.2cm} \left\langle {c^4 \over (p+i\varepsilon \pm q)^2 - r^2 c^2}
	\right\ranglec =
{3 \over (3 - 2 \epsilon)(5 - 2 \epsilon)} \,
\\ && \hspace{1mm}
\times {1 \over (p+i\varepsilon \pm q)^2} \,
F\left({{5 \over 2},1 \atop {7 \over 2} - \epsilon} \Bigg|
	{r^2 \over (p+i\varepsilon \pm q)^2} \right)\;.
\label{avec:c4xx}
\end{eqnarray}
%
%{\bf [jmp]}

In the $+q$ case of (\ref{avec:c4xx}),
the $i\varepsilon$ prescription is unnecessary.
The argument of the hypergeometric function can be written $1 - \beta y$,
where $y = (1-x)/2$.
After using a transformation formula to change the argument to $\beta y$,
we can evaluate the angular average over $x$ to obtain hypergeometric
functions with argument $\beta$.
For example, the average over $x$ of (\ref{avec:c4xx}) is
\begin{eqnarray}&& \nonumber \hspace{-5mm}
\left\langle F\left( {{5 \over 2}, 1 \atop {7 \over 2} - \epsilon}
	\Bigg| {r^2 \over (p+q)^2} \right) \right\ranglex =
- {5-2\epsilon \over 2 \epsilon}
\left[ F\left( { 1-\epsilon , {5 \over 2} , 1
		\atop 2-2\epsilon , 1+\epsilon } \Bigg| \beta \right)
\right.
\nonumber
\\
&& \hspace{-4mm} \left. \;-\;
{ (1)_\epsilon (1)_{-2\epsilon} (2)_{-2 \epsilon} ({5\over2})_{-\epsilon}
	\over (1)_{-\epsilon} (2)_{-3\epsilon} }
\beta^{-\epsilon}
F\left( { 1-2\epsilon , {5\over2}-\epsilon
		\atop 2-3\epsilon } \Bigg| \beta \right)
\right] \, ,
\end{eqnarray}
%
%{\bf [jm]}
where $(a)_b$ is Pochhammer's symbol which is defined in (\ref{Poch}).
Integrating over $\beta$, we obtain hypergeometric
functions with argument 1:
\begin{eqnarray}
&& \hspace{-2mm} s^2 \int_0^1 d\beta \, \beta^{-2\epsilon} (1-\beta)^{-1/2}
\left\langle {c^4 \over (p+q)^2 - r^2 c^2}
	\right\ranglecx \;=\;
- {1 \over 4\epsilon}
\nonumber\\&& \hspace{-3mm} 
\times 
{ (1)_\epsilon (2)_{-2\epsilon} \over (1)_{-\epsilon} }
	\left[ { (1)_{-2\epsilon} (1)_{-\epsilon}
	\over ({5\over2})_{-2\epsilon} (2)_{-2\epsilon} (1)_{\epsilon} }
F\left( { 1-2\epsilon , 1-\epsilon , {5 \over 2} , 1
	\atop {5\over2}-2\epsilon , 2-2\epsilon , 1+\epsilon }
	\Bigg| 1 \right)
\right.
\nonumber
\\
&& \hspace{-3mm} \left. \;-\;
{ (1)_{-3\epsilon} (1)_{-2\epsilon} ({5\over2})_{-\epsilon}
	\over ({5\over2})_{-3\epsilon} (2)_{-3\epsilon} }
F\left( { 1-3\epsilon , 1-2\epsilon , {5\over2}-\epsilon
		\atop {5\over2}-3\epsilon,2-3\epsilon } \Bigg| 1 \right)
\right] \;.
\label{FF:2}
\end{eqnarray}
%
%{\bf [mp]}
Expanding in powers of $\epsilon$,
we obtain
\begin{eqnarray}
&& \hspace{-5mm} \nonumber
s^2 \int_0^1 d\beta \, \beta^{-2\epsilon} (1-\beta)^{-1/2}
\left\langle {c^4 \over (p+q)^2 - r^2 c^2}
	\right\ranglecx 
\\ &&
\hspace{2cm} = {\pi^2 \over 72} (1 + 10.8408 \, \epsilon) \;.
\label{intavecx:2p}
\end{eqnarray}
%
%{\bf [mp]}

In the $-q$ case of (\ref{avec:c4xx}),
the argument of the hypergeometric functions can be written
$(1-\beta y)/(1-\beta \pm i \varepsilon)$, where $y=(1-x)/2$
and the prescriptions $+i \varepsilon$ and $-i \varepsilon$
correspond to the regions $p>q$ and $p<q$, respectively.
These regions correspond to the two terms
inside the average over $x$ in (\ref{int:n2f}).
In  order to obtain an analytic result in terms of hypergeometric functions,
it is necessary to integrate over $\beta$ before averaging over $x$.
The integrals over $\beta$ can be evaluated by first using a
transformation formula to change the argument of the hypergeometric
function to $-\beta(1-y)/(1-\beta)$ and then using
the integration formula (\ref{int-2F1}) to obtain hypergeometric
functions with arguments $y$ or $1-y$:
\begin{eqnarray}
&&\hspace{-3mm} \int_0^1 d\beta \, \beta^{-2\epsilon} (1-\beta)^{-3/2}
F\left( { {3\over2}, 1 \atop {5\over2}-\epsilon }
	\Bigg| {1-\beta y \over 1-\beta + i \varepsilon} \right)
\nonumber
\\
&&  \hspace{-2mm} \;=\;
{3-2\epsilon \over \epsilon} \,
{(1)_{-2\epsilon} \over ({1\over2})_{-2\epsilon}} \,
F\left( { 1-2\epsilon, 1
		\atop 1+\epsilon } \Bigg| 1-y \right)
\nonumber
\\
&& \hspace{-1mm}
- {3-2\epsilon \over \epsilon} \,
{ (1)_{\epsilon} \over ({1\over2})_{\epsilon} } \,
(1-y)^{-1/2}
F\left( { {1\over2}-2\epsilon, 1 \atop {1\over2}+\epsilon }
	\Bigg| 1-y \right)
\nonumber
\\ \nonumber
&& \hspace{-1mm}
+ {3 \over 2\epsilon(1-3\epsilon)} e^{i\pi \epsilon} \,
(1)_{\epsilon} (\mbox{${5\over2}$})_{-\epsilon}\,
(1-y)^{-\epsilon}
%\\ &&
%\times
F\left( { 1-3\epsilon, {3\over2}-\epsilon \atop 2-3\epsilon }
	\Bigg| y \right) \,.
	\\
\end{eqnarray}
%
%{\bf [mp]}
After averaging over $x$, we obtain hypergeometric
functions with argument 1:
\begin{eqnarray}
&& \hspace{-3mm} s^2 \int_0^1 d\beta \, \beta^{-2\epsilon} (1-\beta)^{-1/2}
\left\langle {c^2 \over (p+i\varepsilon - q)^2 - r^2 c^2}
	\right\ranglecx
\nonumber
\\
&&  \hspace{-2mm} \;=\;
{1 \over 4\epsilon}\,
{(1)_{-2\epsilon} \over ({1\over2})_{-2\epsilon}} \,
F\left( { 1-\epsilon, 1-2\epsilon, 1
		\atop 2-2\epsilon, 1+\epsilon } \Bigg| 1 \right)
\nonumber
\\
&& \hspace{-1mm}
- {1 \over 2\epsilon}\,
{ (2)_{-2\epsilon} (1)_{\epsilon} ({1\over2})_{-\epsilon}
	\over (1)_{-\epsilon} ({1\over2})_{\epsilon}
			({3\over2})_{-2\epsilon} } \,
F\left( { {1\over2}-\epsilon,{1\over2}-2\epsilon,1
		\atop {3\over2}-2\epsilon,{1\over2}+\epsilon }
	\Bigg| 1 \right)
\nonumber
\\ \nonumber
&& \hspace{-1mm}
+ {1 \over 8\epsilon(1-3\epsilon)} \, e^{i\pi \epsilon}
{ (2)_{-2\epsilon} (1)_{-2\epsilon} (1)_{\epsilon}({3\over2})_{-\epsilon}
	\over (1)_{-\epsilon} (2)_{-3\epsilon} } \,
\\ &&
\hspace{1cm} \times F\left( { 1-\epsilon, 1-3\epsilon, {3\over2}-\epsilon
		\atop 2-3\epsilon, 2-3\epsilon }
	\Bigg| 1 \right) \;.
\end{eqnarray}
%
%{\bf [jmp]}
%The integral weighted by $c^4$ can be evaluated in a similar way.
Expanding in powers of $\epsilon$ and then taking the real parts,
we obtain
\begin{eqnarray}
\nonumber
&& \hspace{-3mm} {\rm Re} \, s^2 \int_0^1 d\beta \, \beta^{-2\epsilon} (1-\beta)^{-1/2}
\left\langle {c^4 \over (p+i\varepsilon - q)^2 - r^2 c^2}
	\right\ranglecx
\\ && \hspace{2cm} = - {12+\pi^2 \over 72} (1 + 1.10518 \, \epsilon) \;.
\label{intavecx:2m}
\end{eqnarray}
%
%{\bf [mp]}

To evaluate the subtraction in the integral (\ref{intHTL:3f}),
we use the identity
$q^2 = (r^2 + q^2 - p^2 - 2 {\bf p}\cdot{\bf q})/2$.
The integral with $q^2-p^2$ in the numerator is purely imaginary.
Thus the real part of the integral can be expressed as
\begin{eqnarray}
&& \hspace{-3mm} \int_{\bf pq} {s^2 n^2(s) \over p^2 q^2} {q^2 \over r^2}
{\rm Re} \left\langle c^2 {r^2 c^2 - p^2 - q^2 \over
        \Delta(p+i\varepsilon,q,r c)}  \right\ranglec
        \nonumber
        \\ \nonumber
&& \hspace{-1mm}\;=\;
\int_{\bf pq} {s^2 n^2(s) \over p^2 q^2}
        \left( {1\over2} - {{\bf p}\cdot{\bf q} \over r^2} \right)
%\\ &&
{\rm Re} \left\langle c^2 {r^2 c^2 - p^2 - q^2 \over
        \Delta(p+i\varepsilon,q,r c)}  \right\ranglec  \;.
	\\
\label{intHTL:3a}
\end{eqnarray}
%
%{\bf [jmp]}

The first term in Eq.~(\ref{intHTL:3a}) is 
decomposed into 2 terms:
\begin{equation}
{r^2 c^2 - p^2 - q^2 \over
        \Delta(p+i\varepsilon,q,r c)} \;=\;
-{1 \over 2} \sum_\pm {1 \over (p+i\varepsilon \pm q)^2 - r^2 c^2}\;.
\end{equation}
%
%{\bf [jmp]}
The weighted averages over $c$ give hypergeometric functions:
\begin{eqnarray}\nonumber
&& \hspace{-1cm}
\left\langle {c^2 \over (p+i\varepsilon \pm q)^2 - r^2 c^2}
        \right\ranglec 
\\ && \hspace{-8mm}
= {1 \over 3 - 2 \epsilon} \,
{1 \over (p+i\varepsilon \pm q)^2} \,
F\left({{3 \over 2},1 \atop {5 \over 2} - \epsilon} \Bigg|
        {r^2 \over (p+i\varepsilon \pm q)^2} \right)\,,
\label{avec:c2}
\\ && \nonumber \hspace{-1cm}
\left\langle {c^4 \over (p+i\varepsilon \pm q)^2 - r^2 c^2}
        \right\ranglec =
{3 \over (3 - 2 \epsilon)(5 - 2 \epsilon)} \,
\\ &&
\hspace{1mm} \times{1 \over (p+i\varepsilon \pm q)^2} \,
F\left({{5 \over 2},1 \atop {7 \over 2} - \epsilon} \Bigg|
        {r^2 \over (p+i\varepsilon \pm q)^2} \right)\, .
\label{avec:c4}
\end{eqnarray}
%
%{\bf [jmp]}
In the $+q$ case of (\ref{avec:c2}),
the $i\varepsilon$ prescription is unnecessary.
The argument of the hypergeometric function can be written $1 - \beta y$,
where $y = (1-x)/2$.
After using a transformation formula to change the argument to $\beta y$,
we can evaluate the angular average over $x$ to obtain hypergeometric
functions with argument $\beta$.
For example, the average over $x$ of (\ref{avec:c2}) is
\begin{eqnarray}&&\nonumber \hspace{-4mm}
\left\langle F\left( {{3 \over 2}, 1 \atop {5 \over 2} - \epsilon}
        \Bigg| {r^2 \over (p+q)^2} \right) \right\ranglex =
- {3-2\epsilon \over 2 \epsilon}
\left[ F\left( { 1-\epsilon , {3 \over 2} , 1
                \atop 2-2\epsilon , 1+\epsilon } \Bigg| \beta \right)
\right.
\nonumber
\\
&& \hspace{-4mm} \left. -\;
{ (1)_\epsilon (1)_{-2\epsilon} (2)_{-2 \epsilon} ({3\over2})_{-\epsilon}
        \over (1)_{-\epsilon} (2)_{-3\epsilon} }
\beta^{-\epsilon}
F\left( { 1-2\epsilon , {3\over2}-\epsilon
                \atop 2-3\epsilon } \Bigg| \beta \right)
\right] \; ,
\end{eqnarray}
%
%{\bf [jmp]}
where $(a)_b$ is Pochhammer's symbol which is defined in (\ref{Poch}).
Integrating over $\beta$, we obtain hypergeometric
functions with argument 1:
\begin{eqnarray}
&& \hspace{-5mm} s^2 \int_0^1 d\beta \, \beta^{-2\epsilon} (1-\beta)^{-1/2}
\left\langle {c^2 \over (p+q)^2 - r^2 c^2}
        \right\ranglecx 
\\ && \hspace{-4mm} \nonumber
= - {1 \over 4\epsilon}
{ (1)_\epsilon (2)_{-2\epsilon} \over (1)_{-\epsilon} }
\nonumber
\left[ { (1)_{-2\epsilon} (1)_{-\epsilon}
        \over ({3\over2})_{-2\epsilon} (2)_{-2\epsilon} (1)_{\epsilon} }
\right.\\&& \hspace{0cm}\left.
\times F\left( { 1-2\epsilon , 1-\epsilon , {3 \over 2} , 1
        \atop {3\over2}-2\epsilon , 2-2\epsilon , 1+\epsilon }
        \Bigg| 1 \right)
%\right.\nonumber\\ \nonumber&& \hspace{0cm} \left. 
\;-\;\nonumber
{ (1)_{-3\epsilon} (1)_{-2\epsilon} ({3\over2})_{-\epsilon}
        \over ({3\over2})_{-3\epsilon} (2)_{-3\epsilon} }
\right.\\ && \left. \hspace{5mm}
\times F\left( { 1-3\epsilon , 1-2\epsilon , {3\over2}-\epsilon
                \atop {3\over2}-3\epsilon,2-3\epsilon } \Bigg| 1 \right)
\right] \;.
\label{FF:1}
\end{eqnarray}
%
%{\bf [jmp]}
Expanding in powers of $\epsilon$,
we obtain
\begin{eqnarray}\nonumber
&& \hspace{-5mm} s^2 \int_0^1 d\beta \, \beta^{-2\epsilon} (1-\beta)^{-1/2}
\left\langle {c^2 \over (p+q)^2 - r^2 c^2}
        \right\ranglecx 
\\ &&
\hspace{2cm} = {\pi^2 \over 24}+O(\epsilon)\;.% (1 + 3.54518 \, \epsilon) \, ,
\label{intavecx:1p}
\end{eqnarray}
%
%{\bf [jmp]}

In the $-q$ case of (\ref{avec:c2}),
the argument of the hypergeometric functions can be written
$(1-\beta y)/(1-\beta \pm i \varepsilon)$, where $y=(1-x)/2$
and the prescriptions $+i \varepsilon$ and $-i \varepsilon$
correspond to the regions $p>q$ and $p<q$, respectively.
These regions correspond to the two terms
inside the average over $x$ in (\ref{int:n2f}).
In  order to obtain an analytic result in terms of hypergeometric functions,
it is necessary to integrate over $\beta$ before averaging over $x$.
The integrals over $\beta$ can be evaluated by first using a
transformation formula to change the argument of the hypergeometric
function to $-\beta(1-y)/(1-\beta)$ and then using
the integration formula (\ref{int-2F1}) to obtain hypergeometric
functions with arguments $y$ or $1-y$:
\begin{eqnarray}
&&\hspace{-5mm} \int_0^1 d\beta \, \beta^{-2\epsilon} (1-\beta)^{-3/2}
F\left( { {3\over2}, 1 \atop {5\over2}-\epsilon }
        \Bigg| {1-\beta y \over 1-\beta + i \varepsilon} \right)
\nonumber
\\
&&  \hspace{0cm} \;=\;
{3-2\epsilon \over \epsilon} \,
{(1)_{-2\epsilon} \over ({1\over2})_{-2\epsilon}} \,
F\left( { 1-2\epsilon, 1
                \atop 1+\epsilon } \Bigg| 1-y \right)
\nonumber
\\
&& \hspace{6mm}
- {3-2\epsilon \over \epsilon} \,
{ (1)_{\epsilon} \over ({1\over2})_{\epsilon} } \,
(1-y)^{-1/2}
F\left( { {1\over2}-2\epsilon, 1 \atop {1\over2}+\epsilon }
        \Bigg| 1-y \right)
\nonumber
\\ \nonumber
&& \hspace{6mm}
+ {3 \over 2\epsilon(1-3\epsilon)} e^{i\pi \epsilon} \,
(1)_{\epsilon} (\mbox{${5\over2}$})_{-\epsilon}\,
(1-y)^{-\epsilon}
\\ && \hspace{12mm} \times
F\left( { 1-3\epsilon, {3\over2}-\epsilon \atop 2-3\epsilon }
        \Bigg| y \right) \;.
\end{eqnarray}
%
%{\bf [jmp]}
After averaging over $x$, we obtain hypergeometric
functions with argument 1:
\begin{eqnarray}
&& \hspace{-5mm} s^2 \int_0^1 d\beta \, \beta^{-2\epsilon} (1-\beta)^{-1/2}
\left\langle {c^2 \over (p+i\varepsilon - q)^2 - r^2 c^2}
        \right\ranglecx
\nonumber
\\
&&  \;=\;
{1 \over 4\epsilon}\,
{(1)_{-2\epsilon} \over ({1\over2})_{-2\epsilon}} \,
F\left( { 1-\epsilon, 1-2\epsilon, 1
                \atop 2-2\epsilon, 1+\epsilon } \Bigg| 1 \right)
\nonumber
\\
&& \hspace{6mm}
- {1 \over 2\epsilon}\,
{ (2)_{-2\epsilon} (1)_{\epsilon} ({1\over2})_{-\epsilon}
        \over (1)_{-\epsilon} ({1\over2})_{\epsilon}
                        ({3\over2})_{-2\epsilon} } \,
F\left( { {1\over2}-\epsilon,{1\over2}-2\epsilon,1
                \atop {3\over2}-2\epsilon,{1\over2}+\epsilon }
        \Bigg| 1 \right)
\nonumber
\\\nonumber
&& \hspace{6mm}
+ {1 \over 8\epsilon(1-3\epsilon)} \, e^{i\pi \epsilon}
{ (2)_{-2\epsilon} (1)_{-2\epsilon} (1)_{\epsilon}({3\over2})_{-\epsilon}
        \over (1)_{-\epsilon} (2)_{-3\epsilon} } \,
\\ && \hspace{12mm} \times
F\left( { 1-\epsilon, 1-3\epsilon, {3\over2}-\epsilon
                \atop 2-3\epsilon, 2-3\epsilon }
        \Bigg| 1 \right) \;.
\end{eqnarray}
%
%{\bf [jmp]}
Expanding in powers of $\epsilon$ and then taking the real parts,
we obtain
\begin{eqnarray}\nonumber
&& \hspace{-3mm}
{\rm Re} \; s^2 \int_0^1 d\beta \, \beta^{-2\epsilon} (1-\beta)^{-1/2}
\left\langle {c^2 \over (p+i\varepsilon - q)^2 - r^2 c^2}
        \right\ranglecx 
\\ && \hspace{3cm}
= - {\pi^2 \over 24}+O(\epsilon)\;.% (1 + 0.34275 \, \epsilon) \;.
\label{intavecx:1m}
\end{eqnarray}
%
%{\bf [jmp]}
Inserting the sum of the integrals~(\ref{intavecx:1p}) 
and~(\ref{intavecx:1m})
into the thermal integral (\ref{int:n2f}), we obtain

\begin{eqnarray}
\hspace{-5mm}
\int_{\bf pq}
{s^2 n^2_F(s) \over p^2 q^2}
{\rm Re} \left\langle c^2 {r^2 c^2 - p^2 - q^2 \over
        \Delta(p+i\varepsilon,q,r c)}  \right\ranglec &=&
O(\epsilon)\;.
%{T^2 \over (4\pi)^2} \left[ \, 0.133434 \, \right] \, , \;\; 
\label{intHTL:1d}
\end{eqnarray}
%{\bf [jmp]}

%
%Adding this integral to the subtracted integral in
%(\ref{intHTL:1f}), we obtain the final results in Eq.~(\ref{intHTL:1}).
It remains only to evaluate the integral in Eq.~(\ref{intHTL:3a})
with ${\bf p}\cdot{\bf q}$
in the numerator.
We begin by using the identity
\begin{eqnarray}
\nonumber
&&\hspace{-12mm} \left\langle c^2 \, {{\bf p}\cdot{\bf q} \over r^2} \,
{r^2 c^2 - p^2 - q^2 \over
        \Delta(p+i\varepsilon,q,r c)} \right\ranglecx
\\ \nonumber&&\hspace{-1cm}=
- {p^2+q^2 \over (p^2 - q^2 +i\varepsilon)^2} \langle c^2 \rangle_c
        \left\langle {{\bf p}\cdot{\bf q} \over r^2} \right\ranglex
\nonumber
\\
&&\hspace{-8mm}
-\; {1 \over 2} \sum_\pm  {1 \over (p+i\varepsilon \pm q)^2} \,
        \left\langle {{\bf p}\cdot{\bf q} \, c^4
                        \over (p+i\varepsilon \pm q)^2 - r^2 c^2}
                \right\ranglecx .
\label{pqrdelta}
\end{eqnarray}
%
%{\bf [jmp]}
In the first term on the right side,
the average over $c$ is a simple multiplicative factor:
$\langle  c^2\rangle_c = 1/(3-2\epsilon)$.
The average over $x$ gives hypergeometric functions of argument $\beta$:
\begin{equation}
\left\langle { {\bf p}\cdot{\bf q} \over r^2 } \right\ranglex \;=\;
{1 \over 8} \beta
\left[
F\left( { 1-\epsilon, 1 \atop 3-2\epsilon } \Bigg| \beta \right)
- F\left( { 2-\epsilon, 1 \atop 3-2\epsilon } \Bigg| \beta \right)
\right]\;.
\end{equation}
%
%{\bf [jmp]}
The integral over $\beta$ gives hypergeometric functions of argument 1:
\begin{eqnarray}
&& \hspace{-5mm} s^2 \int_0^1 d\beta \, \beta^{-2\epsilon} (1-\beta)^{-1/2}
{p^2 + q^2 \over (p^2 - q^2)^2}
\left\langle { {\bf p}\cdot{\bf q} \over r^2 } \right\ranglex
\nonumber
\\
&& \hspace{-3mm} =\;
- {1 \over 8} \,
{ (2)_{-2\epsilon} \over ({3\over2})_{-2\epsilon} } \,
\left[
F\left( { 2-2\epsilon, 1-\epsilon,1
                \atop {3\over2}-2\epsilon, 3-2\epsilon }
        \Bigg| 1 \right)
\right.\nonumber\\&& \hspace{0cm}\left.
- F\left( { 2-2\epsilon, 2-\epsilon,1
                \atop {3\over2}-2\epsilon, 3-2\epsilon }
        \Bigg| 1 \right) \right]
+ {1 \over 12} \,
{ (3)_{-2\epsilon} \over ({5\over2})_{-2\epsilon} } \,
\left[
F\left( { 1-\epsilon, 1 \atop {5\over2}-2\epsilon } \Bigg| 1 \right)
\right.\nonumber\\&& \hspace{0cm}\left.
\hspace{2cm} - F\left( { 2-\epsilon, 1 \atop {5\over2}-2\epsilon } \Bigg| 1 \right)
\right]
\;.
\end{eqnarray}
%
%{\bf [jmp]}
Expanding in powers of $\epsilon$, we obtain
\begin{equation}
s^2 \int_0^1 d\beta \, \beta^{-2\epsilon} (1-\beta)^{-1/2}
{p^2 + q^2 \over (p^2 - q^2)^2}
\left\langle { {\bf p}\cdot{\bf q} \over r^2 } \right\ranglex \;=\;
- {\pi^2 \over 16}+O(\epsilon)\;.
\label{intavecx:3}
\end{equation}
%
%{\bf [jmp]}
In the second term of (\ref{pqrdelta}), the average over $c$ is given by
(\ref{avec:c4}). In the $+q$ term, the average over
$x= \hat{\bf p} \cdot \hat{\bf q}$ is
\begin{eqnarray}
&&\hspace{-4mm}\left\langle
x F\left( { 1, {5\over2} \atop {7\over2}-\epsilon }
        \Bigg| {r^2 \over (p+q)^2} \right)
\right\ranglex = \,
{5-2\epsilon \over 4\epsilon}\,
\left[
F\left( { 2-\epsilon, 1, {5\over2}
        \atop 3-2\epsilon, 1+\epsilon } \Bigg| \beta \right)
\right.\nonumber\\ \nonumber&& \hspace{0cm}\left.
- F\left( { 1-\epsilon, 1, {5\over2}
        \atop 3-2\epsilon, 1+\epsilon } \Bigg| \beta \right)
\right]
+ {5 \over 4\epsilon}\,
{ (1)_{\epsilon} (1)_{-2\epsilon} (3)_{-2\epsilon} ({7\over2})_{-\epsilon}
        \over (1)_{-\epsilon} (3)_{-3\epsilon} } \,
\beta^{-\epsilon}
\\ && 
\hspace{-3mm} \times \left[
F\left( { 1-2\epsilon, {5\over2}-\epsilon
        \atop 3-3\epsilon } \Bigg| \beta \right)
- {1-2\epsilon \over 1-\epsilon}
F\left( { 2-2\epsilon, {5\over2}-\epsilon
                \atop 3-3\epsilon } \Bigg| \beta \right)
\right] . 
\nonumber \\
\end{eqnarray}
%
%{\bf [jmp]}
Integrating over $\beta$,
we obtain hypergeometric functions of argument 1:
\begin{eqnarray}
&& \hspace{-8mm} \int_0^1 d\beta \, \beta^{-2\epsilon} (1-\beta)^{-1/2} \,
\left\langle {{\bf p}\cdot{\bf q} \, c^4 \over (p+q)^2 - r^2 c^2}
        \right\ranglecx
\nonumber
\\ \nonumber
&& \hspace{-3mm} =\;
{1 \over 4\epsilon(3-2\epsilon)}\,
{(2)_{-2\epsilon} \over ({5\over2})_{-2\epsilon}} \,
\left[
F\left( { 2-2\epsilon, 2-\epsilon, 1, {5\over2}
        \atop {5\over2}-2\epsilon, 3-2\epsilon, 1+\epsilon } \Bigg| 1 \right)
\right.\\ &&\left.
\hspace{1cm} - F\left( { 2-2\epsilon, 1-\epsilon, 1, {5\over2}
        \atop {5\over2}-2\epsilon, 3-2\epsilon, 1+\epsilon } \Bigg| 1 \right)
\right]
\nonumber
\\
&& \hspace{1mm}
+ \; {1 \over 6\epsilon(2-3\epsilon)}\,
{ (1)_{\epsilon} (1)_{-2\epsilon} (3)_{-2\epsilon} ({3\over2})_{-\epsilon}
        \over (1)_{-\epsilon} ({5\over2})_{-3\epsilon} } \,
\nonumber
\\ \nonumber
&& \hspace{0cm} \times \left[
F\left( { 2-3\epsilon, 1-2\epsilon, {5\over2}-\epsilon
                \atop {5\over2}-3\epsilon, 3-3\epsilon }
        \Bigg| 1 \right)
\right.\\ &&\left.
\hspace{8mm} - {1-2\epsilon \over 1-\epsilon}
F\left( { 2-3\epsilon, 2-2\epsilon, {5\over2}-\epsilon
                \atop {5\over2}-3\epsilon, 3-3\epsilon }
        \Bigg| 1 \right)
\right] \;.
\end{eqnarray}
%
%{\bf [mp]}
Expanding in powers of $\epsilon$, we obtain
\begin{eqnarray}\nonumber
&&
\int_0^1 d\beta \, \beta^{-2\epsilon} (1-\beta)^{-1/2}  \,
\left\langle {{\bf p}\cdot{\bf q} \, c^4 \over (p+q)^2 - r^2 c^2}
        \right\ranglecx =
{\pi^2 - 6 \over 18}\;.
\\ &&
% (1-0.0728428 \, \epsilon)\;.
\label{intavecx:4p}
\end{eqnarray}
%
%{\bf [mp]}
In the $-q$ term in the integral of the second term of (\ref{pqrdelta}),
we integrate over $\beta$ before averaging over $x$.
The integral over $\beta$ can be expressed in terms of
hypergeometric functions of type $_2F_1$:
\begin{eqnarray}
&& \hspace{-3mm} s^2 \int_0^1 d\beta \, \beta^{-2\epsilon} (1-\beta)^{-1/2}  \,
{4 {\bf p}\cdot{\bf q} \over (p-q)^2}
\left\langle {c^4 \over (p+i\varepsilon - q)^2 - r^2 c^2}
        \right\ranglec
\nonumber
\\\nonumber
&& \hspace{0mm} \;=\;
- {1 \over 2(3-2\epsilon)\epsilon} \,
{(2)_{-2\epsilon} \over ({1\over2})_{-2\epsilon}} \,
(1-2y) \,
F\left( { 2-2\epsilon, 1 \atop 1+\epsilon } \Bigg| 1-y \right)
\nonumber
\\\nonumber
&& \hspace{3mm}
- {1 \over 4(3-2\epsilon)\epsilon}  \,
{(1)_{\epsilon} \over (-{1\over2})_{\epsilon}}
(1-2y) \, (1-y)^{-3/2}  \,
\\ \nonumber&&
\hspace{1cm} \times F\left( { {1\over2}-2\epsilon, 1 \atop -{1\over2} +\epsilon } \Bigg| 1-y \right)
\nonumber
\\\nonumber
&& \hspace{3mm}
+ {1 \over 8(2-3\epsilon)\epsilon}
e^{\mp i \pi \epsilon} (1)_\epsilon (\mbox{$3\over2$})_{-\epsilon} \,
(1-2y) \, (1-y)^{-\epsilon} \,
\\ && 
\hspace{1cm} \times F\left( { 2-3\epsilon, {5\over2}-\epsilon
        \atop 3-3\epsilon } \Bigg| y \right)
\;.
\end{eqnarray}
%
%{\bf [mp]}
The phase in the last term is $e^{-i \pi \epsilon}$ for the
$f(s_+,s_-,r)$ term of (\ref{int:n2f}), which comes from the $p>q$ region
of the integral, and $e^{i \pi \epsilon}$
for the $f(s_-,s_+,r)$ term, which comes from the $p<q$ region.
The average over $x=\hat{\bf p} \cdot \hat{\bf q}$
can be expressed in terms of
hypergeometric functions of type $_3F_2$ evaluated at 1:
\begin{eqnarray}
&& \hspace{-5mm} s^2 \int_0^1 d\beta \, \beta^{-2\epsilon} (1-\beta)^{-1/2}  \,
\left\langle {4 {\bf p} \cdot {\bf q} \over (p-q)^2} \,
        {c^4 \over (p+i\varepsilon - q)^2 - r^2 c^2}
        \right\ranglecx
\nonumber
\\\nonumber
&& \hspace{0cm} \;=\;
{1 \over 4(3-2\epsilon)\epsilon} \,
{(2)_{-2\epsilon}\over ({1\over2})_{-2\epsilon}}
\left[ F\left( { 1-\epsilon,2-2\epsilon, 1
                \atop 3-2\epsilon,1+\epsilon } \Bigg| 1 \right)
\right.\\ && \left. \nonumber
\hspace{1cm} - F\left( { 2-\epsilon,2-2\epsilon, 1
                \atop 3-2\epsilon,1+\epsilon } \Bigg| 1 \right) \right]
\nonumber
\\
&& \hspace{0cm}
- {1 \over (3-2\epsilon)\epsilon} \,
{ (1)_{\epsilon} (3)_{-2\epsilon} (-{1\over2})_{-\epsilon}
\over (1)_{-\epsilon} (-{1\over2})_{\epsilon} ({3\over2})_{-2\epsilon} }
\left[ F\left( { -{1\over2}-\epsilon, {1\over2}-2\epsilon, 1
        \atop {3\over2}-2\epsilon, -{1\over2}+\epsilon } \Bigg| 1 \right)
\right.
\nonumber
\\\nonumber
&& \hspace{0cm} \left.
\hspace{1cm} + {1+2\epsilon \over 2(1-\epsilon)}
        F\left( { {1\over2}-\epsilon, {1\over2}-2\epsilon, 1
        \atop {3\over2}-2\epsilon, -{1\over2}+\epsilon } \Bigg| 1 \right)
\right]
\nonumber
\\ \nonumber
&& \hspace{0cm}
+ {1 \over 16(2-3\epsilon)\epsilon} \, e^{\mp i \pi \epsilon} \,
{ (1)_{\epsilon} (2)_{-2\epsilon} (2)_{-2\epsilon} ({3\over2})_{-\epsilon}
        \over (1)_{-\epsilon} (3)_{-3\epsilon} }
\\&& \nonumber
\hspace{1cm}
\times\left[ F\left( { 1-\epsilon,  2-3\epsilon, {5\over2}-\epsilon
                \atop 3-3\epsilon, 3-3\epsilon } \Bigg| 1 \right)
\right.\\ && \left.
\hspace{15mm}- {1-\epsilon \over 1-2\epsilon}
        F\left( { 2-\epsilon,  2-3\epsilon, {5\over2}-\epsilon
                \atop 3-3\epsilon, 3-3\epsilon } \Bigg| 1 \right)
\right]  \;.
\end{eqnarray}
%
%{\bf [mp]}
The expansion of the real part of the integral in powers of $\epsilon$ is
\begin{eqnarray}\nonumber
&&\hspace{-3mm}s^2 \int_0^1 d\beta \, \beta^{-2\epsilon} (1-\beta)^{-1/2} \,
\\ && \hspace{-1mm} \times \hspace{1mm} \nonumber
{\rm Re} \left\langle {4 {\bf p} \cdot {\bf q} \over (p-q)^2} \,
        {c^4 \over (p+i\varepsilon - q)^2 - r^2 c^2}
        \right\ranglecx
%\nonumber\\&& \hspace{8cm}
%\\ \nonumber&&
= {9-\pi^2\over18}+O(\epsilon)\,.
\\ &&
\label{intavecx:4m}
\end{eqnarray}
%
%{\bf [mp]}
Inserting (\ref{intavecx:3}), (\ref{intavecx:4p}), and (\ref{intavecx:4m})
into the thermal integral of (\ref{pqrdelta}), we obtain
\begin{eqnarray}\nonumber
&& \hspace{-5mm} \int_{\bf pq}
{s^2 n^2_F(s) \over p^2 q^2} {{\bf p}\cdot{\bf q} \over r^2}
{\rm Re} \left\langle c^2 {r^2 c^2 - p^2 - q^2 \over
        \Delta(p+i\varepsilon,q,r c)}  \right\ranglec 
\\ &&
\hspace{1cm} = {T^2 \over (4\pi)^2}%  \left({\mu\over4\pi T}\right)^{4\epsilon}
{\pi^2-1\over6\pi^2} \left[{\pi^2\over12}-\log2\right] \,.
\end{eqnarray}
%
%{\bf [jmp]}
Inserting this along with (\ref{intHTL:1d}) into (\ref{intHTL:3a}),
we obtain
\begin{eqnarray}\nonumber
&& \hspace{-5mm} \int_{\bf pq}
{s^2 n_F^2(s) \over p^2 r^2}
{\rm Re} \left\langle c^2 {r^2 c^2 - p^2 - q^2 \over
        \Delta(p+i\varepsilon,q,r c)}  \right\ranglec 
\\ &&
\hspace{1cm} = {T^2 \over (4\pi)^2}
{1-\pi^2\over6\pi^2} \left[{\pi^2\over12}-\log2\right] \,.
\label{intHTL:3d}
\end{eqnarray}
%
%{\bf [jmp]}
Adding this integral to the subtracted integral in
(\ref{intHTL:3f}), we obtain the final result in (\ref{f3}).
The subtracted integral appearing in~(\ref{extra}) vanishes due to 
antisymmetry of the integrand. Thus the final result~(\ref{ff4}) is given
by~(\ref{extra}).

The integrals~(\ref{intHTL:4x}) and~(\ref{intHTL:5x}) 
can be computed directly in three dimensions, as described above.
The integrals~(\ref{intHTL:6x})--~(\ref{llll}) 
are divergent and require subtractions to remove the divergences.
We first isolate the divergent part which come from the region 
$q\rightarrow0$. We need one subtraction:
\bqa
\label{nsub-1}
n_B(q)&=&\left(n_B(q)-{T\over q}+{1\over2}\right)
+{T\over q}-{1\over2}\;.
\eqa
%{\bf [jmp]}
In the integral~(\ref{intHTL:7x}), it is convenient to first use the
identity $r_c^2 = p^2 + 2 {\bf p} \cdot {\bf q}/c + q^2/c^2$ to expand
it into 3 integrals, two of which are (\ref{intHTL:4x}) and (\ref{intHTL:6x}).
In the third integral, the subtraction (\ref{nsub-1})
is needed to remove the divergences.

For the convergent terms, the HTL average over $c$ and the angular 
average over $x=\hat{\bf p}\cdot\hat{\bf q}$ can be calculated in 
three dimensions:
\bqa\nonumber
&& \hspace{-3mm}
{\rm Re} \left\langle c
        {r_c^2 - p^2 - q^2 \over \Delta(p+i\varepsilon,q,r_c)} \,
        \right\ranglecx =
{1 \over 6(4p^2 - q^2)}
\\ \nonumber && \hspace{-1mm}
+ {q^2(4p^2+3q^2) \over 3(4p^2 - q^2)^3} \log{2p \over q}
+ {(p+q)(4p^2+2p q+q^2) \over 12 p q(2p+q)^3} \log{p+q \over p}
\\ && \hspace{5mm}
- {(p-q)(4p^2-2p q+q^2) \over 12 p q(2p-q)^3} \log{|p-q| \over p}
\, ,
\\
&& \hspace{-3mm} {\rm Re} \left\langle \hat{\bf p} \cdot \hat{\bf q}
        {r_c^2 - p^2 - q^2 \over \Delta(p+i\varepsilon,q,r_c)} \,
        \right\ranglecx =
{1 \over 6pq}
- {q(12p^2-q^2) \over 6p (4p^2 - q^2)^2} \log{4p \over q}
\nonumber
\\\nonumber
&& \hspace{5mm}
+ {(p+q)(2p^2-2p q-q^2) \over 12 p^2q(2p+q)^2} \log{p+q \over 4p}
\\ && \hspace{5mm}
+ {(p-q)(2p^2+2p q-q^2) \over 12 p^2q(2p-q)^2} \log{|p-q| \over 4p} \;,\\
&&\hspace{-3mm}\mbox{Re}\left\langle\nonumber
{r_c^2-p^2\over q^2}{r^2_c-p^2-q^2\over\Delta(p+i\epsilon,q,r_c)}c^{-1}
-{1\over q^2}c^{-1}+{\log2\over q^2}
\right\rangle_{c,x}
\\
&& \hspace{5mm} = {1\over4pq^2}\left[q\log{p+q\over|p-q|}
+p\log{|p^2-q^2|\over p^2}
\right]\;.
\eqa
%
%{\bf [jmp]}
The remaining 2-dimensional integral over $p$ and $q$
can then be evaluated numerically:
\bqa\nonumber
&& \hspace{-5mm} \int_{\bf pq} {n_F(p)\over p}
\left( {n_B(q) \over q} - {T\over q^2} +{1\over2q}\right)
{p^2 \over q^2} \,
\\ &&\nonumber
\times{\rm Re}\left\langle c
        {r_c^2 - p^2 - q^2 \over \Delta(p+i\varepsilon,q,r_c)} \,
        \right\ranglec \;=\;
{T^2 \over (4 \pi)^2} \left[1.480\times 10^{-2} 
\right]  \; ,
\\ &&
\label{intHTL:6f}
\\\nonumber
&& \hspace{-5mm} \int_{\bf pq} {n_F(p)\over p}
\left( {n_B(q) \over q} - {T\over q^2}+{1\over2q} \right){{\bf p} \cdot {\bf q} \over q^2} \, 
\\ &&\nonumber\times
{\rm Re} \left\langle
        {r_c^2 - p^2 - q^2 \over \Delta(p+i\varepsilon,q,r_c)} \,
        \right\ranglec
={T^2 \over (4 \pi)^2} \left[ -2.832\times 10^{-3} 
\right]
\; ,
\\ &&
\label{intHTL:8f}
\\&&\nonumber
\hspace{-5mm} \int_{\bf pq}{n_F(p)\over p}{n_F(q)\over q}
\mbox{Re}\left\langle
{r_c^2-p^2\over q^2}{r^2_c-p^2-q^2\over\Delta(p+i\epsilon,q,r_c)}c^{-1}
\right.\\ && \hspace{5mm}
\left.-{1\over q^2}c^{-1}+{\log2\over q^2}
\right\ranglec
={T^2\over(4\pi)^2}\left[4.134\times10^{-2}\right]\;,
\label{nnn}
\\&&\nonumber\hspace{-5mm}
\int_{\bf pq}{n_B(p)\over p}{n_F(q)\over q}
\mbox{Re}\left\langle
{r_c^2-p^2\over q^2}{r^2_c-p^2-q^2\over\Delta(p+i\epsilon,q,r_c)}c^{-1}
\right.\\ &&\left. \hspace{5mm}
-{1\over q^2}c^{-1}+{\log2\over q^2}
\right\ranglec
={T^2\over(4\pi)^2}\left[2.530\times10^{-1}\right]\;.
\label{nnnn}
\eqa
%{\bf [jmp]}
The integrals involving the terms subtracted from $n(q)$
in~(\ref{nsub-1})
are divergent, so the HTL average over
$c$ and the angular average over $x = \hat {\bf p} \cdot \hat {\bf q}$
must be calculated in $3-2\epsilon$ dimensions.
The first step in the calculation of the subtracted terms
is to replace the average over $c$ of the integral over $q$
by an average over $c$ and $x$:
\begin{eqnarray}
\nonumber
&&\int_{\bf q} {1 \over q^n} \,
\left\langle f(c)
{r_c^2 - p^2 - q^2 \over \Delta(p+i\varepsilon,q,r_c)} \right\ranglec
\\ \nonumber&&
=(-1)^{n-1} {1\over 8 \pi^2 \epsilon}
{ (1)_{2 \epsilon}\nonumber
 (1)_{-2\epsilon}
        \over ({3\over2})_{-\epsilon} }
(e^\gamma \mu^2)^\epsilon (2p)^{1-n-2\epsilon}
\\\nonumber
&& \hspace{-2mm}
\times \left\langle f(c) \, c^{3-n-2\epsilon}(1-c^2)^{n-2+2\epsilon}
\sum_\pm (x\mp c - i \varepsilon)^{1-n-2\epsilon} \right\ranglecx \! .
\\ &&
\label{intthq}
\end{eqnarray}
%
%{\bf [jmp]}
The integral over $p$ can now be evaluated easily
using either (\ref{int-th:-1}) or
\begin{equation}
\int_{\bf p} n_F(p) \, p^{-2-2\epsilon} =
{1\over 2\pi^2}
{(1)_{-4\epsilon} \over ({3\over2})_{-\epsilon}}
(1-2^{4\epsilon})\zeta(1-4\epsilon)
(e^\gamma \mu^2)^\epsilon T^{1-4\epsilon} \;.
\label{int-th:-2}
\end{equation}
%
%{\bf [jmp]}
It remains only to calculate the averages
over $c$ and $x$.  The averages over $x$ give $_2F_1$ hypergeometric
functions with argument $[(1 \mp c)/2 - i \varepsilon]^{-1}$:
\begin{eqnarray}
&&\nonumber \hspace{-9mm}
\left\langle  (x\mp c - i \varepsilon)^{-n-2\epsilon}
        \right\ranglex =
(1\mp c)^{-n-2\epsilon}
\\ &&
\times F\left( { 1-\epsilon,n+2\epsilon \atop 2-2\epsilon}
        \Bigg| [(1 \mp c)/2 - i \varepsilon]^{-1} \right) \;,
\label{avex:1}
\\\nonumber
&& \hspace{-9mm} \left\langle x (x\mp c - i \varepsilon)^{-n-2\epsilon}
        \right\ranglex =
{1 \over 2} (1\mp c)^{-n-2\epsilon}
\\ && \hspace{-2mm}
\times\left[ F\left( { 1-\epsilon,n+2\epsilon \atop 3-2\epsilon}
        \Bigg| [(1 \mp c)/2 - i \varepsilon]^{-1} \right) \right.
\nonumber
\\
&& \hspace{0cm}
\left. -
F\left( { 2-\epsilon,n+2\epsilon \atop 3-2\epsilon}
        \Bigg| [(1 \mp c)/2 - i \varepsilon]^{-1} \right) \right] \;.
\label{avex:2}
\end{eqnarray}
%
%{\bf [m]}
Using a transformation formula, the arguments
can be changed to $(1 \mp c)/2 - i \varepsilon$.
If the expressions (\ref{avex:1}) and (\ref{avex:2})
are averaged over $c$ with a weight that is an even
function of $c$, the $+$ and $-$ terms combine to give
$_3F_2$ hypergeometric functions with argument 1.
For example,
\begin{eqnarray}
&&\nonumber \hspace{-1cm}
\left\langle (1-c^2)^{2 \epsilon}
	\sum_\pm (x\mp c - i \varepsilon)^{-1-2\epsilon}
	\right\ranglecx 
\\ && \hspace{-6mm}
= {1 \over 3\epsilon}
{ (2)_{-2 \epsilon} (1)_\epsilon ({3\over2})_{-\epsilon}
	\over (1)_{-\epsilon} (1)_{-\epsilon} }
\nonumber
\\
&&  \hspace{-5mm} \times
\left\{ - e^{- i \pi \epsilon}
	{ (1)_{3\epsilon} (1)_{-2 \epsilon}
	\over (1)_{2 \epsilon} (2)_{-\epsilon} }
F\left( { 1-2\epsilon,1-\epsilon,\epsilon
	\atop 2-\epsilon,1-3\epsilon} \Bigg| 1 \right)
\right.
\nonumber
\\
&&  \left. \hspace{-4mm}
+ \;  e^{i 2\pi \epsilon}
	{ (1)_{-3\epsilon} (1)_\epsilon
	\over  (1)_{-4\epsilon} (2)_{2\epsilon} }
F\left( { 1+\epsilon,1+2\epsilon,4\epsilon
	\atop 2+2\epsilon,1+3\epsilon} \Bigg| 1 \right)
\right\} \,.
\end{eqnarray}
%
%{\bf [jm]}
Upon expanding the hypergeometric functions in powers of
$\epsilon$ and taking the real parts, we obtain
\begin{eqnarray}
&& \nonumber \hspace{-1cm}
{\rm Re} \left\langle (1-c^2)^{2 \epsilon}
	\sum_\pm (x\mp c - i \varepsilon)^{-1-2\epsilon}
	\right\ranglecx 
\\ &&
\hspace{18mm} = \pi^2 \left[ - \epsilon + 2 (1-\log 2) \epsilon^2 \right] \, ,
%\;\;[j]
\label{avecx:1}
\\ &&\nonumber \hspace{-1cm}
{\rm Re} \left\langle c^2 (1-c^2)^{2 \epsilon}
	\sum_\pm (x\mp c - i \varepsilon)^{-1-2\epsilon}
	\right\ranglecx 
\\ &&
\hspace{18mm} = \pi^2 \left[ - {1 \over 3} \epsilon
	+ {2 \over 9} (2-3\log 2) \epsilon^2 \right] \, ,
\;\;
\label{avecx:2}
\\ &&\nonumber \hspace{-1cm}
{\rm Re} \left\langle (1-c^2)^{2+2 \epsilon}
	\sum_\pm (x\mp c - i \varepsilon)^{-3-2\epsilon}
	\right\ranglecx 
\\ &&
\hspace{18mm} = \pi^2 \left[ - {8 \over 3} \epsilon^2  \right] \, ,
%\;\;[j]
\label{avecx:3}
\\ && \nonumber \hspace{-1cm}
{\rm Re} \left\langle x (1-c^2)^{1+2 \epsilon}
	\sum_\pm (x\mp c - i \varepsilon)^{-2-2\epsilon}
	\right\ranglecx
\\ &&
\hspace{18mm} = \pi^2 \left[ - {2 \over 3} \epsilon
	+ {2 \over 9} (1-6\log 2) \epsilon^2  \right] \;.
\label{avecx:5}
\end{eqnarray}
%
%{\bf [mp]}

If the expressions (\ref{avex:1}) and (\ref{avex:2})
are averaged over $c$ with a weight that is an odd
function of $c$, they reduce to integrals of $_2F_1$
hypergeometric functions with argument $y$.
For example,
\begin{eqnarray}
&&\hspace{-7mm}\left\langle c (1-c^2)^{1+2 \epsilon}
        \sum_\pm (x\mp c - i \varepsilon)^{-2-2\epsilon}
        \right\ranglecx =
{(2)_{-2\epsilon} ({3\over2})_{-\epsilon}
\over (1)_{-\epsilon} (1)_{-\epsilon}}
\nonumber
\\ \nonumber
&& \hspace{0.cm} \hspace{-3mm} \times \left\{
- 2 e^{-i\pi \epsilon}
{(1)_{3\epsilon} \over (2)_{2\epsilon}}
\int_0^1 dy \, y^{-2\epsilon} (1-y)^{1+\epsilon} |1-2y|
\right.\\ &&\left.
\hspace{2cm} \times F\left( { 1-\epsilon,\epsilon \atop -3\epsilon } \Bigg| y \right)
\right.
\nonumber
\\ \nonumber
&& \hspace{0cm} \left. \hspace{2mm}
- {8 \over 3(1+3\epsilon)} e^{2i\pi \epsilon}
{(1)_{-3\epsilon} \over (1)_{-4\epsilon}}
\int_0^1 dy \, y^{1+\epsilon} (1-y)^{1+\epsilon} |1-2y|
\right.\\ &&\left. \hspace{2cm} \times
F\left( { 2+2\epsilon,1+4\epsilon \atop 2+3\epsilon } \Bigg| y \right)
\right\} \;.
\end{eqnarray}
%
%{\bf [m]}

The resulting expansions for the real parts
of the averages over $c$ and $x$ are
\begin{eqnarray}
&&\nonumber\hspace{-4mm}
{\rm Re} \left\langle c (1-c^2)^{1+2 \epsilon}
        \sum_\pm (x\mp c - i \varepsilon)^{-2-2\epsilon}
        \right\ranglecx 
\\ &&
\hspace{28mm} = -1 + {14(1-\log2) \over3} \epsilon \;,
\label{avecx:4}
\\ \nonumber
&& \hspace{-4mm}
{\rm Re} \left\langle x c(1-c^2)^{2 \epsilon}
        \sum_\pm (x\mp c - i \varepsilon)^{-1-2\epsilon}
        \right\ranglecx =
{2 (1 - \log 2) \over3}
\nonumber
\\
&& \hspace{8mm} + \left( {4\over9} + {8\over 9} \log 2
- {4\over3}\log^22 + {\pi^2 \over 18} \right) \epsilon \;.
\label{avecx:6}
\end{eqnarray}
%
%{[\bf jmp]}
Multiplying each of these expansions by the appropriate
factors from the integral over $q$ in (\ref{intthq}) and
the integral over $p$ in (\ref{int-th:-2}) or (\ref{int-th:-1}),
we obtain
\bqa
&& \hspace{-4mm} \int_{\bf pq} {n_F(p)\over p} {p^2 \over q^3}
{\rm Re}\left\langle c^{1+2\epsilon}
        {r_c^2 - p^2 -  q^2 \over \Delta(p+i\varepsilon,q,r_c)} \,
        \right\ranglec 
 = {T^2 \over (4\pi)^2}  \left({\mu\over4\pi T}\right)^{4\epsilon}
\nonumber
\\
&& \hspace{8mm}
\times\left(-{1 \over 48}\right)
\left[ {1\over\epsilon} - {2\over3} - {4\over3} \log2
        +4 {\zeta'(-1) \over \zeta(-1)} \right] \, ,
\label{intHTL:6db} \\ %\nonumber
&& \hspace{-4mm} \int_{\bf pq} {n_F(p)\over p} {p^2 \over q^4}
{\rm Re}\left\langle c^{1+2\epsilon}
        {r_c^2 - p^2 -  q^2 \over \Delta(p+i\varepsilon,q,r_c)} \,
        \right\ranglec 
	\; = \;  O(\epsilon)\;,
\label{intHTL:6dbext}
\\ \nonumber
&& \hspace{-4mm} \int_{\bf pq} {n_F(p)\over p} {{\bf p} \cdot {\bf q} \over q^3}
{\rm Re} \left\langle   c^{2\epsilon}
        {r_c^2 - p^2 -  q^2 \over \Delta(p+i\varepsilon,q,r_c)} \,
        \right\ranglec =
	{T^2 \over (4\pi)^2}  \left({\mu\over4\pi T}\right)^{4\epsilon}
\\ && \nonumber
\hspace{6mm} \times \left(-{1 \over 36}\right)%\nonumber
\Bigg[ (1-\log2)
 \left( {1\over\epsilon} + {14\over3}-4\log2 \right.
\\ && \hspace{4cm} \left. 
                + 4 {\zeta'(-1) \over \zeta(-1)} \right)
        + {\pi^2 \over 12} \Bigg] \, ,
\label{intHTL:8db}
\\\nonumber
&& \hspace{-3mm} \int_{\bf pq} {n_F(p)\over p} {{\bf p} \cdot {\bf q} \over q^4}
{\rm Re} \left\langle   c^{2\epsilon}
        {r_c^2 - p^2 -  q^2 \over \Delta(p+i\varepsilon,q,r_c)} \,
        \right\ranglec  \\
	&& \hspace{38mm}
	= {T\over (4\pi)^2}\left(-{1\over6}\log2\right)\;.
\label{intHTL:8dbext}
\eqa
%
%{\bf [jmp]}

Adding Eq.~(\ref{intHTL:6dbext}) to the subtracted integral~(\ref{intHTL:6f})
we obtain the final result in Eq.~(\ref{intHTL:6x}).
Combining (\ref{intHTL:8f}) with (\ref{intHTL:8db}) and~(\ref{intHTL:8dbext}), 
we obtain
\begin{eqnarray}
&& \nonumber \hspace{-14mm}
\int_{\bf pq} {n_F(p)n_B(q)\over pq} {{\bf p} \cdot {\bf q} \over q^2}
{\rm Re} \left\langle   c^{2\epsilon}
        {r_c^2 - p^2 -  q^2 \over \Delta(p+i\varepsilon,q,r_c)} \,
        \right\ranglec 
\\ &&
\hspace{-8mm} = {T^2 \over (4\pi)^2}  \left({\mu\over4\pi T}\right)^{4\epsilon}%\left(-1\right)
%\nonumber\\&& 
\left({1-\log2\over 72}\right)
\left[{1\over\epsilon} - 15.2566 \right] \; .
\label{fpq}
\end{eqnarray}
%
%{\bf [jmp]}
The integral~(\ref{intHTL:7x}) is obtained from~(\ref{intHTL:4x}),~(\ref{intHTL:6x}) and~(\ref{fpq}). Finally consider~(\ref{lll}) and~(\ref{llll}).
In order to evaluate them we need two subtractions for each integral
\bqa&&\nonumber \hspace{-1cm}
\int_{\bf pq}{n_F(p)\over p}{n_F(p)\over q}{1\over q^2}
\langle c^{2\epsilon}\rangle_c
={T^2\over(4\pi)^2}\left({\mu\over4\pi T}\right)^{4\epsilon}
\\ && \hspace{-5mm}
\times\left(-{1\over12}\right)
\left[
{1\over\epsilon}+2+2\log2+2\gamma
%\right.  \\ && \left.
\hspace{0cm}
+2{\zeta^{\prime}(-1)\over\zeta(-1)}
\right]\;,
\label{ss1}
\\ \nonumber
&&\hspace{-1cm}\int_{\bf pq}{n_F(p)\over p}{n_F(q)\over q}{1\over q^2}
\langle c^{-1+2\epsilon}\rangle_c
={T^2\over(4\pi)^2}\left({\mu\over4\pi T}\right)^{4\epsilon}
\\ && \hspace{-7mm} \nonumber
\times \left(-{1\over24}\right)
\Bigg[
{1\over\epsilon^2}
+\left.(2+2\gamma+4\log2
%\right.\\&& \left.  \hspace{0cm}
+2{\zeta^{\prime}(-1)\over\zeta(-1)}
\right){1\over\epsilon}
\\ && \hspace{42mm} 
+53.1064
\Bigg]
\label{ss2}
\;, 
\\ \nonumber
&& \hspace{-1cm}
\int_{\bf pq}{n_B(p)\over p}{n_F(q)\over q}{1\over q^2}
\langle c^{2\epsilon}\rangle_c
={T^2\over(4\pi)^2}\left({\mu\over4\pi T}\right)^{4\epsilon}
\\ && \hspace{-4mm} \times\left(-{1\over6}\right)
\left[
{1\over\epsilon}+2+4\log2+2\gamma
+2{\zeta^{\prime}(-1)\over\zeta(-1)}
\right]\;, 
\label{ss3}
\\ \nonumber
&& \hspace{-1cm}
\int_{\bf pq}{n_B(p)\over p}{n_F(q)\over q}{1\over q^2}
\langle c^{-1+2\epsilon}\rangle_c
=
{T^2\over(4\pi)^2}\left({\mu\over4\pi T}\right)^{4\epsilon}
\\ && \nonumber \hspace{-5mm}
\times \left(-{1\over12}\right)
\Bigg[
{1\over\epsilon^2}
+\left(2+2\gamma+6\log2
+2{\zeta^{\prime}(-1)\over\zeta(-1)} \right){1\over\epsilon}
\\ && \hspace{42mm} +69.7097
\Bigg]\;.
\label{ss4}
\eqa
%{\bf [jmp]}
The subtractions can be evaluated directly in three dimensions and the
results are given in Eqs.~(\ref{nnn})--(\ref{nnnn})
The integrals~(\ref{lll}) and~(\ref{llll}) are then given by the
by the sum of the difference terms~(\ref{nnn}) and~(\ref{nnnn}) and
the subtraction terms~(\ref{ss1})--(\ref{ss4}).

\subsection{4-dimensional integrals}

In the sum-integral formula (\ref{int-2loop}),
the second term on the right side involves an integral over
4-dimensional Euclidean momenta. The integrands are functions
of the integration variable $Q$ and $R=-(P+Q)$.
The simplest integrals to evaluate are those whose integrands
are independent of $P_0$:
\bqa
\int_Q {1 \over Q^2 r^2} & = &
{1 \over (4 \pi)^2} \mu^{2 \epsilon} p^{-2\epsilon}
\; 2 \left[{1 \over \epsilon} + 4 - 2 \log 2 \right] \;,
\label{int4:1}
\\
\int_Q {q^2 \over Q^2 r^4} & = &
{1 \over (4 \pi)^2} \mu^{2 \epsilon} p^{-2\epsilon}
\;2 \left[{1 \over \epsilon} + 1 - 2 \log 2  \right]\;,
\\\nonumber
\int_Q {1 \over Q^2 r^4} & = &
{1 \over (4 \pi)^2} \mu^{2 \epsilon} p^{-2-2\epsilon}
\; ( -2 ) \left[1 + 
%\right. \\ && \left.
(-2 - 2 \log 2) \epsilon \right] \,. \\
\eqa
%
%[{\bf jmp}]
%
Another simple integral that is needed
depends only on $P^2=P_0^2+p^2$:
\bqa
&& \hspace{-15mm} \int_Q {1 \over Q^2 R^2}
= 
{1 \over (4 \pi)^2} (e^\gamma \mu^2)^\epsilon (P^2)^{-\epsilon} \;
{1 \over \epsilon} \,
{(1)_\epsilon (1)_{-\epsilon} (1)_{-\epsilon}
        \over (2)_{-2\epsilon}} \, ,
\label{int4:8}
\eqa
%
%[{\bf jmp}]
where $(a)_b$ is Pochhammer's symbol which is defined in (\ref{Poch}).
We need the following weighted averages over $c$ of this function
evaluated at $P = (-i p,{\bf p}/c)$:
\bqa
&& \nonumber \hspace{-1cm}
\left\langle c^{-1+2\epsilon}
        \int_Q {1 \over Q^2 R^2} \bigg|_{P \to (-i p,{\bf p}/c)}
        \right\ranglec
=
{1 \over (4 \pi)^2} \mu^{2 \epsilon} p^{-2\epsilon}
\\ && \hspace{7mm}
\times {1 \over 4}
\left[ {1 \over \epsilon^2} + {2 \log 2 \over \epsilon}
        + 2 \log^2 2 + {3 \pi^2 \over 4} \right]
\, ,
\label{int4:8.1}
\\ && \hspace{-1cm} \nonumber
\left\langle c^{1+2\epsilon}
        \int_Q {1 \over Q^2 R^2} \bigg|_{P \to (-i p,{\bf p}/c)}
        \right\ranglec
=
{1 \over (4 \pi)^2} \mu^{2 \epsilon} p^{-2\epsilon}
\\ &&
\hspace{32mm} \times {1 \over 2}
\left[ {1 \over \epsilon} + 2 \log 2  \right]
\;.
\label{int4:8.2}
\eqa
%
%[{\bf jmp}]

The remaining integrals are functions of $P_0$ that must
be analytically continued to the point $P_0 = -i p + \varepsilon$.
Several of these integrals are straightforward to evaluate:
\bqa
&& \hspace{-7mm}
\int_Q {q^2 \over Q^2 R^2}
        \bigg|_{P_0 = -i p} =  0 \;,
\label{int4:4}
\\ && \nonumber \hspace{-7mm}
\int_Q {q^2 \over Q^2 r^2 R^2}
        \bigg|_{P_0 = -i p} =
{1 \over (4 \pi)^2} \mu^{2 \epsilon} p^{-2 \epsilon}
\\ &&
\times (-1) \left[ {1 \over \epsilon^2} + {1 - 2 \log 2 \over \epsilon} + 10 - 2 \log 2 
\right. \nonumber
\\
&& \hspace{0cm} \left. \hspace{32mm}
	+ 2 \log^2 2 - {7 \pi^2 \over 12} \right] \;,
\label{int4:5}
\\ && \hspace{-7mm} \nonumber
\int_Q {1 \over Q^2 r^2 R^2}
        \bigg|_{P_0 = -i p} =
%\\ &&
{1 \over (4 \pi)^2} \mu^{2 \epsilon} p^{-2 -2 \epsilon}
\; \left[ {1 \over \epsilon} - 2 - 2 \log 2 \right] \,.
\label{int4:6} \\
\eqa
%
%[{\bf jmp}]
We also need a weighted average over $c$ of the integral in (\ref{int4:4})
evaluated at $P = (-i p, {\bf p}/c)$.  The integral itself is
\bqa
&& \hspace{-13mm} \int_Q {q^2 \over Q^2 R^2}
        \bigg|_{P \to (-i p, {\bf p}/c)} =
{1 \over (4 \pi)^2} (e^\gamma \mu^2)^\epsilon p^{2-2\epsilon}
{(1)_\epsilon \over \epsilon}
\nonumber
\\
&& 
\hspace{-9mm}
\times {1\over 4}
{(1)_{-\epsilon} (1)_{-\epsilon}
        \over (2)_{-2\epsilon}}
\left( {1 \over 3 - 2 \epsilon} + c^2 \right)
c^{-2 + 2\epsilon} (1-c^2)^{-\epsilon} \,.
\label{int4:7}
\eqa
%
%[{\bf jmp}]
The weighted averages are
\bqa\nonumber
&& \hspace{-12mm}
\left\langle c^{1+2\epsilon}
        \int_Q {q^2 \over Q^2 R^2} \bigg|_{P \to (-i p,{\bf p}/c)}
        \right\ranglec
=
{1 \over (4 \pi)^2} \mu^{2 \epsilon} p^{2-2\epsilon} \,
\\ && \hspace{4mm} 
\times {1 \over 48}
\left[ {1 \over \epsilon^2} + {2 (10+3\log 2) \over 3\epsilon}
\right.
\nonumber
\\
&& \hspace{8mm} \left.
        + {4\over 9} + {40\over 3}\log 2 + 2 \log^2 2
        + {3 \pi^2 \over 4} \right]
\;.
\label{int4:7a}
\\\nonumber
&& \hspace{-12mm}
\left\langle c^{-1+2\epsilon}
        \int_Q {q^2 \over Q^2 R^2} \bigg|_{P \to (-i p,{\bf p}/c)}
        \right\ranglec =
	{1 \over (4 \pi)^2} \mu^{2 \epsilon} p^{2-2\epsilon} \,
\\ && \hspace{4mm}
\times {1 \over 16}
\left[ {1 \over \epsilon^2} + {2\log 2 \over \epsilon}
        + 2\log^22 + {3 \pi^2 \over 4} \right]
\;.
\label{int4:7aa}
\eqa
%
%{\bf [jm] p mod last}

The most difficult 4-dimensional integrals to evaluate
involve an HTL average of an integral
with denominator $R_0^2 + r^2 c^2$:
\bqa\nonumber
&& \hspace{-9mm}
{\rm Re} \int_Q {1 \over Q^2}
\left\langle {c^2 \over R_0^2 + r^2 c^2} \right\ranglec
=
{1 \over (4 \pi)^2} \mu^{2\epsilon} p^{-2 \epsilon}
\\ && \hspace{-3mm}
\times \left[ {2 - 2\log 2 \over \epsilon}
+ 8 - 4 \log 2 + 4 \log^2 2
        - {\pi^2 \over 2} \right] \;,
\label{int4HTL:1}
\\\nonumber&&
\hspace{-9mm} {\rm Re} \int_Q {1 \over Q^2}
\left\langle {c^2(1-c^2) \over R_0^2 + r^2 c^2} \right\ranglec
{1 \over (4 \pi)^2} \mu^{2\epsilon} p^{-2 \epsilon}
\\ && \hspace{27mm} =
{1 \over 3}
\left[ {1 \over \epsilon} + {20\over3}  -6 \log 2 \right] \, ,
\\ \nonumber&& \hspace{-9mm}
{\rm Re} \int_Q {1 \over Q^2}
\left\langle {c^4 \over R_0^2 + r^2 c^2} \right\ranglec
=
{1 \over (4 \pi)^2} \mu^{2\epsilon} p^{-2 \epsilon}
\\ &&
\hspace{-6mm} \times\left[ {5 - 6\log 2 \over 3 \epsilon}
+ {52 \over 9} - 2 \log 2 + 4 \log^2 2
        - {\pi^2 \over 2} \right] \;,
\label{int4HTL:2}
\\ \nonumber&& \hspace{-9mm}
{\rm Re} \int_Q {1 \over Q^2 r^2}
\left\langle {c^2 \over R_0^2 + r^2 c^2} \right\ranglec
=
{1 \over (4 \pi)^2} \mu^{2\epsilon} p^{-2-2 \epsilon}
\\ &&\hspace{19mm} \times\left( - {1 \over 4} \right)
\left[ {1 \over \epsilon} + {4\over 3} + {2\over3} \log 2 \right] \, ,
\label{int4HTL:3}
\\\nonumber&&\hspace{-9mm}
{\rm Re} \int_Q {q^2 \over Q^2 r^2}
\left\langle {c^2 \over R_0^2 + r^2 c^2} \right\ranglec
=
{1 \over (4 \pi)^2} \mu^{2\epsilon} p^{-2 \epsilon}
\\&&\nonumber
\hspace{-6mm} \times \left[ {13-16\log2 \over 12 \epsilon}
        + {29 \over 9} - {19\over18} \log2
                + {8\over3}\log^22  
%\right.\\ && \left.
- {4\over9} \pi^2 \right] \, , \\
\label{int4HTL:4}
\\ &&\hspace{-9mm}
\left\langle\int_{Q}{q^2-p^2\over Q^2r^2(R_0^2+r^2c^2)}
\right\rangle_c
={1\over(4\pi)^2}\mu^{2\epsilon} p^{-2 \epsilon}
\left[-{\pi^2\over3}\right]\;.
\label{last4d}
\eqa
%
%[{\bf jmp}]
The analytic continuation to $P_0 =-ip+\varepsilon$
is implied in these integrals and in all the 4-dimensional integrals
in the remainder of this subsection.

We proceed to describe the evaluation of the integrals
(\ref{int4HTL:1}) and (\ref{int4HTL:2}).
The integral over $Q_0$ can be evaluated
by introducing a Feynman parameter to combine $Q^2$
and $R_0^2 + r^2 c^2$ into a single denominator:
\bqa
&&\hspace{-7mm}\int_Q {1 \over Q^2 (R_0^2 + r^2 c^2)}
= {1\over4} \int_0^1 dx
\int_{\bf r}\left[ (1-x+xc^2) r^2 
\right.
\nonumber\\&& \hspace{7mm} 
\left.
+ 2(1-x) {\bf r} \!\cdot\! {\bf p}
        + (1-x)^2 p^2 - i \varepsilon \right]^{-3/2} ,
\label{fp:1}
\eqa
%[{\bf jmp}]
%
where we have carried out the analytic continuation to
$P_0 =-ip+\varepsilon$.
Integrating over ${\bf r}$
and then over the Feynman parameter,
we get a ${}_2F_1$ hypergeometric function with argument $1-c^2$:
\bqa
&& \hspace{-6.5mm} \int_Q {1 \over Q^2 (R_0^2 + r^2 c^2)}
\;=\; {1\over (4\pi)^2} (e^\gamma \mu^2)^\epsilon
        p^{-2\epsilon} {(1)_\epsilon \over \epsilon}
\nonumber
\\ 
&& \hspace{-5mm} \times
e^{i \pi \epsilon} {(1)_{-2\epsilon} (1)_{-\epsilon} \over (2)_{-3\epsilon}}
        (1-c^2)^{-\epsilon}
%\\ && \times  
      	F\left( { {3\over2}-2\epsilon , 1-\epsilon
                \atop 2-3\epsilon } \Bigg| 1-c^2 \right) \,. \nonumber
		\\
\label{int4HTL:12Q}
\eqa
%
%[{\bf jmp}]
The subsequent weighted averages over $c$
give ${}_3F_2$ hypergeometric functions
with argument $1$:
\bqa\nonumber
&& \hspace{-4mm} \int_Q {1 \over Q^2}
\left\langle {c^2 \over R_0^2 + r^2 c^2} \right\ranglec
=
{1 \over (4\pi)^2} (e^\gamma \mu^2)^\epsilon
        p^{-2\epsilon} {(1)_\epsilon \over \epsilon}
\\\nonumber
&& \times
{1 \over 3} e^{i \pi \epsilon}
{ ({3\over2})_{-\epsilon} (1)_{-2\epsilon} (1)_{-2\epsilon}
        \over ({5\over2})_{- 2\epsilon} (2)_{-3\epsilon} }
F\left({ 1-2\epsilon , {3\over2}-2\epsilon , 1-\epsilon
        \atop {5\over2}-2\epsilon , 2-3\epsilon } \Bigg| 1 \right) ,
\\
\\ \nonumber&&\hspace{-4mm}
\int_Q {1 \over Q^2}
\left\langle {c^2 (1-c^2) \over R_0^2 + r^2 c^2} \right\ranglec
=
{1 \over (4\pi)^2} (e^\gamma \mu^2)^\epsilon
        p^{-2\epsilon}  {(1)_\epsilon \over \epsilon}
\\\nonumber
&& \times
{2 \over 15} e^{i \pi \epsilon}
{ ({3\over2})_{-\epsilon} (1)_{-2\epsilon} (2)_{-2\epsilon}
        \over ({7\over2})_{- 2\epsilon} (2)_{-3\epsilon} }
F\left( { 2-2\epsilon \;\;{3\over2}-2\epsilon , 1-\epsilon
        \atop {7\over2}-2\epsilon , 2-3\epsilon } \Bigg| 1 \right) .
\\
\eqa
%
%[{\bf jmp}]
After expanding in powers of $\epsilon$, the real part
is (\ref{int4HTL:2}).

The integral (\ref{int4HTL:3})
has a factor of $1/r^2$ in the integrand.
After using (\ref{fp:1}), it is convenient to use a
second Feynman parameter to combine $(1-x+xc^2)r^2$
with the other denominator before integrating over ${\bf r}$:
\bqa
&& \hspace{-3mm} \int_Q {1 \over Q^2 r^2 (R_0^2 + r^2 c^2)}
\;= \; {3\over8} \int_0^1 dx \, (1-x+xc^2) \int_0^1 dy \, y^{1/2}
\nonumber
\\ \nonumber
&& \hspace{4mm}
\times\int_{\bf r}\left[ (1-x+xc^2) r^2 
+ 2y(1-x) {\bf r} \!\cdot\! {\bf p}
\right.\\ && \left. \hspace{32mm}
        + y(1-x)^2 p^2 - i \varepsilon \right]^{-5/2} \;.
\label{fp-2}
\eqa
%
%[{\bf jmp}]
After integrating over ${\bf r}$ and then $y$, we obtain
${}_2F_1$ hypergeometric functions with arguments $x(1-c^2)$.
The integral over $x$ gives a ${}_2F_1$ hypergeometric function
with argument $1-c^2$:
\bqa
\nonumber&& \hspace{-6mm}
\int_Q {1 \over Q^2 r^2 (R_0^2 + r^2 c^2)}
=
{1\over (4\pi)^2} (e^\gamma \mu^2)^\epsilon
p^{-2-2\epsilon} {(1)_\epsilon \over \epsilon}
\\ && \nonumber \hspace{-3mm} \times
\left\{ {(-{1\over2})_{-\epsilon} (1)_{-\epsilon}
        \over ({1\over2})_{-2\epsilon}}
- {3 \over 2(1+2 \epsilon)} e^{i \pi \epsilon}
{(1)_{-2\epsilon} (1)_{-\epsilon} \over (1)_{-3\epsilon}} (1-c^2)^{-\epsilon}
\right.\\ &&\left. \hspace{24mm} \times
        F\left( { {1\over2}-2\epsilon , -\epsilon
                \atop -3\epsilon } \Bigg| 1-c^2 \right)
        \right\} \;.
\label{int4HTL:3Q}
\eqa
%
%[{\bf jmp}]
After averaging over $c$, we get a hypergeometric functions with argument 1:
\bqa&&
\nonumber\hspace{-5mm}
\int_Q {1 \over Q^2 r^2}
\left\langle {c^2 \over R_0^2 + r^2 c^2} \right\ranglec
=
{1 \over (4\pi)^2} (e^\gamma \mu^2)^\epsilon
p^{-2-2\epsilon}
{(1)_\epsilon \over \epsilon}
\\ && \nonumber
\hspace{-3mm}\times\left\{ {1 \over 3-2\epsilon} \,
        { (-{1\over2})_{-\epsilon} (1)_{-\epsilon}
                \over ({1\over2})_{- 2\epsilon} }
\;-\; {1 \over 2} e^{i \pi \epsilon} \,
{ (-{1\over2})_{-\epsilon} (1)_{-2\epsilon} (2)_{-2\epsilon}
        \over ({5\over2})_{- 2\epsilon} (1)_{-3\epsilon} }
\right.\\ && \left.
\hspace{19mm}\times F \left( { 1-2\epsilon , {1\over2}-2\epsilon , -\epsilon
        \atop {5\over2}-2\epsilon , -3\epsilon } \Bigg| 1 \right)
\right\} \;.
\label{int4HTL:3Qc}\\ &&\nonumber \hspace{-5mm}
\int_Q {1 \over Q^2 r^2}
\left\langle {1\over R_0^2 + r^2 c^2} \right\ranglec
=
{1 \over (4\pi)^2} (e^\gamma \mu^2)^\epsilon
p^{-2-2\epsilon}
{(1)_\epsilon \over \epsilon}
\\ &&\nonumber\hspace{-3mm}\times
\left\{ %{1 \over 3-2\epsilon} \,
        { (-{1\over2})_{-\epsilon} ({3\over2})_{-\epsilon}
                \over ({1\over2})_{- 2\epsilon} }
\;-\; {1 \over 2} e^{i \pi \epsilon} \,
{(1)_{-2\epsilon}^2 (1)_{-\epsilon}
        \over ({3\over2})_{- 2\epsilon} (1)_{-3\epsilon} }
\right.\\ && \left. \hspace{19mm} \times
F \left( { 1-2\epsilon , {1\over2}-2\epsilon , -\epsilon
        \atop {3\over2}-2\epsilon , -3\epsilon } \Bigg| 1 \right)
\right\} \;.
\label{int4HTL:3Qcex}
\eqa
%
%{[\bf jmp] mod 2nd eq}
After expanding in powers of $\epsilon$, the real part
is (\ref{int4HTL:3}).

To evaluate the integral (\ref{int4HTL:4}),
it is convenient to first express it as the sum of 3 integrals
by expanding the factor of $q^2$ in the numerator as
$q^2 = p^2 + 2 {\bf p} \cdot {\bf r} + r^2$:
\bqa\nonumber && \hspace{-15mm}
\int_Q {q^2 \over Q^2 r^2 (R_0^2 + r^2 c^2)}
= \int_Q
\left( {p^2 \over r^2} + 2 {{\bf p} \cdot {\bf r} \over r^2} + 1 \right)
\\ && \hspace{23mm} \times
{1 \over Q^2 (R_0^2 + r^2 c^2)}
 \;.
\eqa
%
%[{\bf jmp}]
To evaluate the integral with ${\bf p} \cdot {\bf r}$ in the numerator,
we first combine the denominators using Feynman
parameters as in (\ref{fp-2}).
After integrating over ${\bf r}$ and then $y$, we obtain
${}_2F_1$ hypergeometric functions with arguments $x(1-c^2)$.
The integral over $x$ gives ${}_2F_1$ hypergeometric functions
with arguments $1-c^2$:
\bqa&&\nonumber
\hspace{-8mm} \int_Q {{\bf p} \cdot {\bf r}
        \over Q^2 r^2 (R_0^2 + r^2 c^2)}
=
{1\over (4\pi)^2} (e^\gamma \mu^2)^\epsilon
p^{-2\epsilon} {(1)_\epsilon \over 2\epsilon^2}
\\ && \nonumber
\times \left\{ - {({3\over2})_{-\epsilon} (1)_{-\epsilon}
        \over ({3\over2})_{-2\epsilon}}
%\right.\nonumber\\&& \left.
+ e^{i \pi \epsilon}
{(1)_{-2\epsilon} (1)_{-\epsilon} \over (1)_{-3\epsilon}} (1-c^2)^{-\epsilon}
\right.\\ &&\left. \hspace{2cm} \times
        F\left( { {3\over2}-2\epsilon , -\epsilon
                \atop 1-3\epsilon } \Bigg| 1-c^2 \right)
        \right\} \;.
\label{int4HTL:5Q}
\eqa
%
%[{\bf jmp}]
After averaging over $c$, we get a hypergeometric function with argument 1:
\bqa&& \nonumber \hspace{-6mm}
\int_Q {{\bf p} \cdot {\bf r} \over Q^2 r^2}
\left\langle {c^2 \over R_0^2 + r^2 c^2} \right\ranglec
=
{1 \over (4\pi)^2} (e^\gamma \mu^2)^\epsilon
p^{-2\epsilon}
{(1)_\epsilon \over 2\epsilon^2}
\\ && \nonumber
\times \left\{ - {1 \over 3-2\epsilon} \,
        { ({3\over2})_{-\epsilon} (1)_{-\epsilon}
                \over ({3\over2})_{- 2\epsilon} }
\;+\; {1 \over 3} e^{i \pi \epsilon}
{ ({3\over2})_{-\epsilon} (1)_{-2\epsilon} (1)_{-2\epsilon}
        \over ({5\over2})_{- 2\epsilon} (1)_{-3\epsilon} }
\right.\\ && \left. \hspace{2cm} \times
F \left( { 1-2\epsilon , {3\over2}-2\epsilon , -\epsilon
        \atop {5\over2}-2\epsilon , 1-3\epsilon } \Bigg| 1 \right)
\right\} \;.
\label{int4HTL:5Qc}
\eqa
%
%{\bf [jmp]}
After expanding in powers of $\epsilon$, the real part is
\bqa 
&& \hspace{-5mm} {\rm Re} \int_Q {{\bf p} \cdot {\bf r} \over Q^2 r^2}
\left\langle {c^2 \over R_0^2 + r^2 c^2} \right\ranglec
=
{1 \over (4 \pi)^2} \mu^{2\epsilon} p^{-2 \epsilon}
\left[ {-1 + \log 2 \over 3\epsilon}
\right.
\nonumber
\\
&& \left. \hspace{15mm}
-{20\over9} + {14\over 9} \log2 -{2\over3} \log^22
        + {\pi^2\over 36} \right] \;.
\label{int4HTL:5}
\eqa
%
%[{\bf jmp}]
Combining this with (\ref{int4HTL:1}) and (\ref{int4HTL:2}),
we obtain the integral (\ref{int4HTL:4}).

To evaluate the integral~(\ref{last4d}), we first express the numerator
as a sum of two integrals whose averages have been calculated:
\bqa\nonumber
&&\hspace{-5mm} \left\langle\int_{Q}{q^2-p^2\over Q^2r^2(R_0^2+r^2c^2)}
\right\rangle_x
=
\left\langle\int_{Q}{2{\bf p}\cdot{\bf r}+r^2\over Q^2r^2(R_0^2+r^2c^2)} 
\right\rangle_x \\ \nonumber
&& \hspace{-2mm}=
{1 \over (4\pi)^2} (e^\gamma \mu^2)^\epsilon
p^{-2\epsilon}
{(1)_\epsilon \over \epsilon}
\left\{ -{1\over\epsilon}
        { ({3\over2})_{-\epsilon} (1)_{-\epsilon}
                \over({3\over2})_{- 2\epsilon}}
\right.
\\ \nonumber
&& \left.
+\, e^{i \pi \epsilon} \,
{ (1)_{-\epsilon} (1)_{-2\epsilon}
        \over(1)_{-3\epsilon} }{1\over\epsilon}
(1-c^2)^{-\epsilon}F \left( { -\epsilon , {3\over2}-2\epsilon
        \atop 1-3\epsilon } \Bigg| 1-c^2 \right)
\right.\\ \nonumber
&&
\left.
+\, e^{i \pi \epsilon} \,
{ (1)_{-\epsilon} (1)_{-2\epsilon}
        \over(2)_{-3\epsilon} }
(1-c^2)^{-\epsilon}
\right.\\ && \left. \hspace{22mm} \times
F \left( { 1-\epsilon , {3\over2}-2\epsilon
        \atop 2-3\epsilon } \Bigg| 1-c^2 \right)
\right\}\;.
\eqa
%{\bf [jmp]}
The two hypergeometric functions are now combined into a single
hypergeometric functions, which yields
\bqa\nonumber
\left\langle\int_{Q}{2{\bf p}\cdot{\bf r}+r^2\over Q^2r^2(R_0^2+r^2c^2)} 
\right\rangle_x 
&=&
{1 \over (4\pi)^2} (e^\gamma \mu^2)^\epsilon
p^{-2\epsilon}
{(1)_\epsilon \over \epsilon^2}
\\ && \nonumber
\hspace{-2.5cm}
\times\left\{ -%{1\over\epsilon}
        { ({3\over2})_{-\epsilon} (1)_{-\epsilon}
                \over ({3\over2})_{- 2\epsilon} }
%\right.\\ \nonumber&&\left.\hspace{-3cm}
+e^{i \pi \epsilon} \,
{ (1)_{-\epsilon} (2)_{-2\epsilon}
        \over(2)_{-3\epsilon}}%{1\over\epsilon}
(1-c^2)^{-\epsilon}
\right.
\\ &&\left.
\hspace{-1.5cm}
\times F \left( { -\epsilon , {3\over2}-2\epsilon
        \atop 2-3\epsilon } \Bigg| 1-c^2 \right) \right\}
	\, .
\eqa
%{\bf [jm]}
Averaging over $c$, yields
\bqa\nonumber
&& \hspace{-13mm}
\left\langle\int_{Q}{2{\bf p}\cdot{\bf r}+r^2\over Q^2r^2(R_0^2+r^2c^2)} 
\right\rangle_{c,x} 
=
{1 \over (4\pi)^2} (e^\gamma \mu^2)^\epsilon
p^{-2\epsilon}
\\ &&\times \hspace{1mm}
{1\over\epsilon^2}
{(1)_\epsilon(1)_{-\epsilon}({3\over2})_{-\epsilon} \over
({3\over2})_{-2\epsilon}}
\left[ -1
+e^{i \pi \epsilon}{(1)_{-2\epsilon}\over(1)_{-\epsilon}^2} 
\right]\;.
\eqa
%{\bf [jmp]}
Expansion in powers of $\epsilon$, yields Eq.~(\ref{last4d}).

\subsection{Hypergeometric functions}
\label{app:hyper}

The generalized hypergeometric function of type $_pF_q$
is an analytic function of one variable with $p+q$ parameters.
In our case, the parameters are functions of $\epsilon$,
so the list of parameters sometimes gets lengthy and the standard notation
for these functions becomes cumbersome.  We therefore introduce a more
concise notation:
\begin{equation}
F\left( { \alpha_1,\alpha_2,\ldots,\alpha_p
        \atop \beta_1,\ldots,\beta_q } \Bigg| z \right)
\;\equiv\;
{}_pF_q(\alpha_1,\alpha_2,\ldots,\alpha_p;\beta_1,\ldots,\beta_q;z) \;.
\end{equation}
%{\bf [jmp] }
%
The generalized hypergeometric function has a power series representation:
\begin{equation}
F\left( { \alpha_1,\alpha_2,\ldots,\alpha_p
        \atop \beta_1,\ldots,\beta_q } \Bigg| z \right)
\;=\; \sum_{n=0}^\infty{ (\alpha_1)_n (\alpha_2)_n \cdots (\alpha_p)_n
        \over (\beta_1)_n \cdots (\beta_q)_n n! } z^n
        \, ,
\label{ps-pFq}
\end{equation}
%
%{\bf [jmp]}
where $(a)_b$ is Pochhammer's symbol:
\begin{equation}
(a)_b = {\Gamma(a+b) \over \Gamma(a)} \,.
\label{Poch}
\end{equation}
%
%{\bf [jmp]}
The power series converges for $|z|<1$.
For $z=1$, it converges if ${\rm Re} s > 0$, where
\begin{equation}
s\;=\; \sum_{i=1}^{p-1} \beta_i -  \sum_{i=1}^p \alpha_i\;.
\label{s-def}
\end{equation}
%
%{\bf [pm] }
The hypergeometric function of type $_{p+1}F_{q+1}$
has an integral representation in terms of the hypergeometric function
of type $_pF_q$:
\bqa
&&\nonumber\hspace{-12mm}
\int_0^1 dt \, t^{\nu-1} (1-t)^{\mu-1} \,
F\left( { \alpha_1,\alpha_2,\ldots,\alpha_p
        \atop \beta_1,\ldots,\beta_q  } \Bigg| tz \right)
\\ &&
\hspace{1mm} =\, { \Gamma(\mu) \Gamma(\nu) \over \Gamma(\mu+\nu)} \,
F\left( { \alpha_1,\alpha_2,\ldots,\alpha_p,\nu
        \atop \beta_1,\ldots,\beta_q,\mu+\nu} \Bigg| z \right) \;.
\label{int-pFq}
\eqa
%
%{\bf [jmp] }
If a hypergeometric function has an upper and lower parameter that are
equal, both parameters can be deleted:
\begin{equation}
F\left( { \alpha_1,\alpha_2,\ldots,\alpha_p,\nu
        \atop \beta_1,\ldots,\beta_q, \nu} \Bigg| z \right)
\;=\; F\left( { \alpha_1,\alpha_2,\ldots,\alpha_p
        \atop \beta_1,\ldots,\beta_q } \Bigg| z \right) \;.
\end{equation}
%
%{\bf [jmp] }

The simplest hypergeometric function is the one of type $_1F_0$.
It can be expressed in an analytic form:
\begin{equation}
{}_1F_0(\alpha; \, ;z) \;=\; (1-z)^{-\alpha} \;.
\end{equation}
%
%{\bf [jmp] }
The next simplest hypergeometric functions are those of type $_2F_1$.
They satisfy transformation formulas that allow an $_2F_1$
with argument $z$ to be expressed in terms of an $_2F_1$
with argument $z/(z-1)$ or as a sum of two $_2F_1$'s
with arguments $1-z$ or $1/z$ or $1/(1-z)$.
The hypergeometric functions of type $_2F_1$
with argument $z=1$ can be evaluated analytically in terms of gamma
functions:
\begin{equation}
F\left( { \alpha_1, \alpha_2 \atop \beta_1 } \Bigg| 1 \right)
\;=\; { \Gamma(\beta_1) \Gamma(\beta_1 - \alpha_1 - \alpha_2)
        \over \Gamma(\beta_1 - \alpha_1) \Gamma(\beta_1 - \alpha_2) } \;.
\label{2F1-1}
\end{equation}
%
%{\bf [jmp] }
The hypergeometric function of type $_3F_2$
with argument $z=1$ can be expressed as a $_3F_2$
with argument $z=1$ and different parameters \cite{3F2}:
\bqa\nonumber
&&\hspace{-1cm} F\left( { \alpha_1, \alpha_2, \alpha_3 \atop \beta_1, \beta_2 } \Bigg| 1 \right)
\,=\, { \Gamma(\beta_1) \Gamma(\beta_2) \Gamma(s)
        \over \Gamma(\alpha_1+s) \Gamma(\alpha_2+s) \Gamma(\alpha_3)} \,
\\ &&
\hspace{14mm}
\times
F\left( { \beta_1-\alpha_3, \beta_2-\alpha_3, s
        \atop \alpha_1+s, \alpha_2+s } \Bigg| 1 \right) \;,
\label{3F2-1}
\eqa
%
%{\bf [mp]}
where $s = \beta_1 + \beta_2 - \alpha_1 - \alpha_2 - \alpha_3$.
If all the parameters of a $_3F_2$ are integers and half-odd-integers,
this identity can be used to obtain
equal numbers of half-odd-integers among the upper and lower parameters.
If the parameters of a $_3F_2$
reduce to integers and half-odd-integers in the limit $\epsilon \to 0$ ,
the use of this identity simplifies the expansion of the
hypergeometric functions in powers of $\epsilon$ .

The  most important integration formulas involving $_2F_1$
hypergeometric functions is (\ref{int-pFq}) with $p=2$ and $q=1$.
Another useful integration formula is
\begin{eqnarray}
&&\nonumber\hspace{-8mm}
\int_0^1 dt \, t^{\nu-1} (1-t)^{\mu-1} \,
F\left( { \alpha_1,\alpha_2
        \atop \beta_1 } \Bigg| {t \over 1-t} z \right)
\\ &&\nonumber \hspace{-4mm}
=
 { \Gamma(\mu) \Gamma(\nu) \over \Gamma(\mu+\nu)} \,
F\left( { \alpha_1,\alpha_2,\nu
        \atop \beta_1,1-\mu } \Bigg| -z \right)
\\ \nonumber
&& \hspace{0cm}
\;+\;
{ \Gamma(\alpha_1+\mu) \Gamma(\alpha_2+\mu) \Gamma(\beta_1) \Gamma(-\mu)
        \over \Gamma(\alpha_1) \Gamma(\alpha_2) \Gamma(\beta_1+\mu) } \,
(-z)^\mu \, 
\\ && \hspace{7mm} \times
F\left( { \alpha_1+\mu,\alpha_2+\mu,\nu+\mu
        \atop \beta_1+\mu,1+\mu} \Bigg| -z \right) \;.
\label{int-2F1}
\end{eqnarray}
%
%{\bf [pm]}
This is derived by first inserting the integral representation
for $_2F_1$ in (\ref{int-pFq}) with integration variable $t'$ and then
evaluating the integral over $t$ to get a $_2F_1$ with argument
$1+t'z$.  After using a transformation formula to change
the argument to $-t'z$, the remaining integrals over $t'$ are evaluated
using (\ref{int-pFq}) to get $_3F_2$'s with arguments $-z$.

For the calculation of two-loop thermal integrals involving HTL averages,
we require the expansion in powers of $\epsilon$
for hypergeometric functions of type $_pF_{p-1}$ with argument 1
and parameters that are linear in $\epsilon$.
If the power series representation (\ref{ps-pFq})
of the hypergeometric function is convergent at $z=1$ for $\epsilon=0$,
this can be accomplished simply by expanding the summand
in powers of $\epsilon$ and then evaluating the sums.
If the power series is divergent, we must make subtractions
on the sum before expanding in powers of $\epsilon$.
The convergence properties of the power series at $z=1$
is determined by the variable $s$ defined in (\ref{s-def}).
If $s>0$, the power series converges.
If $s\to 0$ in the limit $\epsilon \to 0$,
only one subtraction is necessary to make the sum convergent:
\begin{eqnarray}
&& \hspace{-2mm} F\left( { \alpha_1,\alpha_2,\ldots,\alpha_p
        \atop \beta_1,\ldots,\beta_{p-1} } \Bigg| 1 \right)
\;=\;
{ \Gamma(\beta_1) \cdots \Gamma(\beta_{p-1}) \over
        \Gamma(\alpha_1) \Gamma(\alpha_2) \cdots \Gamma(\alpha_p) }
\zeta(s+1)
\nonumber
\\\nonumber
&&\;+\; \sum_{n=0}^\infty
\left( { (\alpha_1)_n (\alpha_2)_n \cdots (\alpha_p)_n
        \over (\beta_1)_n \cdots (\beta_q)_n n! }
\right.\\ &&\left. \hspace{12mm}
- { \Gamma(\beta_1) \cdots \Gamma(\beta_{p-1}) \over
        \Gamma(\alpha_1) \Gamma(\alpha_2) \cdots \Gamma(\alpha_p)}
        (n+1)^{-s-1} \right)
        \;.
\end{eqnarray}
If $s\to -1$ in the limit $\epsilon \to 0$,
two subtractions are necessary to make the sum convergent:
\begin{eqnarray}
&&\nonumber\hspace{-3mm}
F\left( { \alpha_1,\alpha_2,\ldots,\alpha_p
        \atop \beta_1,\ldots,\beta_{p-1} } \Bigg| 1 \right)
=
{ \Gamma(\beta_1) \cdots \Gamma(\beta_{p-1}) \over
        \Gamma(\alpha_1) \Gamma(\alpha_2) \cdots \Gamma(\alpha_p)}
\nonumber\\&& \hspace{-1mm}\times
\left[ \zeta(s+1) + t \, \zeta(s+2) \right]
\;+\; \sum_{n=0}^\infty
\left( { (\alpha_1)_n (\alpha_2)_n \cdots (\alpha_p)_n
        \over (\beta_1)_n \cdots (\beta_q)_n n! } \right.
\nonumber
\\ \nonumber
&& \left. - { \Gamma(\beta_1) \cdots \Gamma(\beta_{p-1}) \over
        \Gamma(\alpha_1) \Gamma(\alpha_2) \cdots \Gamma(\alpha_p)}
\left[  (n+1)^{-s-1} 
%\right.\right.\\ &&\left.\left.
+ t \, (n+1)^{-s-2} \right]
\right) \, , \\
\end{eqnarray}
where $t$ is given by
\begin{equation}
t \;=\; \sum_{i=1}^p {(\alpha_i-1)(\alpha_i-2)\over2}
- \sum_{i=1}^{p-1} {(\beta_i-1)(\beta_i-2) \over2}  \;.
\end{equation}

The expansion of a $_pF_{p-1}$ hypergeometric function
in powers of $\epsilon$ is particularly simple
if in the limit $\epsilon \to 0$
all its parameters are integers or half-odd-integers, with equal numbers
of half-odd-integers among the upper and lower parameters.
If the power series representation for such a hypergeometric function
is expanded in powers of $\epsilon$, the terms in the summand will
be rational functions of $n$, possibly multiplied by
factors of the polylogarithm function $\psi(n+a)$ or its derivatives.
The terms  in the sums can often be simplified by using the obvious identity
\begin{equation}
\sum_{n=0}^\infty \left[ f(n) - f(n+k) \right]
\;=\; \sum_{i=0}^{k-1} f(i) \;.
\end{equation}
%
%{\bf [jmp]}
The sums over $n$ of rational functions of $n$ can be evaluated
by applying the partial fraction decomposition and then
using identities such as
\begin{eqnarray}
\sum_{n=0}^\infty \left({1 \over  n+a} - {1\over n+b} \right)
&=& \psi(b) - \psi(a)  \;,
\\
\sum_{n=0}^\infty {1 \over (n+a)^2} &=& \psi'(a) \;.
\end{eqnarray}
%{\bf [jmp] }
%
The sums of polygamma functions of $n+1$ or $n+{1\over2}$
divided by $n+1$ or $n+{1\over2}$ can be evaluated using
\begin{eqnarray} 
&&\hspace{-2mm}\nonumber
\sum_{n=0}^\infty
\left( {\psi(n+1) \over n+1}
        - {\log(n+1) \over n+1} \right)
= - {1\over2} \gamma^2 - {\pi^2 \over 12} 
- \gamma_1 \;,
\\ &&
\\ \nonumber
&&\hspace{-2mm}
\sum_{n=0}^\infty
\left( {\psi(n+1) \over n+{1\over 2}}
        - {\log(n+1) \over n+1} \right)
=  - {1\over2} (\gamma + 2 \log 2)^2 
\\ &&
\hspace{54mm} + {\pi^2 \over 12} - \gamma_1 \;,
\\ \nonumber
&&\hspace{-2mm}\sum_{n=0}^\infty
\left( {\psi(n+{1\over2}) \over n+1}
        - {\log(n+1) \over n+1} \right)
= - {1\over2} \gamma^2 - 4 \log 2 + 2 \log^2 2
\\ && \hspace{54mm}
        - {\pi^2 \over 12} - \gamma_1 \;,
\\ \nonumber
&&\hspace{-2mm}\sum_{n=0}^\infty
\left( {\psi(n+{1\over2}) \over n+{1\over 2}}
        - {\log(n+1) \over n+1} \right)
= 
 - {1\over2} (\gamma + 2 \log 2)^2 
\\ && \hspace{54mm}
- {\pi^2 \over 4} - \gamma_1 \;,
\end{eqnarray}
%
%{\bf [p]}
where $\gamma_1$ is Stieltje's first gamma constant
defined in (\ref{zeta}).
The sums of polygamma functions of $n+1$ or $n+{1\over2}$
can be evaluated using
\begin{eqnarray}
&&\nonumber\hspace{-1cm}
\sum_{n=0}^\infty
\left( \psi(n+1) - \log(n+1)  + {1 \over 2(n+1)} \right)
\\ &&
\hspace{24mm} = {1\over2} + {1\over2} \gamma -{1\over 2} \log(2 \pi) \;,
\\&&
\nonumber\hspace{-1cm}
\sum_{n=0}^\infty
\left( \psi(n+\mbox{$1\over2$}) - \log(n+1)  + {1 \over n+1} \right)
\\&&
\hspace{2cm} = {1\over2}\gamma  - \log2 -{1\over 2} \log(2 \pi) \,.
\end{eqnarray}

We also need the expansions in $\epsilon$
of some integrals of $_2F_1$ hypergeometric functions of $y$
that have a factor of $|1-2y|$.  For example,
the following 2 integrals are needed to obtain (\ref{avecx:4}):
\begin{eqnarray}
&& \nonumber \hspace{-3mm}
\int_0^1 dy \, y^{-2\epsilon} (1-y)^{1+\epsilon} |1-2y| \,
F\left( { 1-\epsilon,\epsilon
        \atop -3\epsilon} \Bigg| y \right) 
\\ && \hspace{26mm}
= {1\over 6} + \left( {2\over9} + {4\over9} \log 2 \right) \epsilon  \;,
\label{Fabs-1}
\\\nonumber
&& \hspace{-3mm} \int_0^1 dy \, y^{1+\epsilon} (1-y)^{1+\epsilon} |1-2y| \,
F\left( { 2+2\epsilon,1+\epsilon
        \atop 2+3\epsilon} \Bigg| y \right) 
\\&& 
\hspace{24mm} = {1\over 4} + \left( {7\over12} + {2\over3} \log 2 \right) \epsilon  \;.
\label{Fabs-2}
\end{eqnarray}
These integrals can be evaluated by expressing them in the form
\begin{eqnarray}
&&\nonumber\hspace{-7mm}\int_0^1 dy \, y^{\nu-1} (1-y)^{\mu-1} |1-2y| \,
F\left( { \alpha_1,\alpha_2
        \atop \beta_1} \Bigg| y \right) 
\\ && \nonumber
\hspace{-3mm}=\, \int_0^1 dy \, y^{\nu-1} (1-y)^{\mu-1}  (2y-1) \,
F\left( { \alpha_1,\alpha_2
        \atop \beta_1} \Bigg| y \right)
\\ \nonumber
&& \hspace{-2mm}
\;+\; 2 \int_0^{1\over2} dy \, y^{\nu-1} (1-y)^{\mu-1}  (1-2y) \;
F\left( { \alpha_1,\alpha_2
        \atop \beta_1} \Bigg| y \right)  \;.
\\ &&
\end{eqnarray}
%
%{\bf [p]}
The evaluation of the first integral on the right side
gives $_3F_2$ hypergeometric functions with argument 1.
The integrals from 0 to {$1\over2$ can be evaluated by expanding
the power series representation (\ref{ps-pFq})
of the hypergeometric function in powers of $\epsilon$.
The resulting series can be summed analytically
and then the integral over $y$ can be evaluated.

\end{document}